%% file: ms.tex
\documentclass[fleqn,usenatbib]{mnras}
\usepackage{newtxtext,newtxmath}
\usepackage[normalem]{ulem}
\usepackage[T1]{fontenc}
\usepackage{ae,aecompl}
\usepackage[x11names, rgb]{xcolor}
\usepackage[utf8]{inputenc}
\usepackage{fleqn}
\usepackage[T1]{fontenc}
\usepackage{ae,aecompl}
\usepackage{color}
\usepackage{pdflscape}
\usepackage{multirow}
\usepackage{makecell}
\usepackage{tikz}
\usetikzlibrary{snakes,arrows,shapes, chains}
\tikzstyle{format}=[rectangle,draw,thin,fill=white,node distance=1cm]  
\tikzstyle{test}=[diamond,aspect=2,draw,thin,node distance=1cm]  
\tikzstyle{point}=[coordinate,on grid,]

\usepackage{amsmath}	% Advanced maths commands
\usepackage{pdflscape}
\usepackage{multirow}
\usepackage{makecell}
\usepackage{tikz}
\usepackage{color}

\usetikzlibrary{arrows,shapes,chains}  

\newcommand{\deleted}[1]{}
\newcommand{\HI}{\ion{H}{i}}
\newcommand{\HII}{\ion{H}{ii}}
\newcommand{\HeI}{\ion{He}{i}}
\newcommand{\HeII}{\ion{He}{ii}}
\newcommand{\HeIII}{\ion{He}{iii}}
\newcommand{\D}{{\rm d}}

\title[Direct Parameter Inference from Global EoR Signal]
{Direct Parameter Inference from Global EoR Signal with Bayesian Statistics}

\author[J.-H. Gu \& J.-Y. Wang]{
Junhua Gu$^{1}$\thanks{E-mail: jhgu@nao.cas.cn}, Jingying Wang$^{2}$\thanks{E-mail: astro.jywang@gmail.com}
\\
$^{1}$National Astronomical Observatories, Chinese Academy of Sciences, 20A Datun Road, Beijing, China\\
$^{2}$Department of Physics and Astronomy, University of the Western Cape, Cape Town 7535, South Africa
}

\date{Accepted XXX. Received YYY; in original form ZZZ}

\pubyear{2019}

\begin{document}
\label{firstpage}
\pagerange{\pageref{firstpage}--\pageref{lastpage}}
\maketitle

% Abstract of the paper
\begin{abstract}
    In the observation of sky-averaged $\HI$ signal from Epoch of Reionization, model parameter inference can be a computation-intensive work, which makes it hard to perform a direct one-stage model parameter inference by using MCMC sampling method in Bayesian framework. 
    Instead, a two-stage inference is usually used, i.e., the parameters of some characteristic points on the EoR spectrum model are first estimated, which are then used as the input to estimate physical model parameters further. 
    However, some previous works had noticed that this kind of method could bias results, and it could be meaningful to answer the question of whether it is feasible to perform direct one-stage MCMC sampling and obtain unbiased physical model parameter estimations.
    In this work, we studied this problem and confirmed the feasibility. 
    We find that unbiased estimations to physical model parameters can be obtained with a one-stage direct MCMC sampling method. 
    We also study the influence of some factors that should be considered in practical observations to model parameter inference. 
    We find that a very tiny amplifier gain calibration error ($10^{-5}$ relative error) with complex spectral structures can significantly bias the parameter estimation; the frequency-dependent antenna beam and geographical position can also influence the results, so that should be carefully handled.
\end{abstract}

% Select between one and six entries from the list of approved keywords.
% Don't make up new ones.
\begin{keywords}
	methods: numerical -- methods: statistical -- cosmology: observations, dark ages, reionization -- diffuse radiation
\end{keywords}

%%%%%%%%%%%%%%%%%%%%%%%%%%%%%%%%%%%%%%%%%%%%%%%%%%

%%%%%%%%%%%%%%%%% BODY OF PAPER %%%%%%%%%%%%%%%%%%
\section{Introduction}
Detection to the cosmic dawn and the following Epoch of Reionization (EoR hereafter) has become the frontier of radio astronomy and observational cosmology.
It is crucial to answer questions such as what sources are responsible for the reionization, how the neutral hydrogen fraction evolved during EoR, how the mass accretion onto dark halos modulates the ionization process, etc. 
The redshifted neutral hydrogen 21 cm signals from the early universe are generally regarded to bring critical information to answer the above questions.
Compared with abundant observations to the last scattering surface (i.e., the farthest universe we can see throw the electromagnetic window) and low redshift (e.g., $z\lesssim6$) universe, much fewer observational evidence about the cosmic dawn and EoR exists by now, which is necessary to discriminate candidate models and constrain cosmological model parameters.
In order to fill this gap, the new generation of radio telescopes have been built or being constructed.
The very long list of these new radio telescopes includes LOFAR \citep{2013A&A...556A...2V}, MWA \citep{2013PASA...30....7T}, 21CMA \citep[e.g., ][]{2013ApJ...763...90W, 2016ApJ...832..190Z, 2016RAA....16...36H}, HERA \citep[][]{2017PASP..129d5001D}, PAPER \citep[][]{2010AJ....139.1468P}, LWA \citep{2012JAI.....150004T}, SKA \citep[][]{2004NewAR..48..979C}, SARAS \citep[][]{2013ExA....36..319P}, BIGHORNS \citep[BIGHORNS][]{2015PASA...32....4S}, EDGES \citep[][]{2016MNRAS.455.3890M}, SCI-HI \citep[][]{2014ApJ...782L...9V}, etc. The detection to the $\HI$ 21 cm signal from the cosmic dawn and EoR is extremely hard.
Several conflicting factors should be compromised, including angular resolution, redshift resolution, signal-to-noise ratio, survey speed, the quantity of data to be processed, and accuracy of instrument calibration.
Different paradigms of detection reflecting the concerning to different factors are invented.

The three major paradigms are 1) imaging, 2) measuring power spectrum, and 3) measuring global 21 cm signal (i.e., the monopole component), which have their own problems respectively to solve and can produce different levels of details about cosmic dawn and EoR.
Direct imaging to cosmic dawn and EoR can obtain the most abundant information among all the above three paradigms, but it requires to solve big problems, including direction-dependent ionosphere correction and wide field imaging difficulties.
Interferometers can directly output power spectrum and can seemingly escape from the difficulties of direct imaging paradigm \citep[e.g.,][]{2012ApJ...758L..24Z}.
However, year scale integration time is still needed \citep{2012ApJ...758L..24Z} to suppress the thermal noise.
Measuring global 21 cm signal needs much less integration time to reach a required signal-to-noise ratio.
Direction-dependent ionosphere correction and wide field imaging can be avoided as well.
Furthermore, the instrument for measuring global 21 cm signal can be as simple as a single dipole, so that the response of antenna can be relatively precisely measured in a microwave darkroom (with near-field measuring technique).
Simplicity is prerequisite for reliability.
Though losing a significant amount of details (only monopole component can be measured), especially before other paradigms obtain reliable results, global 21 cm signal can be a promising method for detecting the cosmic dawn and EoR.

The framework of global 21 cm signal studies can be divided into following three tightly related aspects as the global 21 cm signal modeling, statistical inferring based on observational data, and the obtaining of accurate enough observational data.

For the modeling part, it has been widely accepted that the evolution of observed 21 cm signal brightness temperature contrast to CMB is dominated by the competition between the influence of CMB and gas on the atomic hydrogen ($\HI$) spin temperature \citep[e.g., ][]{2006PhR...433..181F}.
The CMB temperature decreases as it adiabatically expands with the universe and follows a relation to redshift $z$ as $T_{\gamma}\propto(1+z)$.
Meanwhile, the gas kinetic temperature is simultaneously modulated by the universe expansion and heating and cooling of a variety of different mechanisms.
The CMB affects spin temperature mainly through Compton scattering, and gas affects spin temperature through collision and Wouthuysen-Field effect \citep{1952AJ.....57R..31W,1958PIRE...46..240F}.
The strength of above three effects are mainly controlled by gas neutral fraction, gas temperature, and background Ly $\alpha$ intensity, which are all tightly connected to the evolving history of potential ionizing sources, i.e., black holes and stars \citep[see ][ for a quantitative discussion]{2014MNRAS.443.1211M, 2012ApJ...756...94M}.

Inferring model parameters from observational data is another critical aspect of global 21 cm signal detection.
Optimization based model-fitting is a traditional method of parameter inferring; however Markov-Chain Monte Carlo (MCMC) based methods are gradually becoming major statistical inference tools and have been widely used in the studies of global 21 cm signal.
A lot of recent works e.g., \cite{2013ApJ...777..118M}, \cite{2015ApJ...813...11M}, \cite{2016MNRAS.455.3829H}, and \cite{2016MNRAS.461.2847B} have used MCMC-based methods as their major methods of statistical inferences.
Comparing to classical optimization-based methods, MCMC-based methods cannot only solve the optimal model parameters but also inspect the probability distribution of model parameters, so that can offer the information about the structure of the model parameter space.
However, there is still one shortage of MCMC-based methods, i.e., most of such kind of methods require evaluating model values for a vast number of times to ensure the Markov chains to reach equilibrium.
Considering the factor that the model evaluation of the global 21 cm signal itself is computation intensive, it requires a large amount of computing resources to perform a direct model evaluation in MCMC methods.
Currently, few work directly evaluate the model predicted values in the step of parameter inference.
At least two methods are utilized to avoid direct model evaluation in MCMC, and the first one is to approximate the global 21 cm signal model with some simple mathematical expressions \citep[e.g., ][modeled the early reionization stage signal with a Gaussian model]{2016MNRAS.461.2847B}, another method is to split the MCMC inference into two stages \citep[e.g., ][]{2013ApJ...777..118M, 2015ApJ...813...11M, 2016MNRAS.455.3829H}.
In these works, they first infer the frequencies and intensities of some characteristic points of the global 21 cm signal spectrum, and then in the second stage use the distribution of these characteristic points to infer the values of physical model parameters.
The pitfall of the expressing the global 21 cm signal model in a simple mathematical formula is that either the formula cannot fully approximate the precise model values, or the formulas would have too many degrees of freedom, as a result, cannot be well constrained and lead to the over-fitting problem.
As to the two-stage method, as \cite{2016MNRAS.455.3829H} mentioned, there exists a risk to obtain biased results.
In the same paper, the authors raised a question of whether it is feasible to perform direct single-stage parameter inference and whether unbiased parameter estimation can be obtained.

The efforts to obtain the well-calibrated observational data forms the last aspect of the study to the global EoR signal detection.
This problem mainly falls into the category of antenna and electronic system design.
The issue about how to design and calibrate electronic system will not be covered in this work, so we are not going to expand this topic here.
However, there are two interesting issues in the early stages of projects that are aiming to detect global 21 cm signals.
The first issue is how accurate the analog frontend should be calibrated and how the calibration errors propagate in the data reduction chain.
The second one is to guide the design of the antenna by evaluating the influence of frequency-dependent antenna pattern (FDAP) through simulation.
These two issues raise for both of them can cause artificial frequency structures in the global 21 cm signal spectrum.
In some previous works, e.g., \cite{2016MNRAS.461.2847B}, the amplifier calibration error is treated to be uncorrelated between different frequency channels, which, however, does not fully reflect its influence on the measured data.
The frequency-dependent antenna beam issue has been mentioned by e.g., \cite{2013PhRvD..87d3002L}, and is also widely considered in the antenna design of recent projects.
For example, in experiments including SARAS \citep[][]{2013ExA....36..319P}, BIGHORNS \citep[BIGHORNS][]{2015PASA...32....4S}, EDGES \citep[][]{2016MNRAS.455.3890M}, and SCI-HI \citep[][]{2014ApJ...782L...9V}, the antennas are specially designed to make the beam pattern less frequency-dependent.
Nevertheless, the effect of frequency-dependent beam cannot be fully canceled, which must be evaluated and considered in actual data reductions.

The motivation of this work is to inspect what information we can extract from the global $\HI$ 21 cm signal in actual observation conditions.
This goal can be further divided into the following two aspects.
The first aspect of this work is to find out the feasibility of inferring physical model parameters from observation data through a direct physics-based modeling method.
We intend to evaluate the physics-based global 21 cm model in each step of the MCMC sampling and to check whether unbiased parameter estimations can be inferred.
The second motivation is to study in practical observations, how will different factors, including thermal noise, instrument calibration errors, and frequency-dependent antenna beam patterns, affect model parameter inference and model comparison.

So in this work, we investigate the feasibility of constraining the parameters of physical global 21 cm signal model with MCMC sampling method.
We will evaluate the influence of uncorrected instrumental effects and frequency-dependent antenna beam patterns on the results and estimate how well the instruments should be calibrated in order to gain meaningful results.
The paper is organized as follows, we explain the modeling method of the reionization processes in section \ref{sec:simu}, describe the MCMC sampling method that we use to constrain the model and the model parameters in section \ref{sec:mcmc_method}, present the results in section \ref{sec:results}, and give a discussion and conclude this work in section \ref{sec:discussion}.
We adopt the $\Lambda$CDM cosmological parameters released by \cite{2016A&A...594A..13P} ($H_0=67.74$ km s$^{-1}$ Mpc$^{-1}$, $\Omega_{\Lambda, 0}=0.3089$, $\Omega_{m, 0}=0.3089$, $\Omega_{b, 0}=0.0486$, $\sigma_8=0.8159$, and $n_s=0.9667$).

\section{Modeling Sky-Averaged Low Frequency Spectrum}
\label{sec:simu}
In this section, we present the global $\HI$ 21 cm signal model of cosmic dawn and EoR used in this work.
We first briefly review the algorithm for computing the global $\HI$ signal based on previous studies.
Then the parameter values for simulating the reference 21 cm signal are presented.
Last, we show the foreground emission spectrum model.

\subsection{Redshifted 21 cm Signal from the EoR}
\label{ssec:eor_modeling}
The computing of the global $\HI$ signal involves three threads, i.e., 1) solving radiative transfer equation (RTE) to determine the background radiation intensity, 2) updating ionization, recombination, cooling and heating rate coefficients to compose the rate equations, and 3) solving the rate equations for each time slice sequentially.
These threads are twisted and complicated.
Even though it is not the central topic of this paper, and has been well and comprehensively studied by many other authors, we still choose to briefly review the modeling method as follows.

To model the global $\HI$ signal in the low-frequency radio band, we follow the method and algorithm that has been comprehensively described by previous works mainly including \cite{2006PhR...433..181F}, \cite{2012ApJ...756...94M}, and \cite{2014MNRAS.443.1211M}.
Among these three works, \cite{2014MNRAS.443.1211M} described the detailed algorithm to calculate the global 21 cm signal, together with some important practical instructions; \cite{2012ApJ...756...94M} described the one dimensional radiative transfer algorithm used by the former work, and the authors released a \textsc{python} package \textsc{ares}\footnote{\url{https://bitbucket.org/mirochaj/ares}}\citep{2014MNRAS.443.1211M,2012ApJ...756...94M}, that has been being actively developed; \cite{2006PhR...433..181F} summarized equations to calculate the physical quantities that are required to derive the global 21 cm signal.
Although the package \textsc{ares} is well developed and is in principle suitable for solving the problem of this work, MCMC sampling that we will perform requires evaluating the model with different parameter sets for a vast number of times, and the \textsc{python}-based package \textit{AREA} becomes a performance bottleneck for us.
As a result, we take the \textsc{ares} code as a reference and re-implement the algorithm in \textsc{rust}\footnote{\url{https://www.rust-lang.org}} with some minor modifications to gain a required high computing efficiency so that in very limited time, we can draw a large enough sample with the MCMC method. As a necessary comment to the code performance, although our code is implemented with \textsc{rust} and compiled to native code, the computational performance is improved by less than one order of magnitude. It is still practical to perform similar MCMC sampling with the \textsc{ares} code.

Followings are the details about the global 21 cm signal modeling algorithm.
The redshifted global 21 cm signal from cosmic dawn and EoR is calculated as 
\begin{gather}
\delta T_b=0.027(1-\bar{x}_i)\left(\frac{\Omega_{b,0}h^2}{0.023}\right)\left(\frac{0.15}{\Omega_{m,0}h^2}\frac{1+z}{10}\right)^{1/2}\left(1-\frac{T_\gamma}{T_{S}}\right)~{\rm K},\label{eqn:HI_Tb}
\end{gather}
where $\bar{x}_i$ is the mean ionized fraction and $T_{S}$ is the spin temperature.
Both $\bar{x}_i$ and $T_{S}$ are to be solved as follows.

We follow the `two component' scenario utilized by e.g., \cite{2014MNRAS.443.1211M} and \cite{2006PhR...433..181F}.
The two-component scenario divides the intergalactic medium (IGM) into the `bulk IGM' that is mostly neutral, and $\HII$ regions that is fully ionized.
The filling factor of $\HII$ region is denoted by $x_i$ and the ionization fraction in the bulk IGM regions is denoted by $x_{\HII}$, so that in above Equation \ref{eqn:HI_Tb}, the average ionization fraction is calculated as $\bar{x}_i=x_i+(1-x_i)x_{\HII}$.
For the bulk IGM regions, the fraction of $\HII$ can be solved from following rate equation.
\begin{gather}
\frac{\D }{\D t}x_{\HII}=(\Gamma_{\HI}+\gamma_{\HI}+\beta_{\HI}n_e)x_{\HI}-\alpha^{B}_{\HII}n_e x_{\HII},\label{eqn:rate_equation}
\end{gather}
where $n_e$ is the electron density and the meanings of other symbols are listed in Table \ref{tbl:quantities}. The temperature of bulk IGM is solved from the equation 
\begin{gather} 
\frac{\D }{\D t} T_{K}=-2H(z)T_K+\frac{2}{3}\frac{\epsilon_X+\epsilon_{\rm comp}-\mathcal{C}}{k_Bn_{\rm tot}}, \label{eqn:heating} \end{gather} 
where $n_{\rm tot}$ is the total number density of particles, $H(z)$ is the Hubble constant at redshift $z$.
$\mathcal{C}$
represents the contribution of all cooling mechanisms, including Hubble cooling, collisional ionization cooling, collisional excitation cooling, and recombination cooling. The coefficients of the latter three cooling mechanisms are calculated with the equations in \cite{1994MNRAS.269..563F}. The rate equation for the filling factors of the $\HII$ region surrounding galaxies is  
\begin{gather}
\frac{\D x_i}{\D t}=f_{*}f_{\rm esc}N_{\rm ion}\bar{n}_{\rm    b}^0\frac{\D f_{\rm coll}}{\D t}(1-x_{\HII})-\alpha^{\rm A}n_ex_iC(z)\label{eqn:HII_rate_equation}
\end{gather} 
where $f_{\rm coll}$ is the fraction of gas in collapsed halos more massive than some minimum mass (see below), $f_{*}$ is the star formation efficiency, $\bar{n}_{b}^0$ is the baryon number density today. We set the clumping factor $C(z)\equiv 1$, same as \cite{2014MNRAS.443.1211M}.
Ionization on the boundary of $\HII$ regions is assumed to be mainly driven by photons with energy between 13.6 eV and 200 eV, the escape fraction of which is denoted by $f_{\rm esc}$.
$N_{\rm ion}$
is the number of photons per baryon particle in star formation and is set to be 4000 \citep[Same as][]{2014MNRAS.443.1211M}. 
The photon ionization coefficient $\Gamma_{\HI}$, ionization rate due to fast secondary electrons $\gamma_{\HI}$, and X-ray heating rate $\epsilon_X$ in Equation \ref{eqn:heating} are determined by background radiation intensity $\hat{J}_{\nu}(z)$ \citep[See Equation 9, 10, and11 of][$\hat{J}_{\nu}$ means the intensity in the unit of photon numbers]{2014MNRAS.443.1211M}, which is obtained from the solution of the cosmological radiative transfer equation \citep[see the Equation 3 in][]{2014MNRAS.443.1211M} as 
\begin{gather}
\hat{J}_\nu(z)=\frac{c}{4\pi}(1+z)^2\int_z^{z_{\rm
            f}}\frac{\hat{\epsilon}_{X,\nu^\prime}(z^\prime)}{H(z^\prime)}e^{-\bar{\tau}_\nu}\D
z^\prime,\label{eqn:J_nu}
\end{gather} 
where the photon depth $\bar{\tau}_{\nu}$ is the mean optical depth (see below), $z_{\rm f}$ is some `first light redshift', when the ionizing sources turn on.
The emissivity $\hat{\epsilon}_{X, \nu}(z)$ (again in the unit of photon numbers) depends on the accretion rate and model of first generation of luminous celestial objects as 
\begin{gather}
\hat{\epsilon}_{X,\nu}(z)=\bar{\rho}_{\rm b}^{0} \frac{C_{X}}{\bar{E}_{\gamma}} f_{*}f_{\rm X}\frac{\D f_{\rm coll}}{\D t}I_{\nu},\label{eqn:X_emissivity}
\end{gather} 
where $\bar{\rho}_{\rm b}^{0}$ is today's baryon matter density, $I_\nu$ is the source spectrum that is normalized as $\int I_\nu \D \nu=1$, $C_{X}=2.6\times10^{39}$ ${\rm erg}~{\rm s}^{-1}~{\rm M}_{\odot}^{-1}~{\rm yr}$ is the energy of X-ray emission produced when unit mass of matter is accreted $\bar{E}_\gamma$ is the mean photon energy within the total band, and $f_{\rm X}$ reflects the uncertainty of $C_{X}$.
In the process of heating and ionization in bulk IGM regions, we follow the treatment of \cite{2014MNRAS.443.1211M} that only X-ray radiation within $0.2-30$ keV is considered and assume the first generation of luminous celestial objects to be accreting black holes, with a power-law spectrum with a fixed slope of $-1.5$.
The energy range of the X-ray responsible for reionization and the slope of the power-law spectrum are both fixed because allowing these parameters to vary will introduce degeneracy in parameter space and make the Markov chain hard to reach an equilibrium.
Studying which kind of source is responsible for the reionization is not the major topic of this paper.
In our future work, we will study the feasibility of discriminating different ionizing sources by using the global $\HI$ signal.
We note that more sophisticated methods have been developed recently \citep[e.g.,][based on luminosity functions]{2018MNRAS.478.5591M,2019MNRAS.484..933P} to model ionizing sources more accurately.
These new modeling methods offer the possibility of inferring more detailed information about the cosmic dawn and EoR, and we will consider including them in our future work.
In this work, we still use the method that we have described above.

In Equation \ref{eqn:J_nu}, the emitting frequency $\nu^\prime$ of a photon emitted at redshift $z^\prime$ that is observed at frequency $\nu$ at redshift $z$ is 
\begin{gather} 
    \nu^\prime=\nu\left(\frac{1+z^\prime}{1+z}\right).
\end{gather}
The optical depth $\bar{\tau}_\nu$ is calculated as 
\begin{gather}
    \bar{\tau}_\nu(z,z^\prime)=\sum_j\int_z^{z^\prime}n_j(z^{\prime\prime})\sigma_{j,\nu^{\prime\prime}}\frac{\D
        l}{\D z^{\prime\prime}}\D z^{\prime\prime}, \label{eqn:tau}
\end{gather} 
where $\D l/\D z=c/H(z)/(1+z)$ is the proper cosmological line element, and $\sigma_{j,\nu}$ is the bound-free absorption cross-section of species $j=\HI, \HeI, \HeII$ with number density $n_j$.
Same as \cite{2014MNRAS.443.1211M}, we approximate $x_{\HeIII}$ to be 0 and assume $x_{\HeII}=x_{\HII}$.
Note that $n_j$ is a function of redshift, which is to be solved in Equation \ref{eqn:rate_equation}.
Fortunately, $n_j$ at any redshift $z_i$ relies only on previous redshift slices with $z>z_j$.
Nevertheless, a strict calculation requires solving Equation \ref{eqn:rate_equation} and Equation \ref{eqn:tau} alternatively.
This issue has been addressed in section 5.1 of \cite{2014MNRAS.443.1211M}. 
In their work and the code \textsc{ares}, a priori ionization history was assumed and used to generate tabulated optical depth $\bar{\tau}$.
In our implementation, we do not use the priori ionization history; instead, we solve Equation \ref{eqn:rate_equation} and  Equation \ref{eqn:tau} in turn.
With the radiation intensity $\hat{J}_\nu$ obtained, we compute the ionization and heating coefficients by using Equations 9-11 in \cite{2014MNRAS.443.1211M}.

In Equations \ref{eqn:HII_rate_equation} and \ref{eqn:X_emissivity}, the coefficients are related to the fraction of gas in collapsed halos more massive than a redshift-dependent minimum mass $m_{\min}(z)$ 
\begin{gather}
    f_{\rm coll}=\rho_{\rm m}^{-1}\int_{m_{\min}}^{\infty}m n(m) dm,\label{eqn:fcoll_vs_z}
\end{gather}
where $\rho_{\rm m}$ is the mean comoving density of the Universe, $n(m)\D m$ is the comoving number density of halos with mass between $m$ and $m+\D m$.
In \cite{2014MNRAS.443.1211M}, $f_{\rm coll}$ is computed with the \textsc{hmf-calc} code \citep{2013A&C.....3...23M}.
In this work, we use an analytic equation 
\begin{gather}
    f_{\rm coll}=\frac{1}{2}\left [1+{\rm erf}\left(\frac{z_0-z}{z_{w}\sqrt{2}}\right)\right],\label{eqn:fcoll_analytic}
\end{gather} 
where the two parameters $z_{0}$ and $z_{w}$ can vary to represent different evolving histories.
So that 
\begin{gather}
    \frac{\D f_{\rm coll}}{\D t}=\frac{\D f_{\rm coll}}{\D z}\frac{\D z}{\D t},
\end{gather} 
and 
\begin{gather}
    \frac{\D f_{\rm coll}}{\D z}=-\frac{1}{z_{w}\sqrt{2\pi}}e^{-\frac{(z-z_0)^2}{2z_w^2}}.
\end{gather} 

\begin{figure*}
    \includegraphics[width=0.4\textwidth]{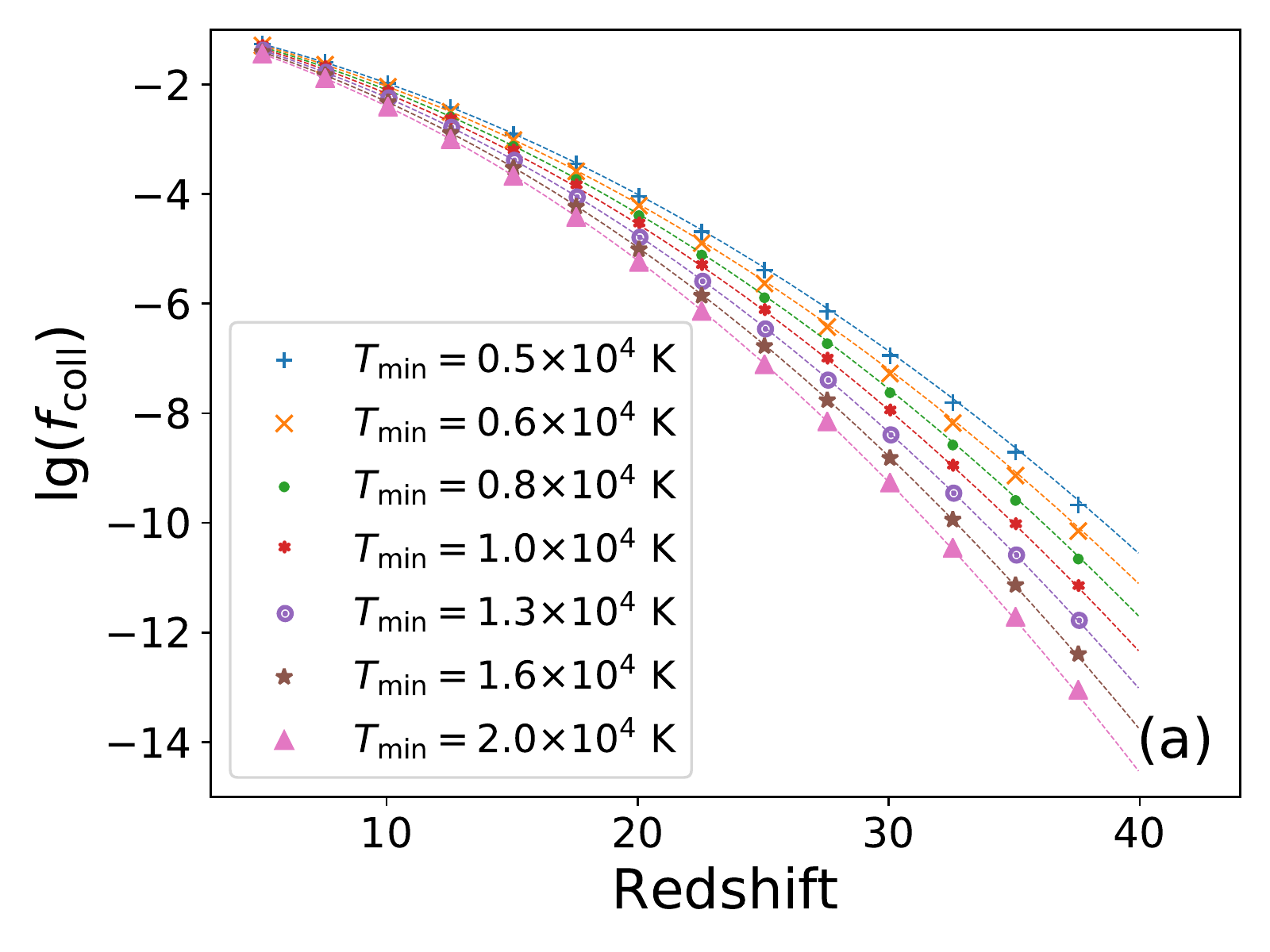}
    \includegraphics[width=0.4\textwidth]{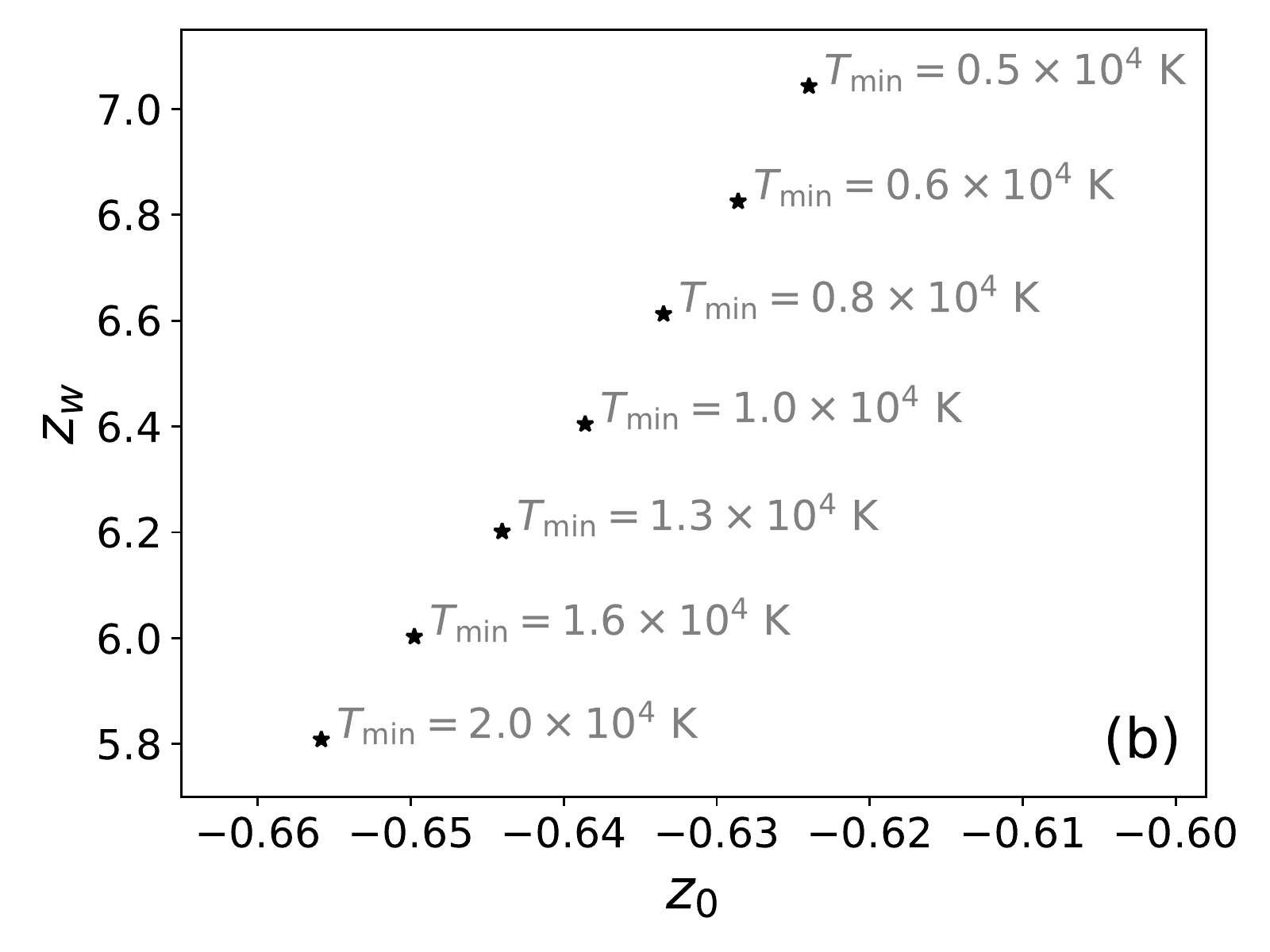}
    \caption{\label{fig:hmf_fitting}
        (a): The data points denote the $f_{\rm coll}(z)$ data computed with \textsc{hmf-calc} and the dashed lines denote $f_{\rm coll}(z)$ values computed with Equation \ref{eqn:fcoll_analytic} by using corresponding $z_w$'s and $z_0$'s obtained with model-fittings. (b): Values of $z_w$'s and $z_0$'s corresponding to different $T_{\rm min}$'s.}
\end{figure*}

In order to check how well this formula can approximate the numerically computed $f_{\rm coll}(z)$ with \textsc{hmf-calc}, we perform a series of ordinary least squares model-fittings by using the above Equation \ref{eqn:fcoll_analytic} to the data calculated with \textsc{hmf-calc} of different $T_{\min}$'s. 
The model-fitting results are shown in Figures \ref{fig:hmf_fitting} (a) and (b).
The fitting results show that our analytic $f_{\rm coll}(z)$ model is a good approximation to the data computed with \textsc{hmf-calc}. We choose $z_0=-0.64$ and $z_w=6.4$ as the parameter of our reference model, which roughly corresponds to $T_{\min}=10^4$ K.

If necessary, $f_{\rm coll}$ can be replaced by other analytic expressions in the future.
This significantly enhances flexibility and computing performance in MCMC sampling.

On the other hand, the spin temperature $T_{\rm S}$ is determined as 
\begin{gather}
    T_{S}^{-1}\approx \frac{T_{\gamma}^{-1}+x_c T_K^{-1}+x_\alpha
    T_c^{-1}}{1+x_c+x_\alpha},\label{eqn:spin_temperature}
\end{gather} 
where the color temperature of UV radiation field $T_c$ is approximated to be $\simeq T_K$.
The coupling coefficient $x_c$ and $x_\alpha$ are calculated by using Equations 24 and 40 of \cite{2006PhR...433..181F}, respectively.
The $x_\alpha$ depends on the background intensity of Ly$\alpha$, which is calculated as 
\begin{gather}
    \hat{J}_\alpha=\frac{c}{4\pi}(1+z)^2\sum_{n=2}^{n_{\max}}f_{\rm rec}^{(n)}
    \int_{z}^{z_{\max}^{(n)}}\frac{\hat{\epsilon}_{\nu^\prime}(z^\prime)}{H(z^\prime)}dz^\prime, \label{eqn:Ja}
\end{gather}
and
\begin{gather}
    \hat{\epsilon}_{\nu}=\bar{\rho}_{\rm b}^{0} C_{\rm LW} f_{*}f_{\rm esc, LW}\frac{\D f_{\rm coll}}{\D t}I_{\nu},\label{eqn:emissivity_LW}
\end{gather} 
where $C_{\rm LW}$ is fixed to be $4.72\times10^{27}~{\rm g}^{-1}$, corresponding to 9690 photons per baryon particle, $f_{\rm esc, LW}$ is the corresponding escape fraction.
Same as \cite{2014MNRAS.443.1211M}, $n_{\max}$ is truncated to 23, and the $f_{\rm rec}^{(n)}$ is the `recycling fraction' \citep{2006MNRAS.367.1057P}.

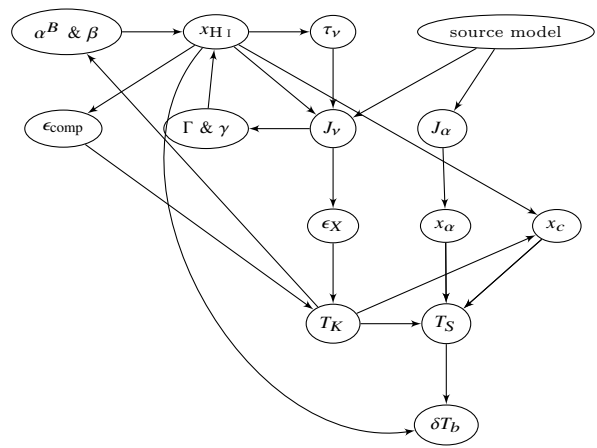
\begin{figure}
    \centering
    \input{fig_dependences}
    \caption{\label{fig:dependences}
        The dependency graph of the quantities needed to calculate the global $\HI$ 21 cm spectrum. Any one arrow points from one quantity to another one, the computing of which depends on the former one.
    }
\end{figure}

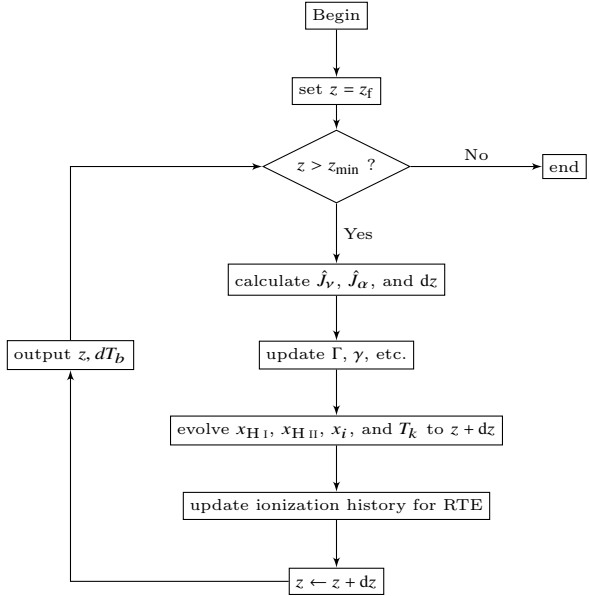
\begin{figure}
    \centering
    \input{fig_flow_chart}
    \caption{\label{fig:flowchart}
        The flowchart of the algorithm for computing the global $\HI$ 21 cm spectrum.
    }
\end{figure}

Finally, by assembling above computing together, we can obtain the global $\HI$ 21 cm signal of high redshifts.
Looking back above still very incompletely summarized procedures, it is obvious that the rate Equations \ref{eqn:rate_equation} and \ref{eqn:HII_rate_equation}, temperature evolving Equation \ref{eqn:heating}, and radiative transfer Equation \ref{eqn:J_nu} twist with each other.
This can be shown a bit more clearly in Figure \ref{fig:dependences}.
The dependence relations between different quantities form two loops, i.e., $x_{\HI}\to $ $\bar{\tau}_\nu \to$ $J_{\nu} \to $ $\Gamma~\&~\gamma\to x_{\HI}$ and $x_{\HI}\to$ $\epsilon_{\rm comp}\to$ $T_{K}\to$ $\alpha^B\&\beta\to x_{\HI}$.
The most natural way is to solve them by turns from a high enough initial redshift to lower redshifts iteratively.
We summarize the algorithm of computing the global $\HI$ 21 cm signal in Figure \ref{fig:flowchart}.

\begin{table*}
    \centering
    \caption{\label{tbl:quantities}
        The References of How Following Quantities are Calculated} 
        \begin{tabular}{clcr} \hline Quantity & Meaning   & Related Eq. in this work           & Ref.\\
        \hline
        $\alpha^A$,$\alpha^B$  & case-A and -B recombination coefficients                      & \ref{eqn:rate_equation}, \ref{eqn:HII_rate_equation}                    & Eq. B5---B9 of \cite{1994MNRAS.269..563F}\\ 
        $\beta$                & collisional ionization coefficients                           & \ref{eqn:rate_equation}                              & Eq. B1---B4 of \cite{1994MNRAS.269..563F} \\
        $\epsilon_{\rm comp}$  & Compton heating coefficient                                   & \ref{eqn:heating}                                    & Eq. 54 of \cite{2000ApJS..128..407S}\\
        $\epsilon_{X}$         & X-ray heating coefficient                                     & \ref{eqn:heating}                                    & Eq. 11  of \cite{2014MNRAS.443.1211M}\\
        $\gamma$               & photon ionization coefficients                                & \ref{eqn:rate_equation}                              & Eq. 10 of \cite{2014MNRAS.443.1211M}\\
        $\Gamma$               & secondary photon ionization coefficients                      & \ref{eqn:rate_equation}                              & Eq. 9 of \cite{2014MNRAS.443.1211M}\\
        $J_\nu$                & background intensity                                          & \ref{eqn:J_nu}                                       & Eq. 4 or 25 of \cite{2014MNRAS.443.1211M}\\
        $J_\alpha$             & Ly$\alpha$ background intensity                               & \ref{eqn:Ja}                                         & Eq. 7 or 25 of \cite{2014MNRAS.443.1211M}\\
        $\tau_\nu$             & optical depth                                                 & \ref{eqn:J_nu}, \ref{eqn:tau}                        & Eq. 6 of \cite{2014MNRAS.443.1211M}\\
        $T_K$                  & gas kinetic temperature                                       & \ref{eqn:heating}, \ref{eqn:spin_temperature}        & Eq. 14 of \cite{2014MNRAS.443.1211M}\\
        $T_{S}$                & $\HI$ spin temperature                                        & \ref{eqn:HI_Tb}, \ref{eqn:spin_temperature}          & Eq. 20 of \cite{2006PhR...433..181F} \\
        $x_\HI$, $x_\HII$      & fraction of $\HI$ and $\HII$ in bulk IGM region, respectively & \ref{eqn:rate_equation}                              & Eq. 12 of \cite{2014MNRAS.443.1211M}\\
        $x_i$                  & volume filling fraction of $\HII$ regions                     & \ref{eqn:HII_rate_equation}                          & Eq. 13 of \cite{2014MNRAS.443.1211M}\\
        $x_c$                  & collisional coupling coefficient                              & \ref{eqn:spin_temperature}                           & Eq. 24 of \cite{2006PhR...433..181F}\\
        $x_\alpha$             & Wouthuysen-Field coupling coefficient                         & \ref{eqn:spin_temperature}                           & Eq. 40 of \cite{2006PhR...433..181F}\\
        \hline
    \end{tabular}
\end{table*}

\subsection{Reference Model and Frequency Range to be Used}
\label{ssec:ref_model}
\begin{figure*} 
    \centering 
    \includegraphics[width=.3\textwidth]{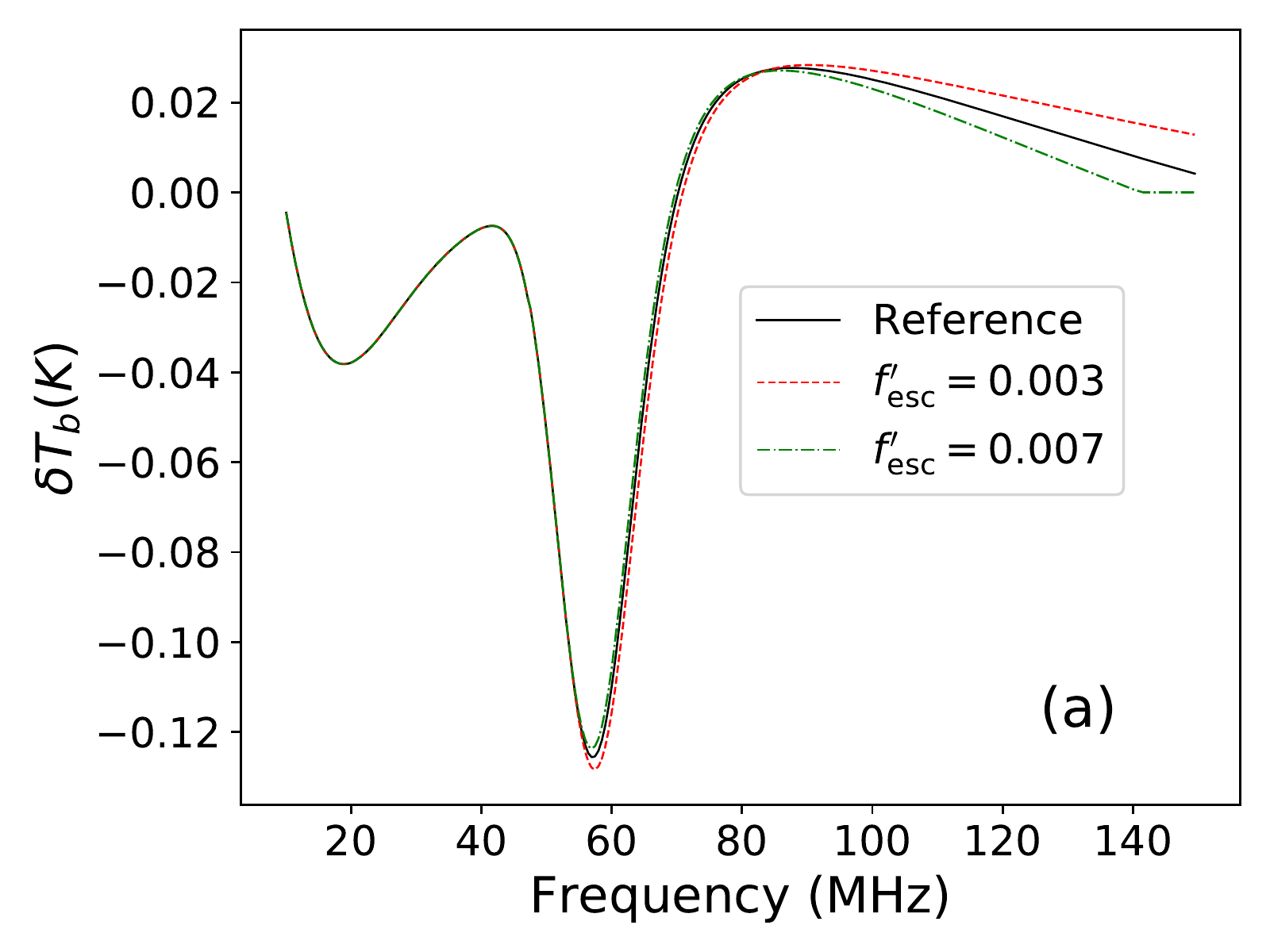} 
    \includegraphics[width=.3\textwidth]{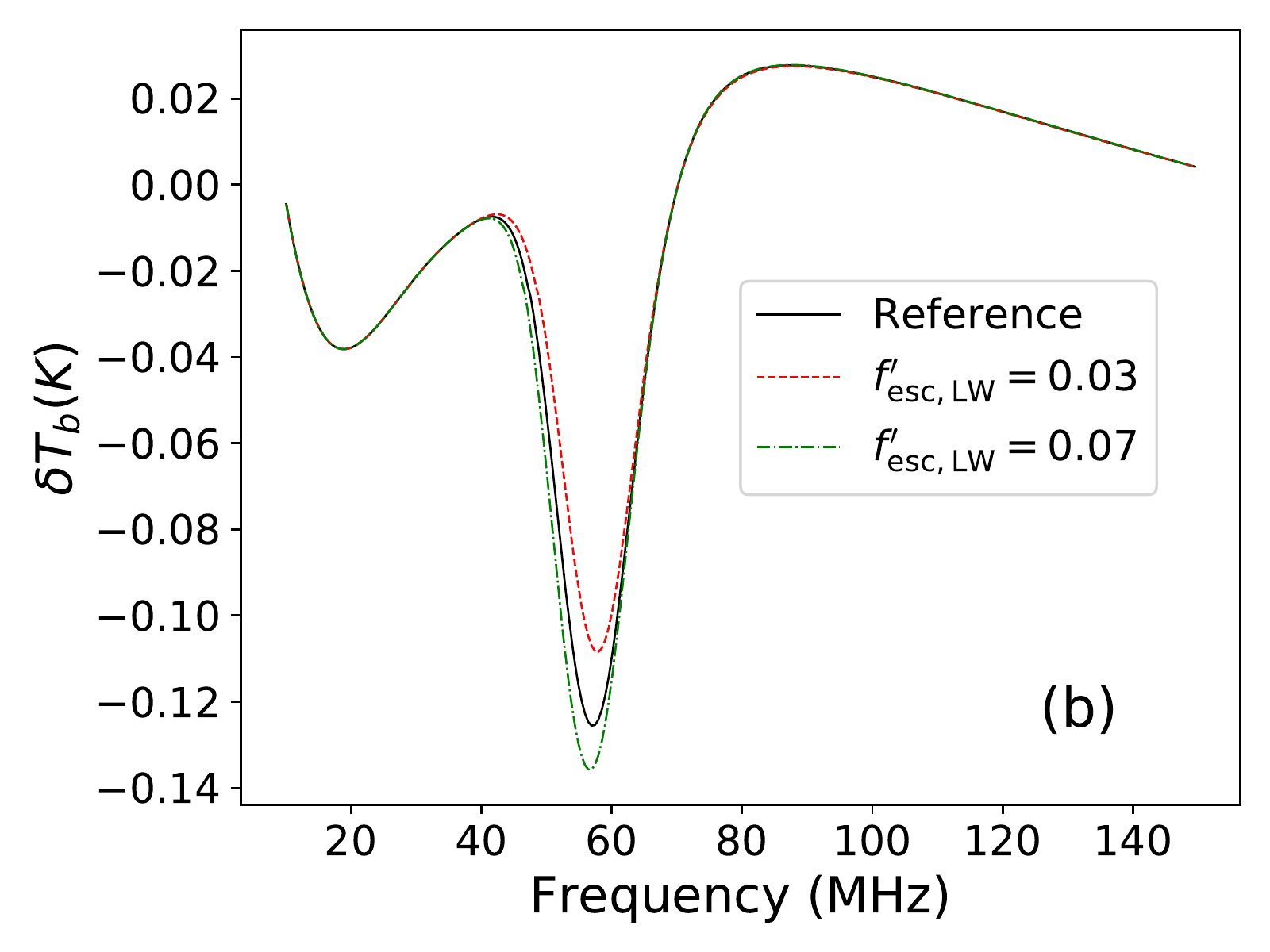} 
    \includegraphics[width=.3\textwidth]{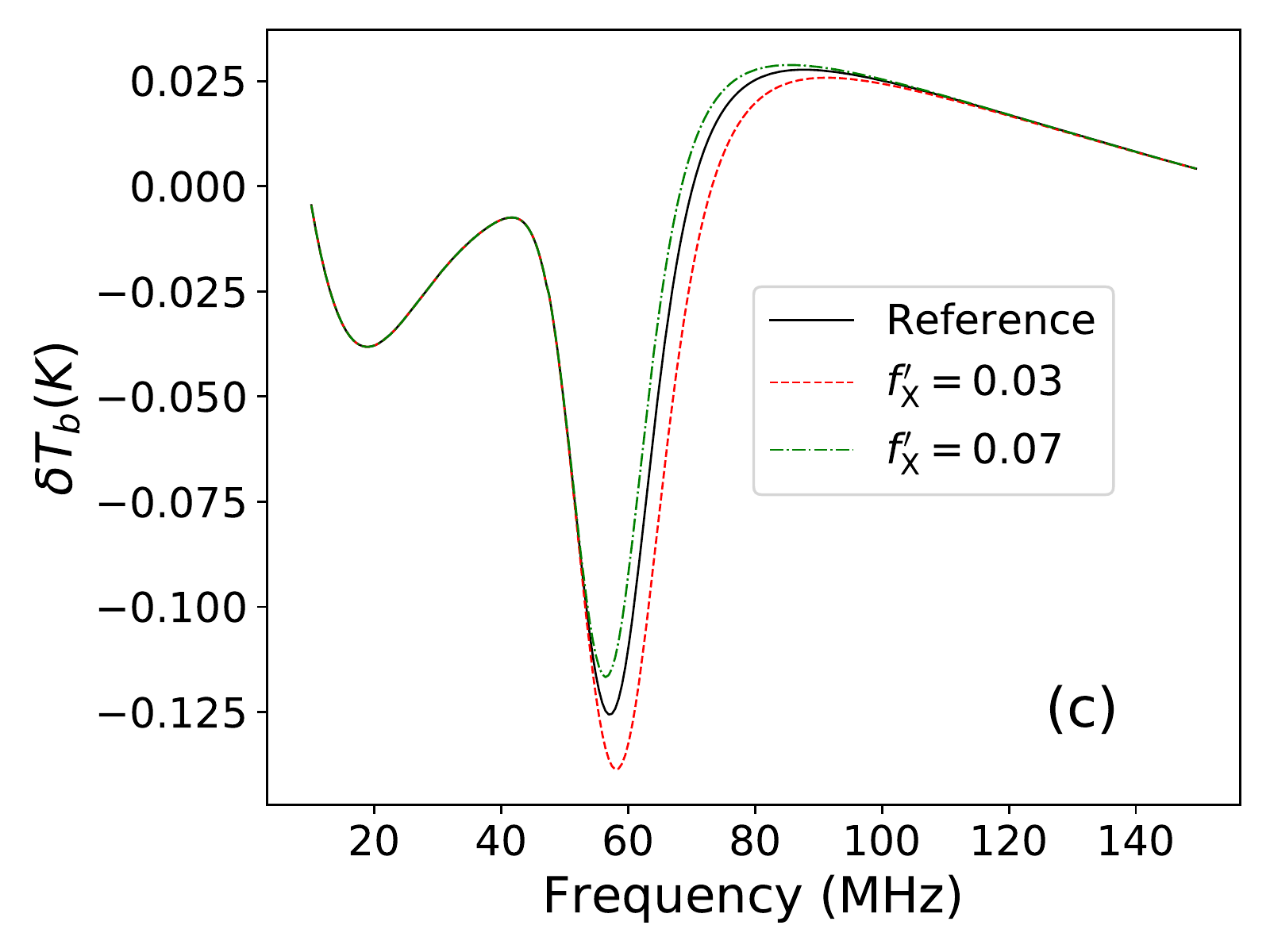} 
    \caption{\label{fig:ref_model} 
    Brightness temperature contrast to CMB temperature calculated with the reference and varied model parameters.
    In (a), (b), and (c), we show the influence of varying $f^{\prime}_{\rm esc}$, $f^{\prime}_{\rm esc, LW}$, and $f^{\prime}_{\rm X}$, respectively.
    }
\end{figure*}

We choose the parameter values listed in Table \ref{tbl:ref_model_param}, and assume the ionizing sources to be black holes with a power-law spectrum to calculate the reference model values.
The power-law index $\alpha$ is set to be $-1.5$.
Note that among $f_{*}$, $f_{\rm esc}$, $f_{\rm esc, LW}$, and $f_{\rm X}$, only three can be determined independently, because $f_{*}$ always appears together with one of the other three factors in above Equations \ref{eqn:HII_rate_equation}, \ref{eqn:X_emissivity}, and \ref{eqn:emissivity_LW}.
Consequently, it becomes an issue that needs to be discussed how to choose the free parameters to be inferred.
The difference in selecting different sets of free parameters can be nontrivial, given that the corresponding priori distributions are not adjusted accordingly.
We regard it more natural and intrinsic to use 
\begin{gather}
    f_{\rm esc}^\prime\equiv f_{*}f_{\rm esc},\\
    f_{\rm esc, LW}^\prime\equiv f_{*}f_{\rm esc, LW},
\end{gather}
and
\begin{gather}
    f_{\rm X}^\prime\equiv f_{*}f_{\rm X}
\end{gather}
as the free parameters to infer.
Note that there can be better choice when more priori knowledge is available.
In Figure \ref{fig:ref_model}, we compare $\delta T_b(\nu)$'s calculated with the reference model parameters and different $f^\prime_{\rm esc}$, $f^\prime_{\rm esc, LW}$, and $f^\prime_{\rm X}$ values. 
It is not surprising that below 40 MHz, the signal changes little no matter how the parameters vary. 
At redshift higher than $z\approx 35$ (corresponding to $40$ MHz), most of the current models predict that ionizing sources had not yet appeared, the brightness temperature of $\HI$ signal is mainly modulated by the adiabatic expansion progress of CMB radiation and gas. 
Because the ionosphere will significantly block electromagnetic wave with frequency $\lesssim 30$ MHz, ground-based experiments cannot arbitrarily extend the frequency range to lower frequencies.
So it is reasonable to set the lower limit of the frequency range to be 40 MHz. 
On the other hand, as we describe above, $\HI$ brightness temperature model assumes that the IGM is divided into the `bulk IGM' and $\HII$ regions, and this assumption can be violated in the late stage. 
As a result, we set the upper limit of the frequency to be 120 MHz, corresponding to $z=10.8$.
At this redshift, the reference model predicts an average $\HI$ fraction still higher than $50\%$, which means it is not in the very late stage of reionization.
As a conclusion, we will assume the data to be acquired within the frequency range of $40-120$ MHz in most of the following analyses.

\begin{table}
    \centering
    \caption{\label{tbl:ref_model_param}
        Reference Model Parameters} \begin{tabular}{cc} \hline Parameter         & Value   \\ 
        \hline 
    $f_{*}$           & 0.05    \\
    $f_{\rm esc}$     & 0.1     \\
    $f_{\rm esc, LW}$ & 1.0     \\
    $f_{\rm X}$ & 1.0\\
    $z_0$             & -0.64     \\
    $z_w$             & 6.4     \\ 
    $N_{\rm poly}$    & 1       \\ 
    $T_{100}$         & $806$ K \\ 
    $p_{1}$           & $-2.27$ \\ 
\hline 
\end{tabular} \end{table} 

\subsection{Foreground Emission}
\label{ssec:fg}
Because the foreground is dominated by the synchrotron radiation from the Milky Way, we naturally model the foreground as a simple power-law spectrum when simulating the signals. 
Then in the stage of parameter inference, i.e., MCMC sampling, we treat the foreground spectral model as a 7th-degree log-polynomial, to emulate any uncorrected instrumental effects on foreground spectrum, just like the treatment in \cite{2016MNRAS.461.2847B}, though we do not think it can adequately compensate for the instrumental effects. 
We express the foreground spectrum as 
\begin{gather}
T_{\rm fg}(\nu)=T_{100}\exp\left [\sum_{n=1}^{N_{\rm poly}}p_n\left(\log\frac{\nu}{100~{\rm MHz}}\right)^{n}\right ]. \label{eqn:fg}
\end{gather} 
When simulating the foreground, we set the polynomial degree $N_{\rm poly}$ to be $1$, and $p_1=-2.27$.
The brightness temperature at 100 MHz $T_{100}$ is set to be $806$ K (see Table \ref{tbl:ref_model_param}), which is in agreement with the value used in \cite{2016MNRAS.461.2847B}. 
In the parameter inference stage, we set $N_{\rm poly}$ to be 7, and allow all the coefficients in Equation \ref{eqn:fg} to vary.

\section{MCMC Sampling}
\label{sec:mcmc_method}
\subsection{Posterior Probability Distribution of Model Parameters}
\label{ssec:posterior_distribution}
According to the Bayes' theorem, given a set of measurements that follow an underlying model, the posterior distribution of the model's parameters can be derived with 
\begin{gather}
p(\mathbf{\theta}|\mathbf{T}_{b}^{\rm obs})=p(\mathbf{\theta})p(\mathbf{T}_{b}^{\rm obs}|\mathbf{\theta}),
\end{gather} 
where $p(\mathbf{\theta})$ is the priori distribution of the model parameters, $p(\mathbf{T}_{b}^{\rm obs}|\mathbf{\theta})$ is the likelihood. 
We assume that there is no intrinsic fluctuation in the model; all deviation of the measurements from the model value is caused by thermal noise. 
In this condition, we can model the distribution of the noise value as normal distribution and safely regard the noise in different frequency channels to be uncorrelated. 
So that the likelihood can be written as 
\begin{gather}
p(\mathbf{T}_{b}^{\rm obs}|\mathbf{\theta})=\prod_i\frac{1}{\sqrt{2\pi \sigma^2_{\nu_i}}}e^{-\frac{(T_{b}^{\rm obs}(\nu_i)-T_{b}^{\rm m}(\nu_i))^2}{2\sigma^2_{\nu_i}}},\label{eqn:likelihood}
\end{gather}
where $\mathbf{T}_{b}^{\rm obs}$ is a vector of measured antenna temperature in all used frequency channels, $\mathbf{T}_{b}^{\rm m}$ is the corresponding model value, $T_{b}^{\rm obs}$ and $T_{ b}^{\rm m}$ are respectively the value for a certain frequency channel, $\sigma_{\nu_i}$ is the measurement uncertainty caused by thermal noise in the $i$-th channel.
According to the Gaussian white noise assumption, the noise is 
\begin{gather}
\sigma_{\nu_i}=\frac{T_{b}^{\rm obs}}{\sqrt{\Delta\nu\tau}}, \label{eqn:noise}
\end{gather} 
where $\Delta\nu$ is the channel width, $\tau$ is the integration time.

According to the foreground spectrum model Equation \ref{eqn:fg}, the antenna temperature at the lowest frequency 40 MHz is 6451 K. 
If we assume a channel bandwidth of $1$ MHz\footnote{Most systems have a much higher spectral resolution to excise RFI. Because the predicted global $\HI$ signal varies slowly with frequency, it is feasible to combine fine raw frequency channels to coarse channels to improve the computational performance.} and an integration time of ten days, the noise level at 40 MHz is $\sigma_{40~{\rm MHz}}\approx6.94$ mK.
Apparently, the signal-to-noise-ratio (SNR) of global $\HI$ signal detections can reach a rather high value with very limited integration time.

The rest part of the posterior distributions is the priori distribution of unknown parameters.
We assume parameters $f^\prime_{\rm esc}$, $f^\prime_{\rm esc, LW}$, and $f^\prime_{\rm X}$ to independently follow a uniform distribution in the interval $(0,100)$.
Note that although these parameters have the meaning of `escape fraction' and in principle should be limited between 0 and 1, we are using them to compensate for the uncertainties of corresponding physical parameters (i.e., $N_{\rm ion}$, $C_{\rm LW}$, and $C_{X}$), so that we do not impose the limit between 0 and 1.

The priori distribution of $z_0$ is 
\begin{gather}
\frac{\D p(z_0)}{\D z_0}=\frac{1}{\sigma_{z_0}\sqrt{2\pi}}e^{-\frac{(z_0-\hat{z}_0)^2}{2\sigma_{z_0}^2}},
\end{gather} 
and the priori distribution of $z_w$ is 
\begin{gather}
\frac{\D p(z_w)}{\D z_w}=\frac{1}{\sigma_{z_w}\sqrt{2\pi}}e^{-\frac{(z_w-\hat{z}_w)^2}{2\sigma_{z_w}^2}},
\end{gather} 
where the hyperparameters $\hat{z}_0=-0.64$ and $\hat{z}_w=6.4$.
In order to set the values of hyperparameters $\sigma_{z_0}$ and $\sigma_{z_w}$ as a part of the priori knowledge, we examine Figure \ref{fig:hmf_fitting}(b).
A reasonable range of $z_0$ corresponding to $T_{\min}\in [0.5\times 10^4, 2\times 10^4]$ can be $z_0\in[-0.66, -0.62]$, so that it is safe enough to set $\sigma_{z_0}$ to be $0.2$, which allows $z_0$ to vary within a slightly larger range.
Similarly, $z_w$ values corresponding $T_{\min}\in [0.5\times 10^4, 2\times 10^4]$ can be safely enclosed within the interval $[5.8, 7.1]$, so that we set $\sigma_{z_w}$ to be $1$.
It is also possible that the $f_{\rm coll}(z)$ calculated with other parameters or other codes can lead to larger reasonable ranges, but they will not differ by order of magnitudes.
If necessary, in actual parameter inference, $\sigma_{z_0}$ and $\sigma_{z_w}$ can be replaced by other values.

\subsection{Sampling Algorithm}
\label{ssec:sampling_algorithm}
With the posterior distribution of the model parameters in hand, we then move on to sampling realizations from the posterior distribution. 
There have been a dozen of different sampling method available. We decided to test the affine invariant sampling method proposed by \cite{goodman2010ensemble} and its parallel tempering version proposed by \cite{liang2011advanced}. 
The most important advantage of the affine invariant sampling method is that it does not need to be finely tuned to fit into a concrete problem, and is efficient enough to sample from very complex posterior distribution function. 
The computing of the global $\HI$ signal spectra in this work happens to be such kind of problem. 
Even after being reimplemented with \textsc{rust}, the program still needs about 4 seconds for one CPU core to compute one spectrum. 
To reach a convergence, we set a `burn-in' of 30,000 steps, the total length of the Markov chain is set to be 60,000. 
In the beginning, we use the origin affine invariant sampling method and find that the chains are very hard to converge when the initial parameter is not close enough to the optimal value. 
So that we finally use the parallel tempering method to accelerate the convergence. We reimplement the \textsc{python} package \textsc{emcee} \footnote{\url{https://github.com/dfm/emcee}}\citep{2013PASP..125..306F} that performs both the affine invariant sampling method and its parallel tempering version with \textsc{rust} and run the sampling program on a high performance computing cluster, utilizing parallel computing technique.
For the parallel tempering method, we set six `temperature's and for each temperature, 28 walkers (hence totally 168 chains) are used. 
The $\beta$ values for the six temperatures are set to be $2^{-i}$ where $i=0, 1,\ldots, 5$.
Totally 168 CPU cores are allocated for the MCMC program. 
Using more CPUs will not help reduce the time needed to compute one spectrum, but will allow us to set more chains, thus may help speed up the convergence of each chain.

\subsection{The Standard We Use to Diagnose if the Inference is Biased}
\label{ssec:bias_standard}
Strictly speaking, in order to diagnose if the inference is biased, an analytical expression of the posterior distribution probability of interesting parameters is needed to deduce the expectations and compare them with input values. 
However, it is hard or even impossible in the condition of detection to global $\HI$ signal, because $\HI$ signal spectra are computed through massive numerical methods.
We have to choose an alternative (not so strict) way--compare the input values with the mean sampled values and their dispersions.
In practical parameter inference, when the distribution of sampled parameters is not heavily multimodal, we can express the inference result as a central value $x$ together with its standard deviation $\Delta x$ (for normal distribution, it is the 68\% confidence level error) in the form of $x\pm \Delta x$.
If the interval $[x-\Delta x, x+\Delta x]$ encloses the true parameter values (i.e., input values here), we can practically regard the inference method to be good enough, so that practically unbiased.
Similar tests can also be performed based on the plots of joint distributions of pairs of parameters.
The above standard may not be appropriate, especially when one trying to examine it from a strict mathematics-based point of view.
Some readers may not agree with our above standard, so in the following sections, we present objective results, including inferred parameter values and some of the plots of the parameter distributions we obtained, and readers can make their own decisions.

\section{Studied Cases and Results}
\label{sec:results}

In this section, we study the influence of a variety of factors on the inference of model parameters. 
The factors include thermal noise, amplifier gain calibration error, and frequency-dependent antenna beam pattern. 
As is mentioned above, we choose $f^\prime_{\rm esc}$, $f^\prime_{\rm esc, LW}$, $f^\prime_{\rm X}$, $z_0$, and $z_{w}$ as our interesting parameters, which can reflect the detailed physical images of the reionization process. 
We choose $1$ MHz spectral resolution, which is achievable with most practical data acquisition systems, and simultaneously sufficiently high for inferring most of the interesting parameters. 
In order to rule out the effect of the fluctuation from the random noise itself, we use the same random number generator seed to simulate the noise in the following sections.

\subsection{The Property of Parameter Space when Ignoring Foreground and Noise}
\label{ssec:space_property}
Before testing the conditions that emulate practical observations, we first inspect the property of parameter space with the foreground component absented. 
The purpose is to discover potential intrinsic multimodal features and degeneracy in parameter space and separate them from the influence of the existence of foreground components. 
We simulate the total signal that is composed of only the global $\HI$ signal by using the parameters listed in Table \ref{tbl:ref_model_param} and thermal noise by using Equation \ref{eqn:noise}. 
When simulating the noise, the foreground signal is included in the $T_{b}^{\rm obs}$ of Equation \ref{eqn:noise} temporarily and observation time of ten days is assumed so that the realization of the noise component is identical to what will be used in the following sections. 
In the parameter inference stage, we test two models, one is the pure global $\HI$ signal model, and the other is the sum of the global $\HI$ signal model and a 7th-degree log-polynomial foreground component (Equation \ref{eqn:fg}). 
Because the nominal value of $T_{100}$ is $0$ K here, other foreground coefficients $p_i$'s degenerates in the parameter space.
If no constraint is applied to $p_i$'s, they can be an arbitrary value and cause the chain hard to converge.
Hence we assume all $p_i$'s to independently follow uniformed distributions between $-1$ and $+1$.
The hyperparameters $\sigma_{z_0}$ and $\sigma_{z_w}$ here are both set to be $10$ temporarily, which are fairly loose constraints, in following sections they will be set to $\sigma_{z_0}=0.2$ and $\sigma_{z_w}=1.0$.
The sampling of both models reach equilibrium very fast, and the joint distribution of parameters show neither multimodal feature nor degeneracy. 
This phenomenon implies the feasibility of inferring parameters of global $\HI$ models with the MCMC sampling algorithm. 
The detailed inference result is unimportant here and cannot be compared to following testing conditions, so we omit them.

\subsection{The Influence of Thermal Noise}
\label{ssec:thermal_noise}
As has been mentioned in section \ref{ssec:posterior_distribution}, an observation with very limited integration time can already obtain a high enough SNR. 
Nevertheless, it is still interesting to study how well the parameters can be constrained in an ideal observation condition (i.e., only inevitable thermal noise is considered). 
By using the above Equation \ref{eqn:noise}, we generate the noise signal for each frequency channel and add it to the total signal.

\begin{figure}
    \centering
    \includegraphics[width=\columnwidth]{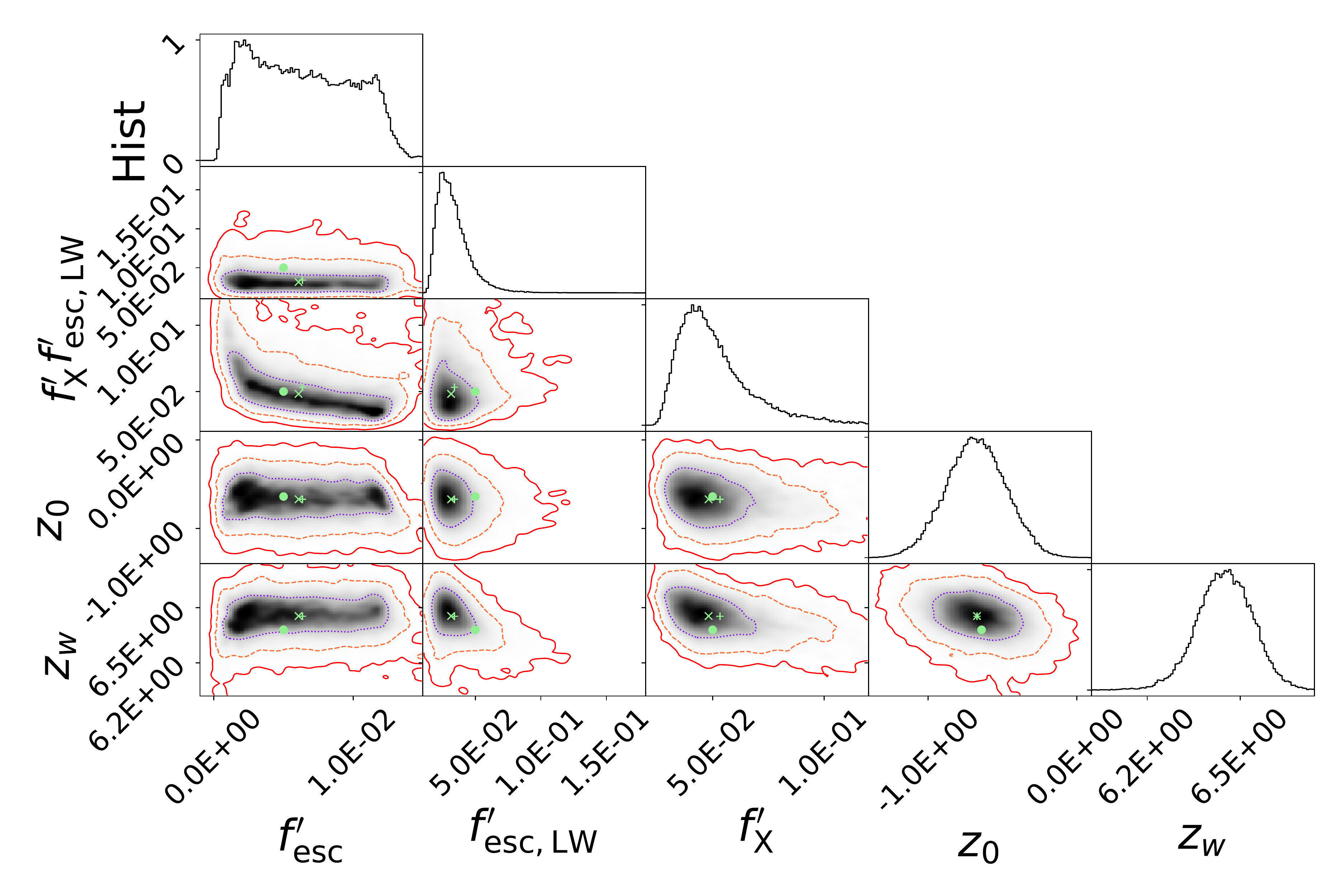}
    \caption{\label{fig:bl7}
    Probability distributions of single parameters (histograms, panels on the diagonal) and of joint parameters (68\%, 95\%, and 99.7\% probability contours, remaining panels) for the test case BL7 in section \ref{ssec:thermal_noise} with the thermal noise calculated by assuming an integration time of ten days, $\sigma_{z_0}=0.2$, and $\sigma_{z_w}=1.0$.
    The instruments are assumed to be perfectly calibrated.
    The `+' symbols denote the estimation to the expected values by averaging the sample.
    The `x' symbols denote the estimations by using median values.
    The dots denote the parameter values used to simulate the data.
    }
\end{figure}

\begin{figure}
    \centering
    \includegraphics[width=.95\columnwidth]{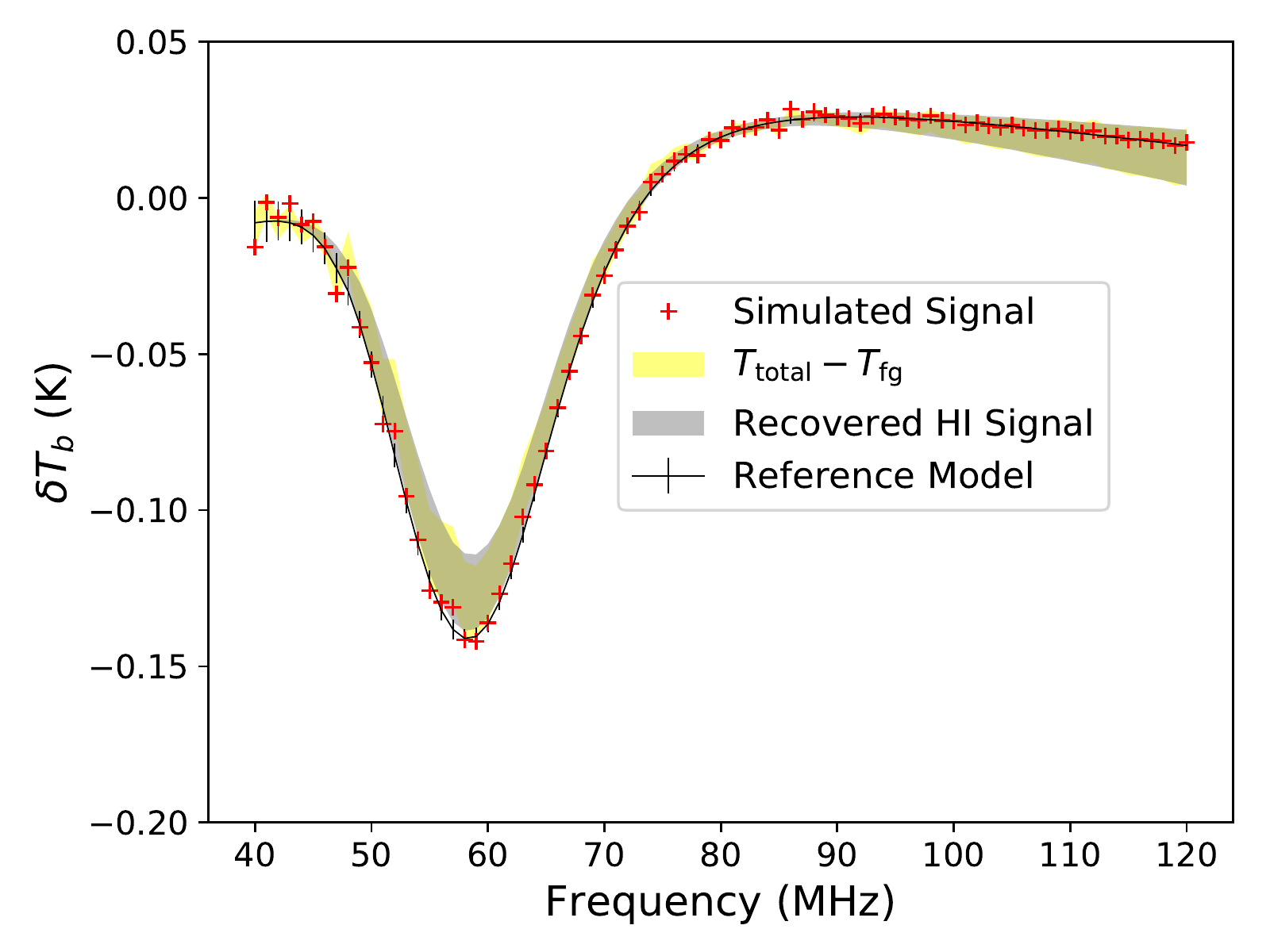}
    \caption{\label{fig:signal_bl7}
    Comparisons between input $\HI$ 21 cm signal and the recovered signal for test case BL7 described in section \ref{ssec:thermal_noise}. 
    Solid line denotes the input global $\HI$ signal, the error bars denote the noise level calculated with Equation \ref{eqn:noise}.
    Red crosses denote the sum of input signal and the actually simulated noise.
    The grey band denotes the $68\%$ confidence interval of the recovered $\HI$ signal.
    The yellow band denote the $68\%$ confidence interval	of the difference between the total signal and the recovered foreground.
    }
\end{figure}

With the posterior distribution we set in the previous section \ref{ssec:posterior_distribution}, we perform the sampling.
We plot the parameter distribution in Figure \ref{fig:bl7} (i.e., the `corner plot', only the $\HI$ signal model parameters, foreground model parameters are not shown here).
The numeric results are listed in Table \ref{tbl:params}, labeled as BL7 (`BL' means baseline, and the `7' denotes the degree of the polynomial for the foreground component).
By using the averaged values as an unbiased estimator, we find that the estimated values are all enclosed by 90\% probability contours. 
The parameter $f^\prime_{\rm esc, LW}$ is slightly underestimated, and 68\% probability contours cannot enclose the input value, so that seems to be biased.
We also recover the global $\HI$ signal by using two methods: 1) directly computing the signal with sampled $\HI$ signal parameters, 2) subtracting the sampled foreground component (computed from the sampled foreground parameters) from the total signal.
The comparison between the input and recovered signals are shown in Figure \ref{fig:signal_bl7}, from which we find that the input signal does not pass through the center of the bands that marks 68\% confidence range.
This phenomenon also suggests that the inference seems to be biased.
We suspect that this can be caused either by the thermal noise or by the over-fitting problem. 
Also, we note there seems to exist a multimodal feature in the corner plot, especially in the histogram of the $f^\prime_{\rm esc}$.
So that we further do two more tests: 1) removing the noise component from total signal (but still taking into account the noise when calculating the likelihood function, i.e., the parameters $\sigma_{\nu_i}$); 2) using lower-degree polynomials to model the foreground in the sampling (the noise component is removed as well).

\begin{figure}
    \centering
    \includegraphics[width=\columnwidth]{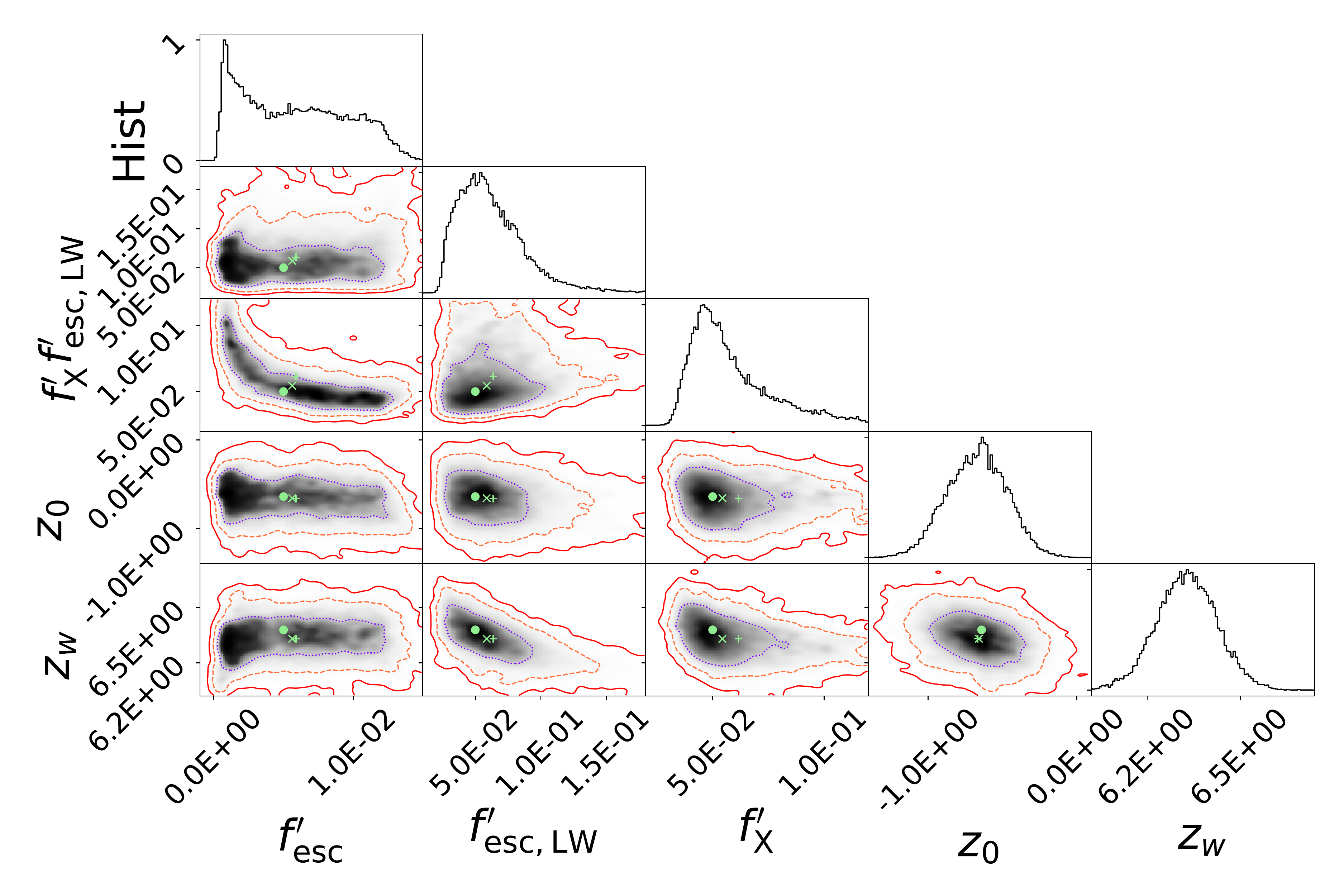}
    \caption{\label{fig:bl7nn}
    Same as Figure \ref{fig:bl7}, except that the noise component is removed from the total signal, corresponding to test case BL7NN.
    }
\end{figure}

\begin{figure}
    \centering
    \includegraphics[width=\columnwidth]{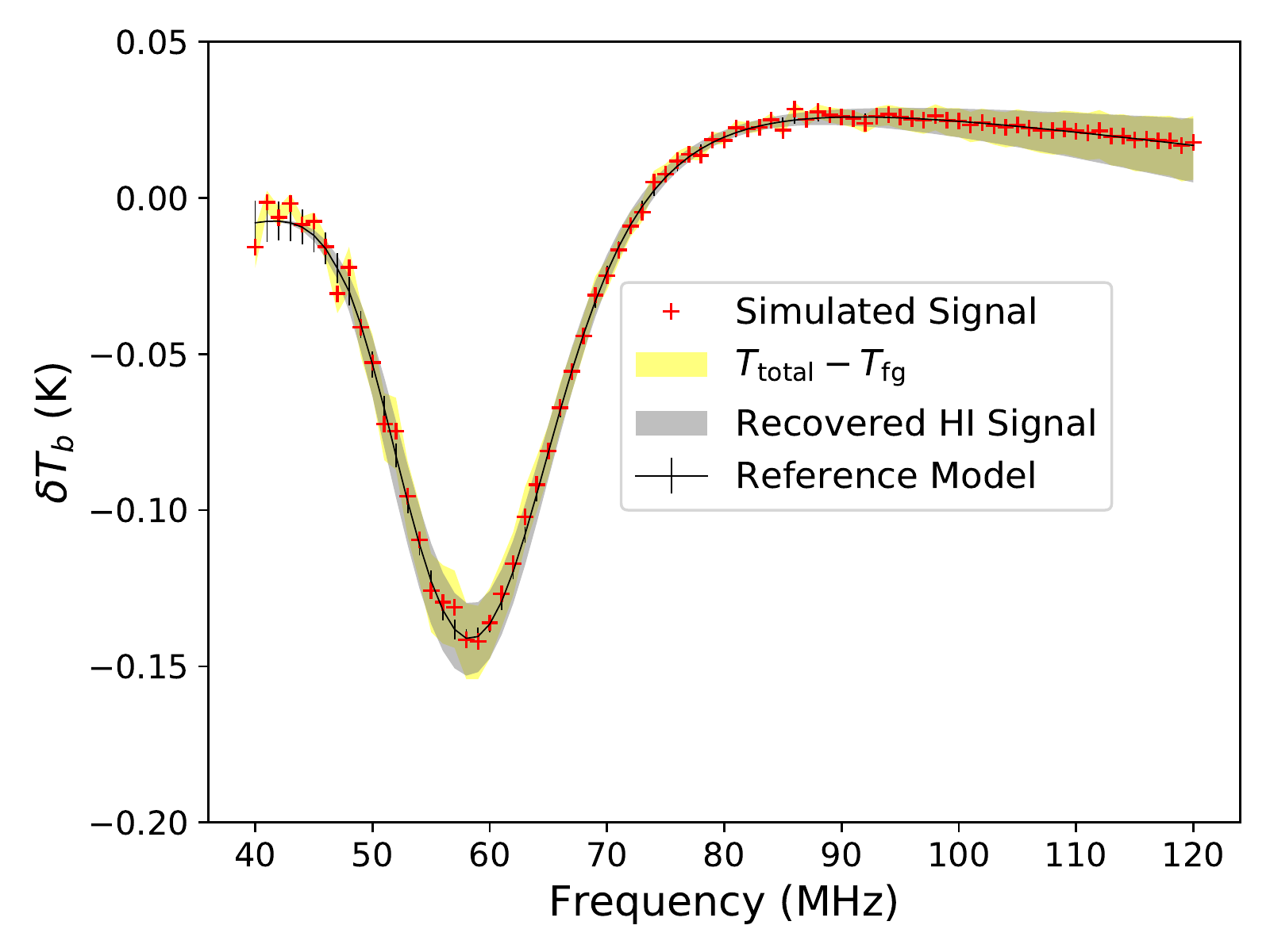}
    \caption{\label{fig:signal_bl7nn}
    Same as Figure \ref{fig:signal_bl7}, except that the noise component is removed from the total signal, corresponding to test case BL7NN.
    }
\end{figure}

The distributions of parameters sampled with noise component removed (labeled as BL7NN) are shown in Figure \ref{fig:bl7nn}, and the numeric results are listed in Table \ref{tbl:params}. 
Obviously, the input parameter values can be all enclosed within 68\% contours.
In Figure \ref{fig:signal_bl7nn}, we note that the input signal right passes through the center of the bands that marks 68\% confidence range of the recovered signals.
Hence we can conclude that the underestimation of parameter $f^\prime_{\rm esc, LW}$ is caused by the certain realization of the noise component, and statistically speaking, the inference is unbiased.
About the slightly multimodal feature, we suspect it is caused by the foreground model with too many degrees of freedom.
We leave the test of using lower-degree polynomials to model the foreground in section \ref{ssec:overfitting}.

\begin{table*}
    \centering
    \caption{\label{tbl:params}Estimation to the model parameters obtained with the MCMC sampling method. Symbols $\hat{f}_*$,
    $\hat{f}_{\rm esc}$, and $\hat{f}_{\rm esc, LW}$ denote the parameter values used to simulate the signal.
    The signal is simulated with the reference model (Section \ref{ssec:ref_model}), and the parameters are sampled assume the ionization sources to be black holes with power-law spectra.
    }
    \scriptsize
    \begin{tabular}{ccccccccc}
        \hline
        Label & $f^\prime_{\rm esc}/\hat{f}^\prime_{\rm esc}$ & $f^\prime_{\rm esc, LW}/\hat{f}^\prime_{\rm esc, LW}$ & $f^\prime_{\rm X}/\hat{f}^\prime_{\rm X}$ & $z_0$  & $z_w$&$\log Z$ & Related Figures & Description \\
        \hline
        \multicolumn{9}{c}{Only thermal noise is considered, no other instrumental effect, see section \ref{ssec:thermal_noise} and \ref{ssec:overfitting}} \\
        \hline
        BL7     & $1.27\pm 0.72$ & $0.68\pm 0.25$ & $1.07\pm 0.41$ & $-0.67\pm 0.20$ & $6.46\pm 0.08$ & $402.09$ & \ref{fig:bl7},\ref{fig:signal_bl7}& $N_{\rm poly}=7$ for foreground \\
        BL7NN  & $1.18\pm 0.76$ & $1.27\pm 0.54$ & $1.23\pm 0.47$ & $-0.67\pm 0.20$ & $6.36\pm 0.08$ & $437.08$ & \ref{fig:bl7nn}, \ref{fig:signal_bl7nn} & No noise; $N_{\rm poly}=7$ for foreground \\
        BL3NN  & $1.01\pm 0.30$ & $1.03\pm 0.12$ & $1.05\pm 0.15$ & $-0.67\pm 0.20$ & $6.40\pm 0.05$ & $439.64$ & \ref{fig:bl3nn}, \ref{fig:signal_bl3nn} & No noise; $N_{\rm poly}=3$ for foreground \\
        BL4NN  & $1.13\pm 0.74$ & $1.11\pm 0.20$ & $1.23\pm 0.54$ & $-0.68\pm 0.20$ & $6.38\pm 0.07$ & $439.55$ & \ref{fig:bl4nn}& No noise; $N_{\rm poly}=7$ for foreground \\
        BL3  & $0.82\pm 0.28$ & $0.97\pm 0.11$ & $0.97\pm 0.11$ & $-0.67\pm 0.20$ & $6.44\pm 0.05$ & $403.26$ & & $N_{\rm poly}=3$ for foreground \\
        \hline
        \multicolumn{9}{c}{Frontend gain calibration error considered, $N_{\rm poly}=7$ for foreground, see section \ref{ssec:gain_error}}\\
        \hline
        G7     & $1.21\pm 0.74$ & $ 0.70\pm 0.26$ & $1.10\pm 0.43$ & $-0.68\pm 0.20$ & $6.46\pm 0.08$ & $402.08$ &\ref{fig:HI_signal_gain_err_g7}, \ref{fig:residual_vs_gain} & $n=7$\\
        G8     & $4.63\pm 0.74$ & $30.75\pm 14.02$ & $5.22\pm 1.43$ & $-0.65\pm 0.24$  & $5.73\pm 0.11$ & $376.20$ &\ref{fig:HI_signal_gain_err_g8}, \ref{fig:residual_vs_gain}& $n=8$\\
        \hline
        \multicolumn{9}{c}{Frequency-dependent antenna pattern considered, in free space, $N_{\rm poly}=7$ for foreground, see section \ref{sssec:antenna_in_space}}\\
        \hline
        L100   & $1.28\pm 0.73$ & $0.73\pm 0.27$ & $1.05\pm 0.33$ & $-0.69\pm 0.20$              & $6.46\pm 0.07$ & $402.35$&\ref{fig:eq_g_vs_freq}, \ref{fig:dipole_gain_coeff} & $L=100$ cm\\
        L200   & $1.80\pm 0.88$ & $2.08\pm 0.93$ & $1.32\pm 0.41$ & $-0.65\pm 0.20$              & $6.26\pm 0.08$ & $400.84$&\ref{fig:eq_g_vs_freq}, \ref{fig:dipole_gain_coeff} & $L=200$ cm\\
        \hline
        \multicolumn{9}{c}{Frequency-dependent antenna pattern considered, above the ground, $N_{\rm poly}=7$ for foreground; see section \ref{ssec:antenna_on_ground}}\\
        \hline
        L100GN & $1.14\pm 0.74$ & $0.52\pm 0.18$ & $1.21\pm 0.60$ & $-0.66\pm 0.20$              & $6.48\pm 0.09$ & $401.92$&\ref{fig:eq_g_vs_freq}, \ref{fig:dipole_gain_coeff} & $L=100$ cm, $H=50$ cm, latitude=$45^\circ N$ \\
        L200GN  & $0.04\pm 0.17$& $9.97\pm 2.34$ & $10.34\pm 1.82$ & $0.03\pm 0.37$              & $5.89\pm 0.07$ &        &\ref{fig:eq_g_vs_freq}, \ref{fig:dipole_gain_coeff}& $L=200$ cm, $H=100$ cm, latitude=$45^\circ N$ \\
        L100GS & $2.10\pm 0.86$ & $1.77\pm 0.86$ & $1.17\pm 0.37$ & $-0.68\pm 0.20$              & $6.29\pm 0.09$ & $400.25$ &\ref{fig:eq_g_vs_freq}, \ref{fig:dipole_gain_coeff}& $L=100$ cm, $H=50$ cm, latitude=$45^\circ S$ \\
        L200GS & $16.10\pm 1.82$& $0.08\pm 0.12$ & $0.07\pm 0.002$ & $0.36\pm 0.21$              & $3.53\pm 0.03$ &          &\ref{fig:eq_g_vs_freq}, \ref{fig:dipole_gain_coeff}& $L=200$ cm, $H=100$ cm, latitude=$45^\circ S$ \\
        \hline
    \end{tabular}
\end{table*}

\subsection{The Influence of Over-fitting}
\label{ssec:overfitting}
    It is meaningful to inspect if this choice of using 7th-degree polynomials to model the foreground component introduces over-fitting problems so that we perform the sampling with different degrees of polynomials and check the distribution of sampled parameters.
    
    We first simulate the total signal, including both the $\HI$ signal and the foreground emission by using the parameters listed in Table \ref{tbl:ref_model_param}, but without the thermal noise component.
    Then we run the sampling program and draw the parameter distributions.

\begin{figure}
    \centering
    \includegraphics[width=\columnwidth]{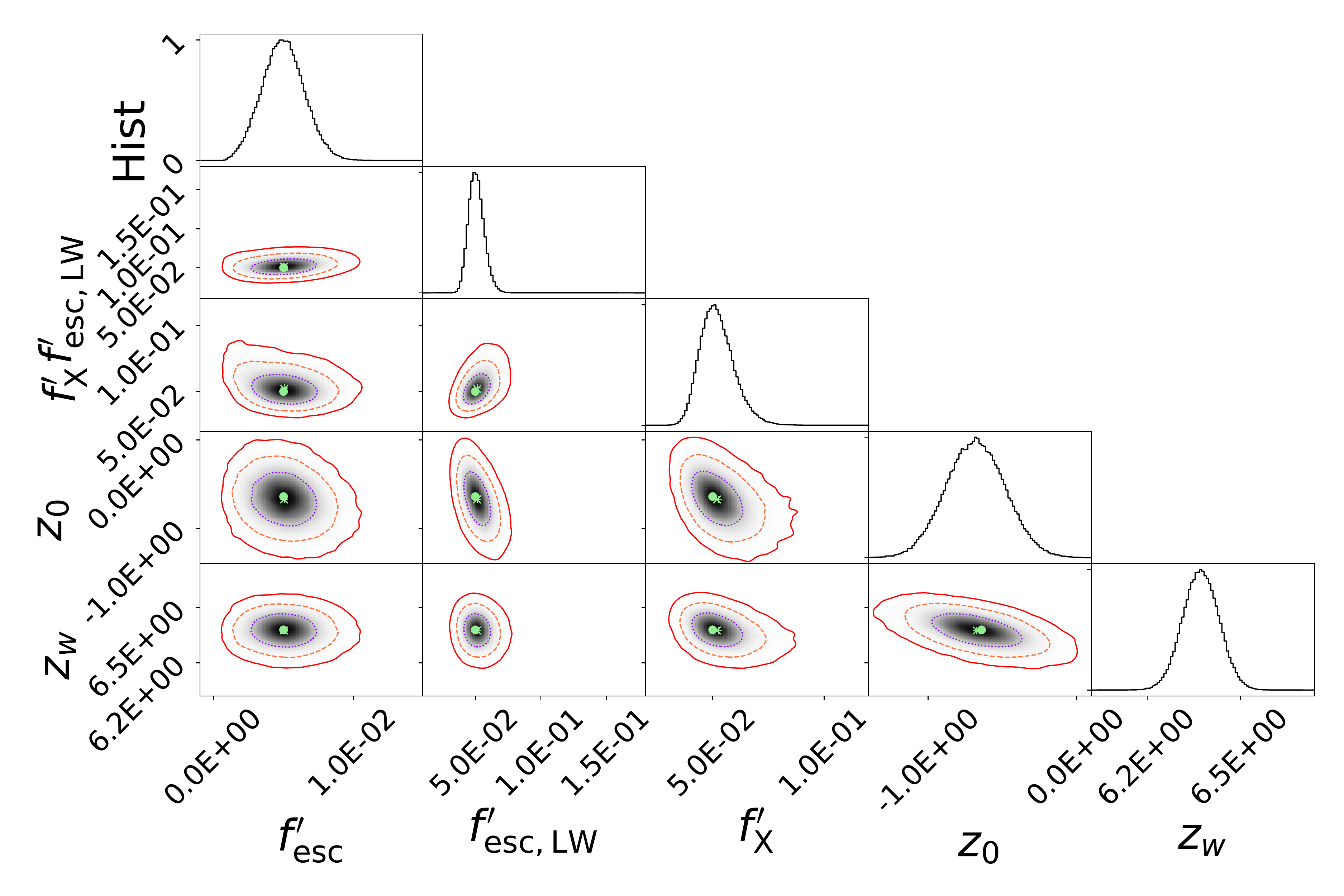}
    \caption{\label{fig:bl3nn}
    Same as Figure \ref{fig:bl7}, except that the noise component is removed from the total signal and a 3rd-degree polynomial is used to model the foreground component, corresponding to test case BL3NN.
    }
\end{figure}

\begin{figure}
    \centering
    \includegraphics[width=.95\columnwidth]{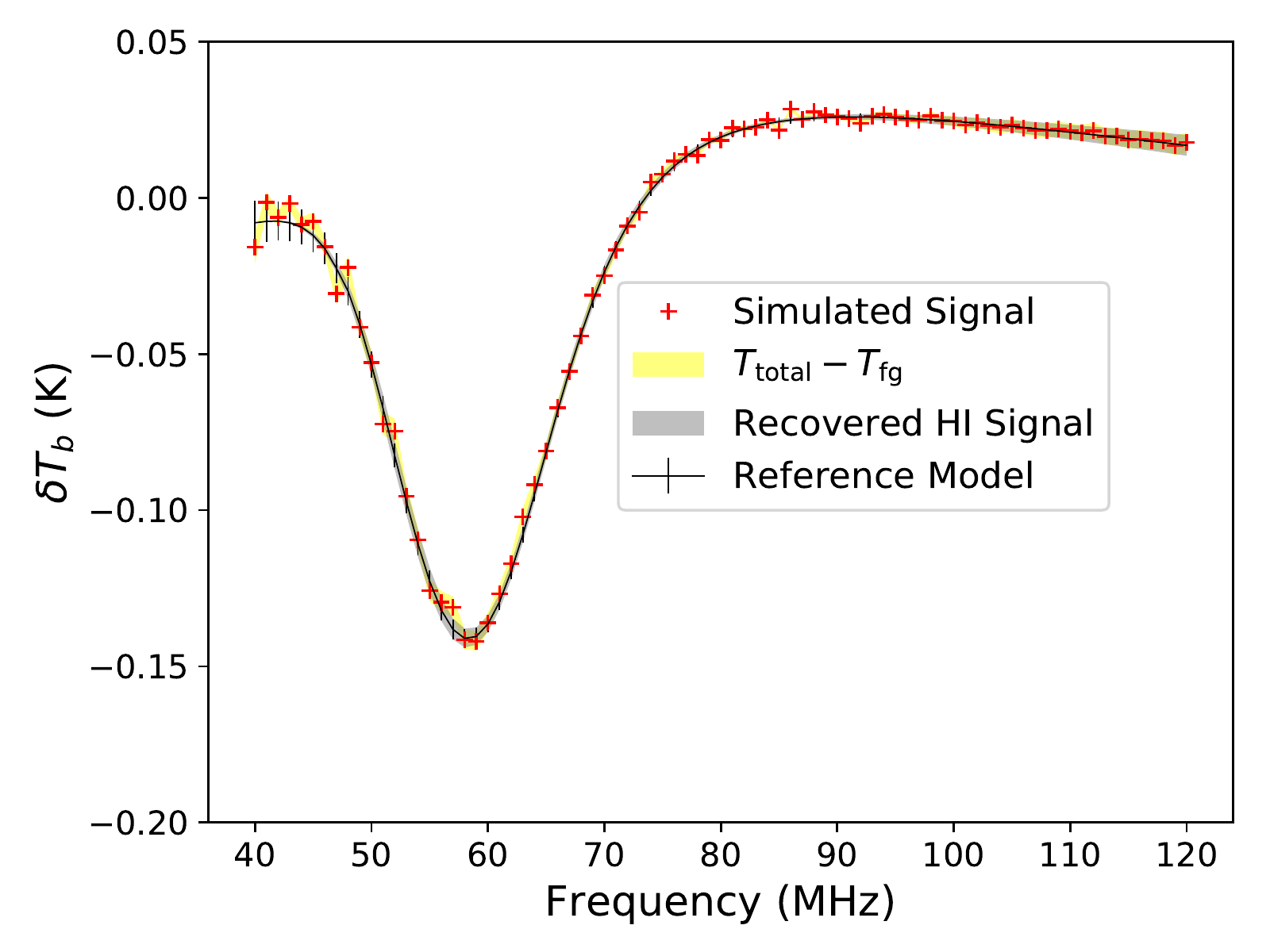}
    \caption{\label{fig:signal_bl3nn}
    Same as Figure \ref{fig:signal_bl7}, except that the noise component is removed from the total signal and a 3rd-degree polynomial is used to model the foreground component, corresponding to test case BL3NN.
    }
\end{figure}

\begin{figure}
    \centering
    \includegraphics[width=\columnwidth]{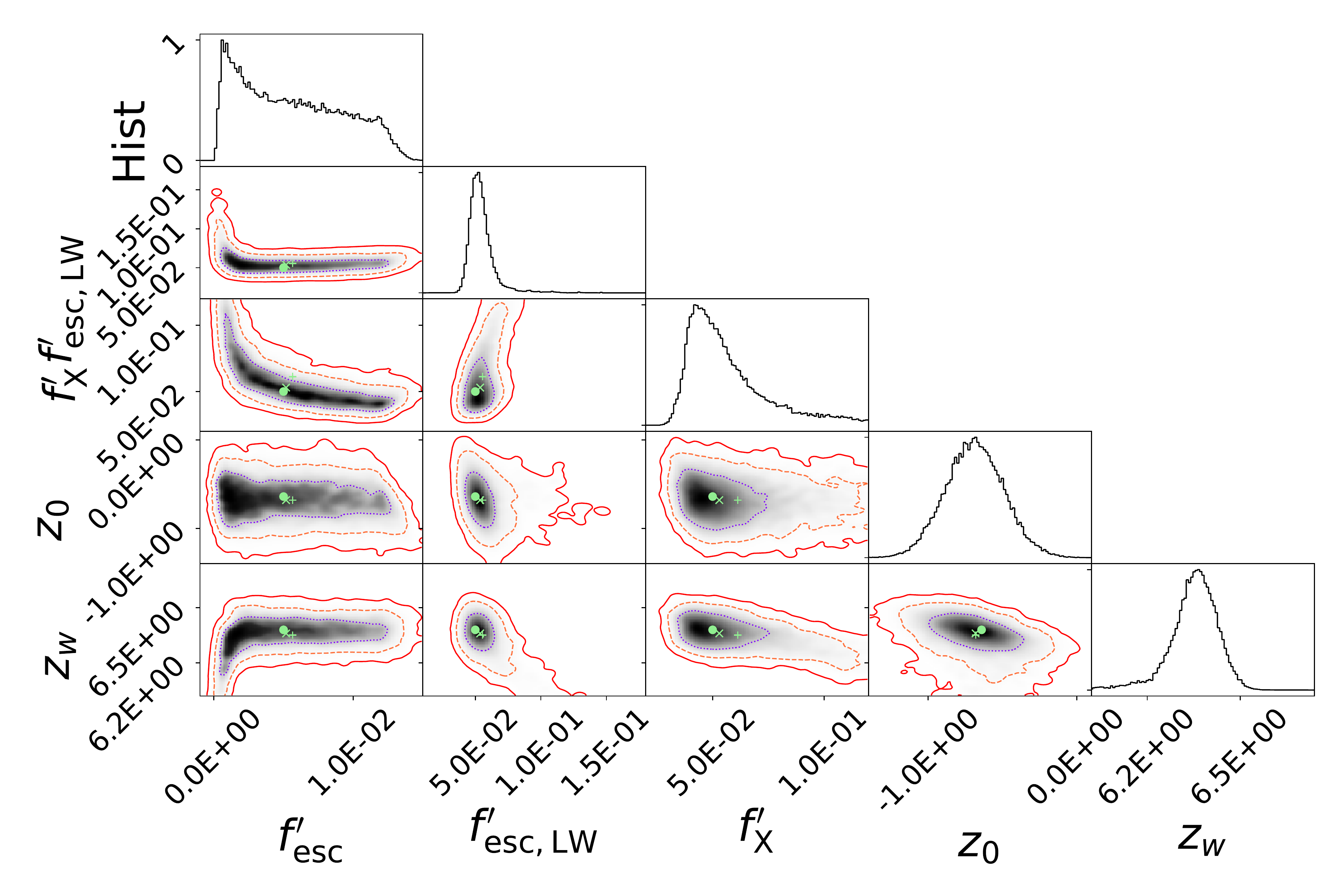}
    \caption{\label{fig:bl4nn}
    Same as Figure \ref{fig:bl7}, except that the noise component is removed from the total signal and a 4th-degree polynomial is used to model the foreground component, corresponding to test case BL4NN.}
\end{figure}

We find that when using a 3rd-degree polynomial to represent the foreground component (labeled as BL3NN), the distributions of parameters are close to the no-foreground condition test as performed in \ref{ssec:space_property}, i.e., close to a multivariate normal distribution, which can be seen in the plot of parameter distributions shown in Figure \ref{fig:bl3nn}.
Figure \ref{fig:signal_bl3nn} shows that the $\HI$ signal can also be recovered unbiasedly.
When the degree of the polynomial is increased to 4th (labeled as BL4NN), the distributions of parameters (Fig. \ref{fig:bl4nn}) begin to deviate from multivariate normal-like distributions. 
The recovered $\HI$ signal resembles the result of test case BL7NN, which is still unbiased and but constrained better. So we do not show it here.
We also test the case when noise is included, and $N_{\rm poly}=3$ (labeled as BL3), and find that both the distribution of parameters and the recovered $\HI$ signal are similar to the test case BL3NN.

So we may reach the conclusion that the over-fitting problem does exist when the degree of freedom of the foreground model reaches a certain value ($N_{\rm poly}=4$ in our test cases).
A foreground component model with too many degrees of freedom may cause the parameter inference too sensitive to thermal noise, and reducing the degree of freedom of foreground model can suppress the over-fitting and the influence of thermal noise.
However, in actual observation data reductions, it is infeasible to use a too simple foreground model.
In the following sections, we will show that the foreground model with proper degrees of freedom is essential to compensate for instrumental gain calibration errors. 
As we will point out in the following sections \ref{ssec:gain_error} and \ref{ssec:antenna_pattern}, a properly modeled foreground will not only help represent the foreground component itself but can also compensate for inaccurately calibrated amplifier gain and frequency-dependent antenna patterns. So we still use a $N_{\rm poly}=7$ polynomial to model the foreground in the following sections.

\subsection{The Influence of Frontend Gain Calibration Error}
\label{ssec:gain_error}

Instrument calibration is crucial to precisely deduce the underlying physical parameters from the observed sky-averaged brightness temperature $T_{b}^{\rm obs}(\nu)$.
The brightness temperature of foreground emission is around $10^3$ K level, while the interesting $\HI$ signal is around $10^{-1}$ K level.
The deviation of \textit{apparent} foreground spectra from a log-polynomial form can easily exceed $10^{-1}$ K level when the precision of amplifier gain flatness calibration is worse than $10^{-4}$.
However, there is still opportunity to recover $\HI$ signal if the foreground model (here we use the 7th-degree log-polynomial) can fit the spectral structure induced by the gain calibration error, given that no significant over-fitting to the global $\HI$ signal happens.
So in this section, we study how the parameter inference can be affected by instrument calibration errors.
Here, we are not going to limit our work to any concrete instrument, but try to propose a universal method to model the frequency-dependent instrument gain calibration error.

Suppose after calibration, the actual frontend gain is 
\begin{gather}
    g(\nu)\equiv 1+\Delta g(\nu),
\end{gather} 
and the observed total spectrum is 
\begin{gather}
    T_{b}^{\rm obs}=T_{b}^{\rm true}g(\nu)\\
    =\delta T_{b}(\nu)(1+\Delta g(\nu))+T_{\rm fg}(1+\Delta g(\nu)).
\end{gather} 
We assume $\Delta g(\nu) \ll 1$, and $\delta T_{b}(\nu)\ll T_{b}^{\rm fg}(\nu)$, so that $\delta T_{b}(\nu)\Delta g(\nu)$ can be ignored.
As a result, 
\begin{gather}
    T_{\rm b}^{\rm obs}\approx \delta T_{b}(\nu)+T_{\rm fg}(1+\Delta g(\nu)).
\end{gather} 
We can define the gain error distorted foreground as 
\begin{gather}
    T_{\rm fg}^{\prime}\equiv T_{\rm fg}(1+\Delta g(\nu)).
\end{gather} 
We substitute Equation \ref{eqn:noise} into above equation and obtain 
\begin{gather} 
    T_{\rm fg}^{\prime}=T_{100}\exp\left [\sum_{n=1}^{N_{\rm poly}}p_n\left(\log\frac{\nu}{100~{\rm MHz}}\right)^{n}\right ]\exp[\log(1+\Delta g(\nu))]\\ 
    \approx T_{100}\exp\left [\sum_{n=1}^{N_{\rm poly}}p_n\left(\log\frac{\nu}{100~{\rm MHz}}\right)^{n}\right ]\exp(\Delta g(\nu))\\ 
    =T_{100}\exp\left [\sum_{n=1}^{N_{\rm poly}}p_n\left(\log\frac{\nu}{100~{\rm MHz}}\right)^{n} + \Delta g(\nu)\right ].
\end{gather}
Theoretically, if $\Delta g(\nu)$ can be expanded with a polynomial about $\log(\nu)$ with a degree not higher than $N_{\rm poly}$, then the gain error distorted foreground $T_{\rm fg}^{\prime}$ can be still fitted by the Equation \ref{eqn:fg}, otherwise, a larger $N_{\rm poly}$ is required. 
However, a too large $N_{\rm poly}$ may lead to the over-fitting problem. 
According to our tests in sections \ref{ssec:thermal_noise} and \ref{ssec:overfitting}, an $N_{\rm poly}=7$ though higher than $1$, which we use to simulate the foreground signal, will cause slight over-fitting, but still acceptable.
Using $N_{\rm poly}>7$ to model the foreground can cause more severe over-fitting.
Practical gain calibration error can have rather complicated spectral structures.
For example, considering Chebyshev filter is used in the frontend amplifier module and this kind of filter has ripple-like frequency response. 
If a frequency-independent relative gain calibration error exists, the ripple-like feature will be kept after calibration.
Such kinds of calibration error have complicated spectral structure and it is easy to verify that they cannot be well represented by a polynomial about $\log(\nu)$ with a degree not higher than $7$.
So the question now is how precise the instrument needs to be calibrated in order to perform a meaningful detection to the global $\HI$ signal, and how serious the gain calibration error can interfere with the parameter inference.

A model is required to describe the gain calibration error as a function of frequency.
The calibration error as a function of frequency can be expressed as the linear combination of a series of Legendre Polynomials as
\begin{gather}
	\Delta g(\nu)=\sum_{n}^{\infty} a_n P_n(x) \label{eqn:gain_model}
\end{gather}
where 
\begin{gather}
	x=\frac{2\log(\nu/\nu_{\min})}{\log(\nu_{\max}/\nu_{\min})}-1,\label{eqn:x}
\end{gather}  
\begin{gather}
	P_n(x)=\frac{1}{2^n}\sum_{k=0}^n \binom {n} {k} (x+1)^k
\end{gather} 
is the Legendre polynomial of order $n$, $a_n$ is the corresponding combination coefficient.
The most important reason why we do not simply use a general form of polynomial $\sum_{n=1}^N a_n x^n$ is that the Legendre polynomials are quasi-periodic on the definition domain, and the `period' decreases as the order increases.
This property makes the Legendre polynomials a suitable formula to express the spectral structures induced by gain errors at different complexity levels.
If we want to find some actual examples with similar features from practical devices, one may recall the ripple structure in the frequency response of Chebyshev filters, if the frequency response were not properly calibrated such kind of quasi-periodic oscillating spectral structure will appear in the final data.

Two conditions with $a_n=10^{-5}\delta_{nm}$ ($\delta$ is the Kronecker delta), where $m=7$ (labeled as G7) or $m=8$ (labeled as G8) are tested, respectively. 
For the G7 test case
\begin{gather}
    \Delta g_7(\nu)=\frac{10^{-5}}{16}(429x^7-693x^5+315x^3-35x),
\end{gather}
and for the G8 test case
\begin{gather}
    \Delta g_8(\nu)=\frac{10^{-5}}{128}(6435x^8-12012x^6+6930x^4-1260x^2+35),
\end{gather}
where $x$ is defined as above Equation \ref{eqn:x}.
For both test cases, a $N_{\rm poly}=7$ polynomial is used to model the foreground.

\begin{figure}
    \centering
    \includegraphics[width=.95\columnwidth]{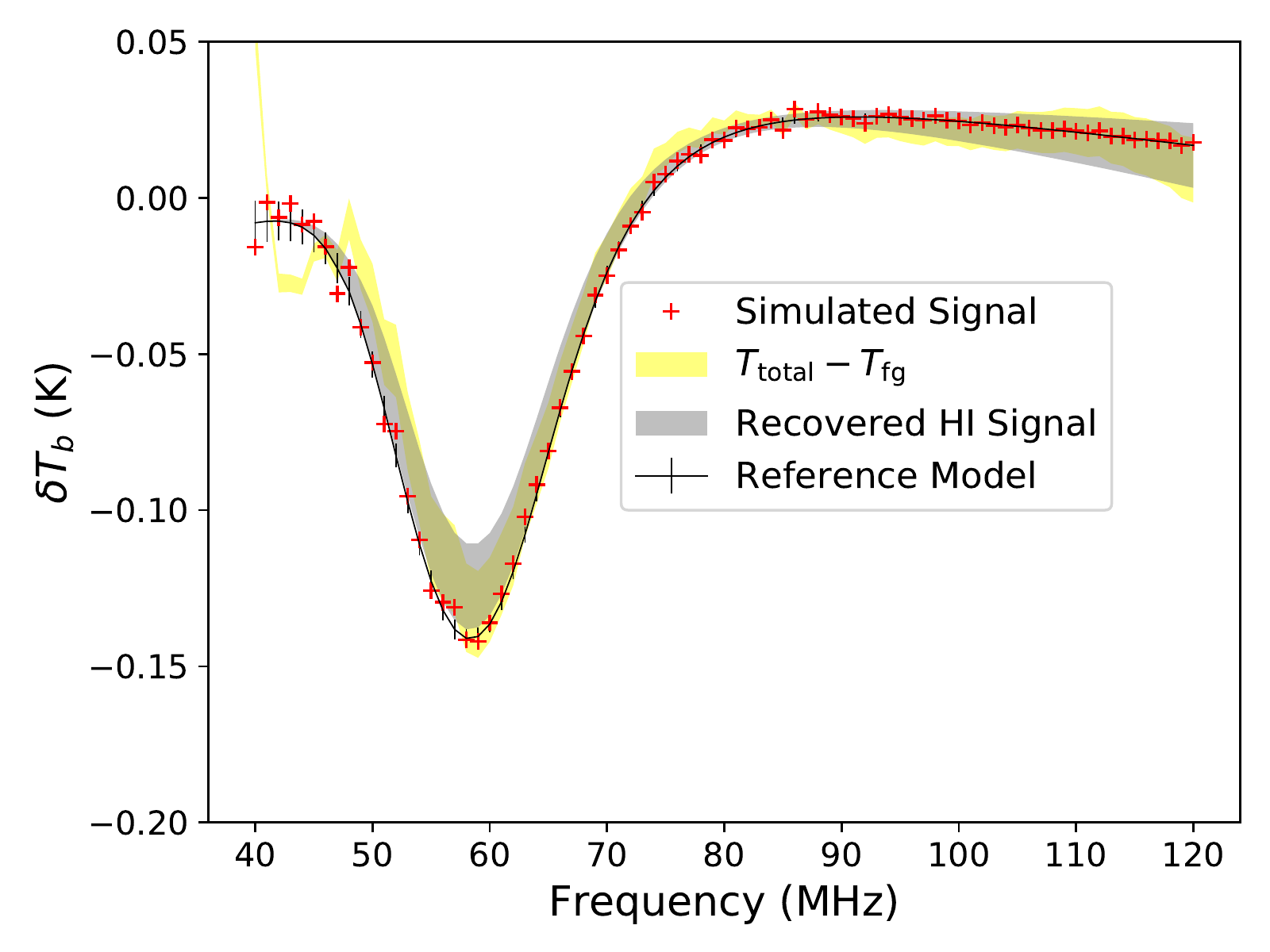}
    \caption{\label{fig:HI_signal_gain_err_g7}
    Same as Figure \ref{fig:signal_bl7}, except that the noise component is removed from the total signal, corresponding to test case G7.
    }
\end{figure}

\begin{figure}
    \centering
    \includegraphics[width=.95\columnwidth]{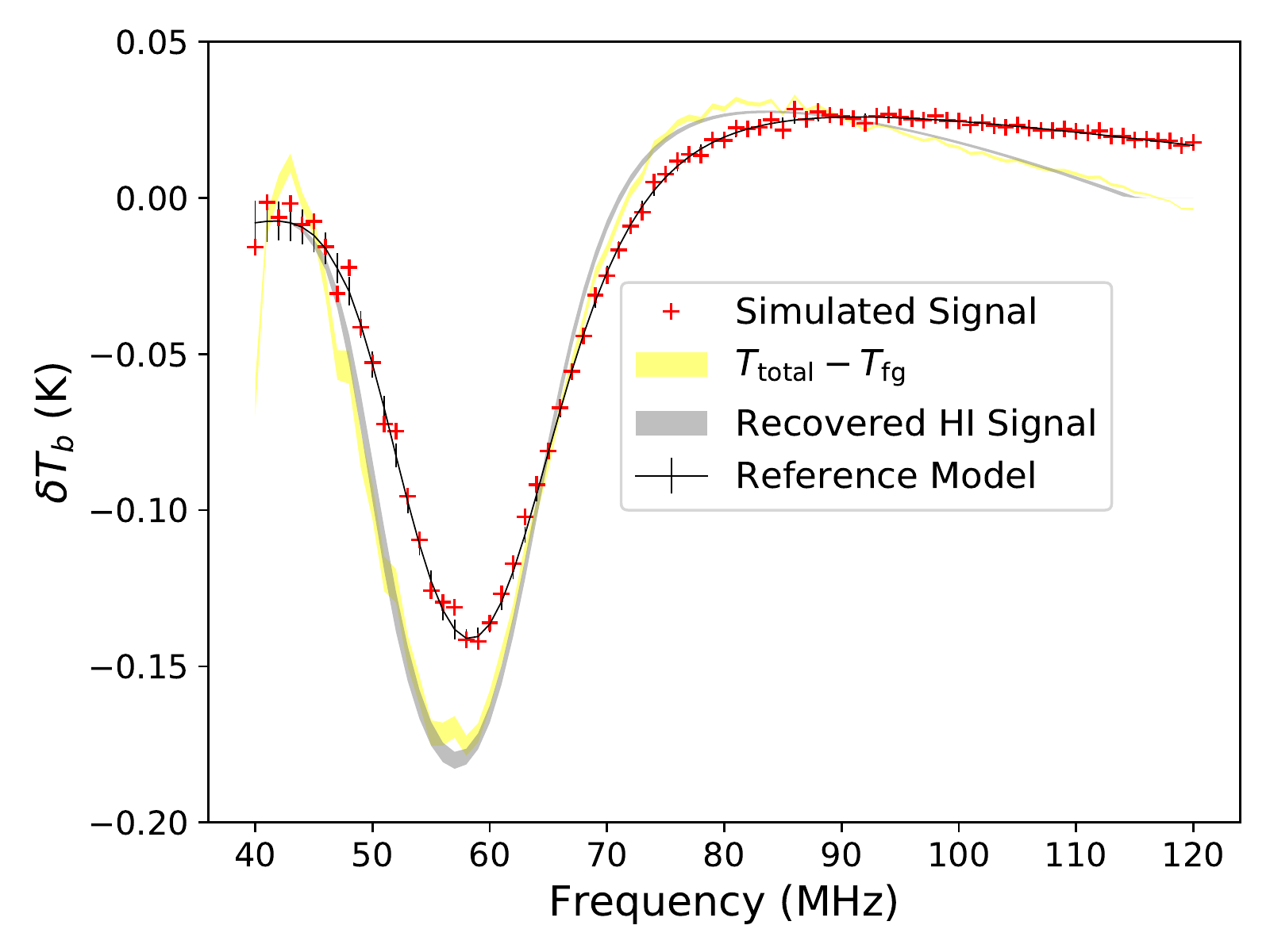}
    \caption{\label{fig:HI_signal_gain_err_g8}
    Same as Figure \ref{fig:signal_bl7}, except that the noise component is removed from the total signal, corresponding to test case G8.
    }
\end{figure}

About the determination of the value of $a_{n}$, we do not have any nominate parameter of actual amplifiers used in global $\HI$ signal detection experiments, and can only give an estimation based on the performance of commercial products. 
Our estimations are based on following facts: 1) typical gain error of commercial radio frequency amplifier chips (e.g., Analog Device AD8079) is around $10^{-3}$; 2) the tolerance of high accuracy resistors can reach $5\times 10^{-5}$; 3) the tolerance of high accuracy capacitor can reach 1\%; 4) typical measurement error of general-purpose commercial measuring receivers (e.g., Agilent 8902A, R\&S FSMR series) is around 1\%.
Considering there can be a significant potential of improving the measurement accuracy for specially designed receivers, we set $a_n$ to be $10^{-5}\delta_{nm}$, which is higher than above commercial measuring receivers by about 3 orders of magnitude. 
We expect the $N_{\rm poly}=7$ foreground model can handle the G7 test case, while the G8 test case will significantly bias the results.
The results are summarized in Table \ref{tbl:params}.
Same as section \ref{ssec:thermal_noise}, we plot the recovered $\HI$ signals together with the input signals in Figure \ref{fig:HI_signal_gain_err_g7} and \ref{fig:HI_signal_gain_err_g8}.

\subsection{The Influence of Frequency-Dependent Antenna Pattern}
\label{ssec:antenna_pattern}
In most of the current work about detecting global averaged $\HI$ signal, it is assumed that a simple dipole-like antenna (not necessarily to be exactly dipole antenna) is used to receive the signal from very large sky area.
According to e.g., \cite{2009tra..book.....W}, given the normalized power pattern of an antenna $P_{\rm n}(\mathbf{n})$ and the sky brightness temperature distribution $T_{\rm b}(\mathbf{n})$, both of which are the functions of direction $\mathbf{n}$, the instant antenna temperature 
\begin{gather}
    T_{\rm A}(\nu)=\frac{\int T_{\rm b}(\mathbf{n}, \nu)P_{\rm n}(\mathbf{n}, \nu)\D \Omega}{\int P_{\rm n}(\mathbf{n}, \nu)\D\Omega}, \label{eqn:T_A}
\end{gather} 
which is actually the weighted mean value of the sky brightness temperature.
When the weight changes, the output antenna temperature changes accordingly.
The power patterns for most antennas are frequency-dependent so that it will create artificial spectral structures.
When ground effect cannot be ignored, more significant frequency-dependent feature will appear.
As a result, the effect of FDAP is potentially nontrivial.
This issue has been pointed out by previous works \citep[e.g.,][]{2013PhRvD..87d3002L}, here we perform some numeric computations, in order to offer practical guides about the antenna design for such kinds of experiments.
In more general conditions, when the orientation of the antenna changes with time, mainly because of the diurnal motion, the time-averaged antenna temperature 
\begin{gather}
    \bar{T}_{\rm A}(\nu)=\frac{1}{\tau}\int\frac{\int T_{\rm b}(\mathbf{M}(t)\mathbf{n}, \nu)P_{\rm n}(\mathbf{n}, \nu)\D \Omega}{\int P_{\rm n}(\mathbf{n}, \nu)\D\Omega}\D t\\
    =\frac{\tau^{-1}\iint T_{\rm b}(\mathbf{M}(t)\mathbf{n}, \nu)P_{\rm n}(\mathbf{n},\nu)\D\Omega \D t}{\int P_{\rm n}(\mathbf{n}, \nu)\D\Omega} \label{eqn:T_A_rotating}
\end{gather} 
is actually used, where $\mathbf{M}(t)$ is the rotation matrix.
When the antenna is fixed in space, $\mathbf{M}$ is a constant matrix, so that Equation \ref{eqn:T_A_rotating} degenerates into Equation \ref{eqn:T_A}.

Note that as the $\HI$ signal is regarded to be isotropic on large scales, the influence of FDAP to the global $\HI$ spectrum can be ignored, so we only need to calculate its influence to the foreground signal.
Some all-sky survey map is required to evaluate Equations \ref{eqn:T_A} and \ref{eqn:T_A_rotating}. We use the map $T_{408}^{\rm Haslam}(\mathbf{n})$ produced by \cite{1981A&A...100..209H} as a sky template. In order to ensure that in the condition of a perfect omnidirectional antenna, the calculated global averaged foreground spectrum degenerates to Equation \ref{eqn:fg}, we renormalize it to 
\begin{gather}
    T_{\rm fg}(\mathbf{n}, 408~{\rm MHz})=4\pi T_{100}\left(\frac{408~{\rm MHz}}{100~{\rm MHz}}\right)^{p_1}\frac{T_{408}^{\rm Haslam}(\mathbf{n})}{\int T_{408}^{\rm Haslam}(\mathbf{n})\D\Omega},
\end{gather}
then
\begin{gather}
    T_{\rm fg}(\mathbf{n}, \nu)=4\pi T_{100}\left(\frac{\nu}{100~{\rm MHz}}\right)^{p_1}\frac{T_{408}^{\rm Haslam}(\mathbf{n})}{\int T_{408}^{\rm Haslam}(\mathbf{n})\D\Omega},\label{eqn:all_sky}
\end{gather}
so that when the antenna beam is frequency-independent and perfectly isotropic, i.e., $P_{\rm n}(\mathbf{n}, \nu)\equiv 1$, the sky averaged foreground spectrum
\begin{gather}
    T_{\rm fg}(\nu)=\frac{\int T_{\rm fg}(\mathbf{n},\nu)\D\Omega}{\int \D\Omega}\equiv T_{100}\left(\frac{\nu}{100~{\rm MHz}}\right)^{p_1}, 
\end{gather}
which is actually Equation \ref{eqn:fg} with $N_{\rm poly}=1$.

With Equation \ref{eqn:all_sky} in hand, let us evaluate the above Equations \ref{eqn:T_A} for an antenna with fixed orientation in free space and \ref{eqn:T_A_rotating} for an antenna on the ground.
For an antenna with fixed orientation in free space, 
\begin{gather}
    T_{\rm A}(\nu)=T_{100}\left(\frac{\nu}{100~{\rm MHz}}\right)^{p_1}4\pi\frac{\int P_{\rm n}T_{408}^{\rm Haslam}(\mathbf{n})\D\Omega}{\int T_{408}^{\rm Haslam}(\mathbf{n})\D\Omega\int P_{\rm n}\D\Omega}\\
    =T_{100}\left(\frac{\nu}{100~{\rm MHz}}\right)^{p_1}\frac{\int P_{\rm n}T_{408}^{\rm Haslam}(\mathbf{n})\D\Omega}{\int P_{\rm n}\D\Omega}\left(\frac{\int T_{408}^{\rm Haslam}(\mathbf{n})\D\Omega}{4\pi} \right)^{-1}\\
    \equiv T_{100}\left(\frac{\nu}{100~{\rm MHz}}\right)^{p_1}g_A^\prime(\nu),
\end{gather}
where the equivalent gain $g_A^\prime(\nu)$ is defined as
\begin{gather}
    g_A^\prime(\nu)\equiv\frac{\int P_{\rm n}(\mathbf{n}, \nu)T_{408}^{\rm Haslam}(\mathbf{n})\D\Omega}{\int P_{\rm n}(\mathbf{n}, \nu)\D\Omega}\left(\frac{\int T_{408}^{\rm Haslam}(\mathbf{n})\D\Omega}{4\pi} \right)^{-1}.
\end{gather}
For an antenna on the ground,
\begin{gather}
    \bar{T}_{\rm A}(\nu)=T_{100}\left(\frac{\nu}{100~{\rm MHz}}\right)^{p_1}\frac{\tau^{-1}\iint T_{408}^{\rm Haslam}(\mathbf{M}(t)\mathbf{n})P_{\rm n}\D\Omega \D t}{(4\pi)^{-1}\int T_{408}^{\rm Haslam}(\mathbf{n})\D\Omega\int P_{\rm n}\D\Omega}\\
    =T_{100}\left(\frac{\nu}{100~{\rm MHz}}\right)^{p_1}\times \notag\\
    \frac{\tau^{-1}\iint T_{408}^{\rm Haslam}(\mathbf{M}(t)\mathbf{n})P_{\rm n}\D\Omega \D t}{\int P_{\rm n}\D\Omega}
    \left(\frac{\int T_{408}^{\rm Haslam}(\mathbf{n})\D\Omega}{4\pi}\right)^{-1}\\
    \equiv T_{100}\left(\frac{\nu}{100~{\rm MHz}}\right)^{p_1}g_A^\prime(\nu),
\end{gather}
where the equivalent gain $g_A^\prime(\nu)$ is defined as
\begin{gather}
    g_A^\prime(\nu)\equiv\frac{\tau^{-1}\iint P_{\rm n}(\mathbf{n}, \nu)T_{408}^{\rm Haslam}(\mathbf{M}(t)\mathbf{n})\D\Omega \D t}{\int P_{\rm n}(\mathbf{n}, \nu)\D\Omega}
    \left(\frac{\int T_{408}^{\rm Haslam}(\mathbf{n})\D\Omega}{4\pi}\right)^{-1}.
\end{gather}
As a result, the influence of FDAP to foreground spectra can be uniformly expressed as an equivalent gain and dealt within the mathematical framework that we described in section \ref{ssec:gain_error}.
It is also notable that when calculating above equivalent gain, we do not need the sky template in every single frequency, and its normalization is not important, either.

Before applying the equivalent gain to the foreground spectra, we further renormalize it as 
\begin{gather}
    g_A(\nu)=\frac{g^\prime_A(\nu)}{B^{-1}\int g^\prime_A(\nu)\D\nu},\label{eqn:equiv_gain}
\end{gather} 
where $B$ is the total bandwidth. 
The purpose of this step is to shift the mean equivalent gain over the whole bandpass to $1$ so that the produced foreground spectra is comparable to our baseline test cases (e.g., BL7).
In other words, although the normalization of the foreground spectrum can be affected by FDAP, here we decide to eliminate this factor to ensure that the differences in results are caused by FDAP itself purely, rather than the change of relative strength of foreground and $\HI$ signals.
Following the above steps, the corresponding equivalent $g_A(\nu)$ is calculated for each test case and multiplied to the foreground spectrum.

\subsubsection{A Dipole in Free Space with Fixed Orientation}
\label{sssec:antenna_in_space}
As a relatively simple beginning, we assume a dipole in free space (for example, on a satellite) with the orientation fixed. 
The normalized power pattern of a dipole antenna is 
\begin{gather}
P_{\rm n}(\theta)=\frac{1}{P_{\max}}\left[\frac{\cos(\frac{kL}{2}\cos\theta)-\cos(\frac{kL}{2})}{\sin\theta}\right]^2,\label{eqn:dipole_pattern}
\end{gather} 
where the polar angle $\theta$ is $0$ or $\pi$ on the axis of the dipole, $k$ is the wave number, and $L$ is the length of the dipole. 
We substitute above Equation into Equation \ref{eqn:T_A}, then use Equation \ref{eqn:equiv_gain} to calculate the equivalent gain and apply to the foreground signal spectra.

\begin{figure}
    \centering
    \includegraphics[width=.48\columnwidth]{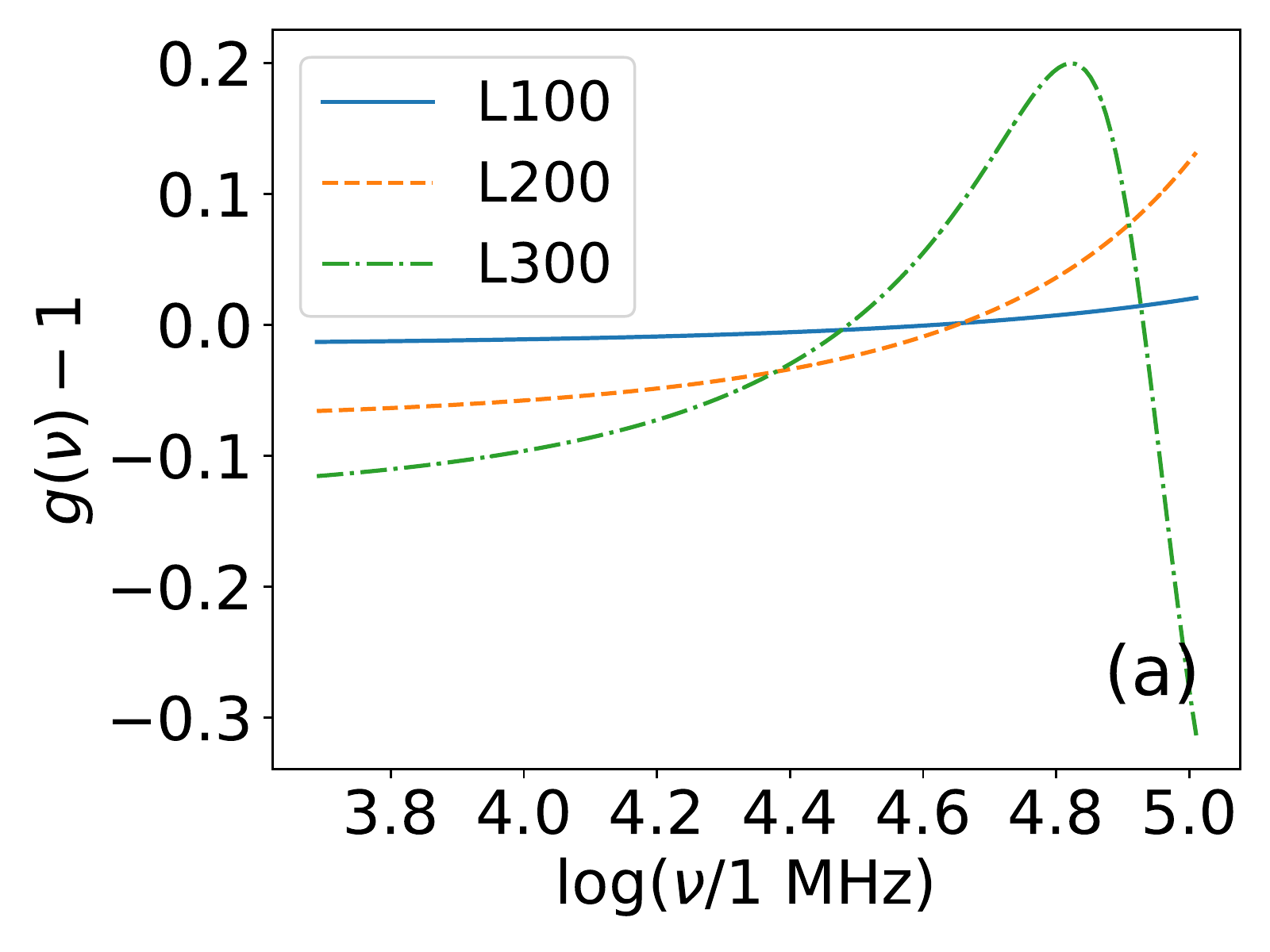}
    \includegraphics[width=.48\columnwidth]{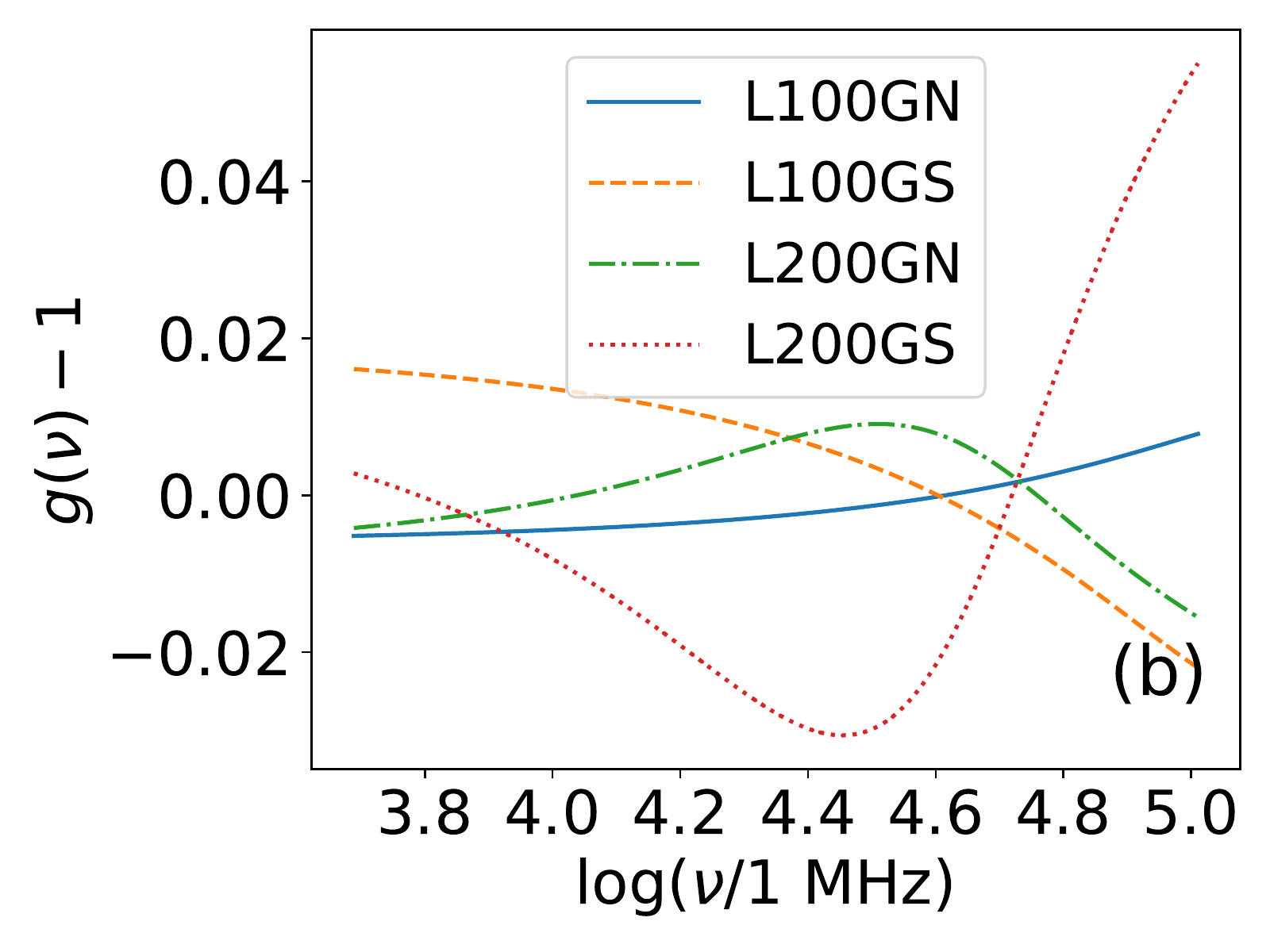}
    \caption{\label{fig:eq_g_vs_freq}
        Equivalent gain of dipole antennas with different length fixed in free space (a), and fixed on the earth (b). 
        The results are derived with Equation \ref{eqn:equiv_gain}.
    }
\end{figure}

We study dipole antennas with $L=100,~200$, and $300$ cm, respectively. 
We first plot the equivalent gain of dipole antennas with different length calculated with Equation \ref{eqn:equiv_gain} in Figure \ref{fig:eq_g_vs_freq} (a). 
Apparently, we can directly rule out the $L=300$ cm dipole. 
For $L=100$ cm and $L=200$ cm dipole antennas, though the deviation of their equivalent gain from $1$ has reached $10^{-3}$ to $10^{-2}$ level, it is possible that they can still be handled by the $N_{\rm poly}=7$ foreground model. 
So we perform following tests for these two dipole antennas.

\begin{figure}
    \centering
    \includegraphics[width=.48\columnwidth]{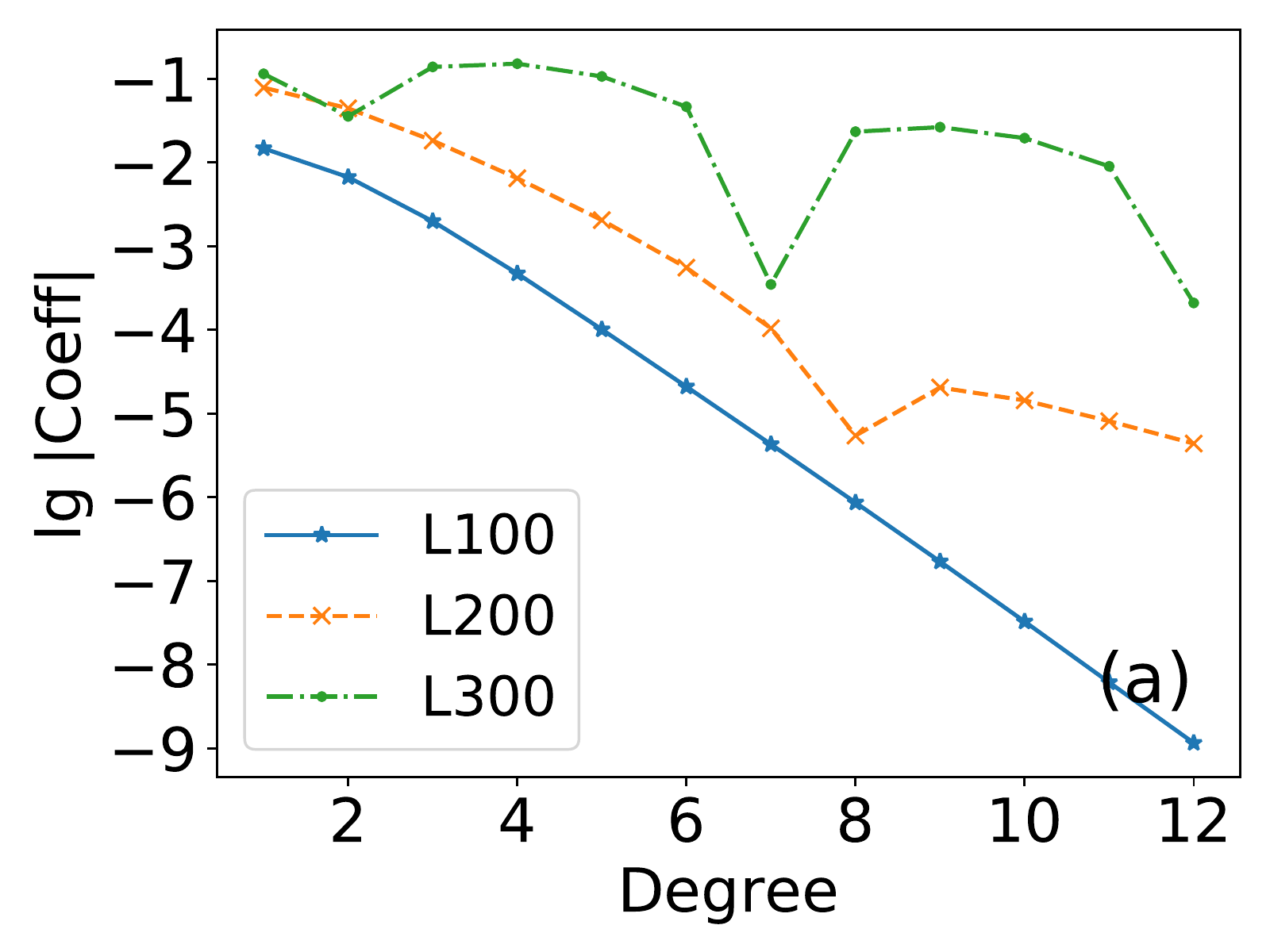}
    \includegraphics[width=.48\columnwidth]{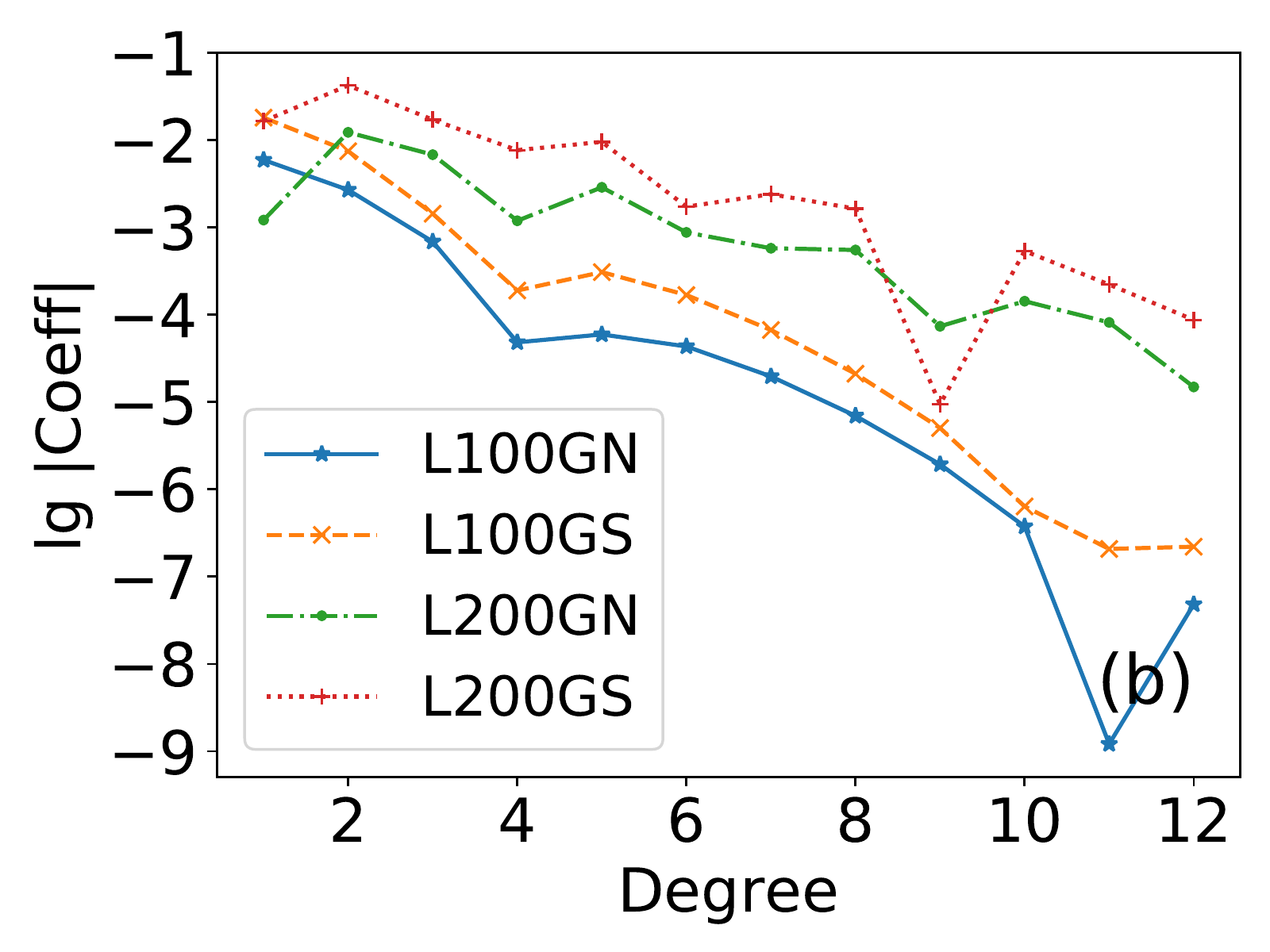}
    \caption{\label{fig:dipole_gain_coeff}
    The common logarithm of the absolute values of the coefficients calculated by expanding the equivalent gain of dipole antennas with different lengths that are fixed in free space (a) or fixed on the earth (b) with Legendre Polynomials.
    }
\end{figure}

We generate the new total sky averaged signal by using the above equivalent gain caused by FDAP, which is then fed into the MCMC sampling program. 
We find that for both $L=100$ cm and $L=200$ cm conditions (denoted as L100 and L200, respectively), the MCMC can quickly reach a convergence. 
The inferred parameters are listed in Table \ref{tbl:params}.
It is not surprising that the inferred parameters of the $L=100$ cm condition are slightly closer to the input values than the $L=200$ condition.
As we stated above, by expressing FDAP into equivalent gain makes it easier to attribute the bias of result to the artificial spectral structures and compare it to the amplifier gain caused deviation that is described in the previous section. 
So we expand the equivalent gain with Legendre polynomials and inspect the amplitude of coefficients of different degrees.
The plots of coefficients of different degrees are shown in Figure \ref{fig:dipole_gain_coeff} (a). 
Obviously, the coefficients of L300 condition for degrees larger than 7 is significantly higher than that of L100 and L200.
The coefficients of L200 is only slightly higher than that of L100.
This agrees well with our conclusion above that the required order of polynomials to approximate the equivalent gain can strongly affect whether the parameter inference is biased.

\subsubsection{A Dipole Antenna on the Ground}
\label{ssec:antenna_on_ground}
Then let us consider another more practical condition---ground-based experiments. 
We assume a dipole antenna horizontally placed at some certain height above the ground. 
This brings two changes compared to an antenna in free space: 
1) the earth will block the signal of the hemisphere below the horizon, and 2) ground effects involve. 
The first change can be handled rather straight forward, while for the ground effect, a ground model is needed. 
Here we assume an infinitely large perfect conductive ground plane so that the antenna pattern can be calculated by simply assuming a mirror antenna. 
Because of the diurnal motion, the sky above the horizon changes with time, and Equation \ref{eqn:T_A_rotating} should be used. 
The effective area of the antenna is zero in directions below the horizon.
Obviously, the received signal is related to the latitude of the point where the antenna is placed, so two representative latitudes are considered: $45^{\circ}00^\prime00^{\prime\prime}N$ and $45^{\circ}00^\prime00^{\prime\prime}S$.
The orientation of the dipole antennas are assumed to be along the meridian, and two lengths $L=100$ cm and $L=200$ cm are assumed.
The height $H$ is assumed to be $L/2$ correspondingly.
The equivalent gains of the four conditions are calculated and shown in Figure \ref{fig:eq_g_vs_freq} (b).
We list the results in Table \ref{tbl:params}, labeled as L100GN, L200GN, L100GS, and L200GS, respectively. 

From Figure \ref{fig:eq_g_vs_freq} (b), we can conclude that the equivalent gain error of conditions L100GS and L200GS are larger than conditions L100GN and L200GN. 
On the other hand, we find that the result of L100GN is best among all the ground-based conditions, L100GS is worse, and L200GN and L200GS conditions give result significantly biased from the input values.
This phenomenon implies that the latitude of the survey site can significantly affect the results. 
Then just like what we have done above, we expand the equivalent gain with Legendre polynomials and plot the coefficients verse the polynomial degree in Figure \ref{fig:dipole_gain_coeff} (b).
By checking the amplitudes of coefficients of degree higher than 7, we find the relative amplitudes of different conditions agree well with the qualities of parameter inference: better inference quality corresponds to smaller coefficients.
This again confirms our above hypothesis that the spectral complexity is a more critical factor than the amplitude of gain calibration error that influences the quality of parameter inference.

\section{Discussion}
\label{sec:discussion}
\subsection{Are Directly Inferred Parameter Unbiased?}
As is mentioned in e.g., \cite{2016MNRAS.455.3829H}, the two-stage fitting method can obtain biased results, and it would be interesting to see if biases in IGM properties would persist even if the signal were fit with the exact model used to generate it.
According to our tests shown in sections \ref{ssec:thermal_noise} and \ref{ssec:overfitting}, the answer to this question is not as straightforward as it looks.
As we have presented in section \ref{ssec:bias_standard}, without the analytical expression of posterior distribution probability, it is hard to answer this question strictly.
We have to make some compromises by comparing input parameters with the inferred confidence range (represented as standard deviations for single parameters and contours for joint distributions of pairs of parameters).
In section \ref{ssec:thermal_noise}, we note that the parameter $f^\prime_{\rm esc, LW}$ seems to be underestimated when thermal noise is involved (test case BL7).
Then by removing thermal noise from the total signal (but still keep $\sigma_{\nu_i}$'s when computing posterior probability) to form test case BL7NN, we find that the inference is actually unbiased, and the seemingly biased results are caused by certain noise realization.
We also note that there are slightly multimodal features in the corner plots (Fig. \ref{fig:bl7} and \ref{fig:bl7nn}).
From test cases BL3NN and BL4NN, we find that reducing $N_{\rm poly}$ from 4 to 3 will significantly change inferred parameter distributions from slightly multimodal to multivariate normal-like distribution.
Furthermore, when $N_{\rm poly}=3$, even if noise is brought back to the total signal, the inferred parameters are still very close to input values, which means the degrees of freedom of foreground model will affect the sensitivity to thermal noise.

As a result, we can conclude that using 7th-degree polynomials to model foreground and directly inferring parameters with the exact model used to generate the signal can obtain unbiased results if only thermal noise is involved.

\subsection{Factors that Should be Considered in Practical Global $\HI$ Signal Detection}
\subsubsection{Instrumental Gain Calibration}
\label{ssec:instru_err}
Through the tests described in Section \ref{ssec:gain_error}, we have shown when the instrumental gain error as a function of frequency cannot be well approximated by a low degree polynomial, very tiny gain calibration error (e.g., $10^{-5}$ as we have tested) can cause both the underlying physical model parameter and the recovered $\HI$ signal to significantly bias from the true values. Higher degree polynomials will be needed, but a too high polynomial degree can lead to the over-fitting problem (see section \ref{ssec:overfitting}). The reason why this issue must be considered seriously is that we cannot simply treat the instrumental calibration error same as thermal noise, which is independent in each frequency channel and can be added to the $\sigma^2$ term in Equation \ref{eqn:likelihood}. Our result still cannot clarify the requirement of instrumental calibration precision, but we have shown that even at the level of $10^{-5}$, the instrumental calibration error can be nontrivial.

\begin{figure}
    \centering
    \includegraphics[width=.48\columnwidth]{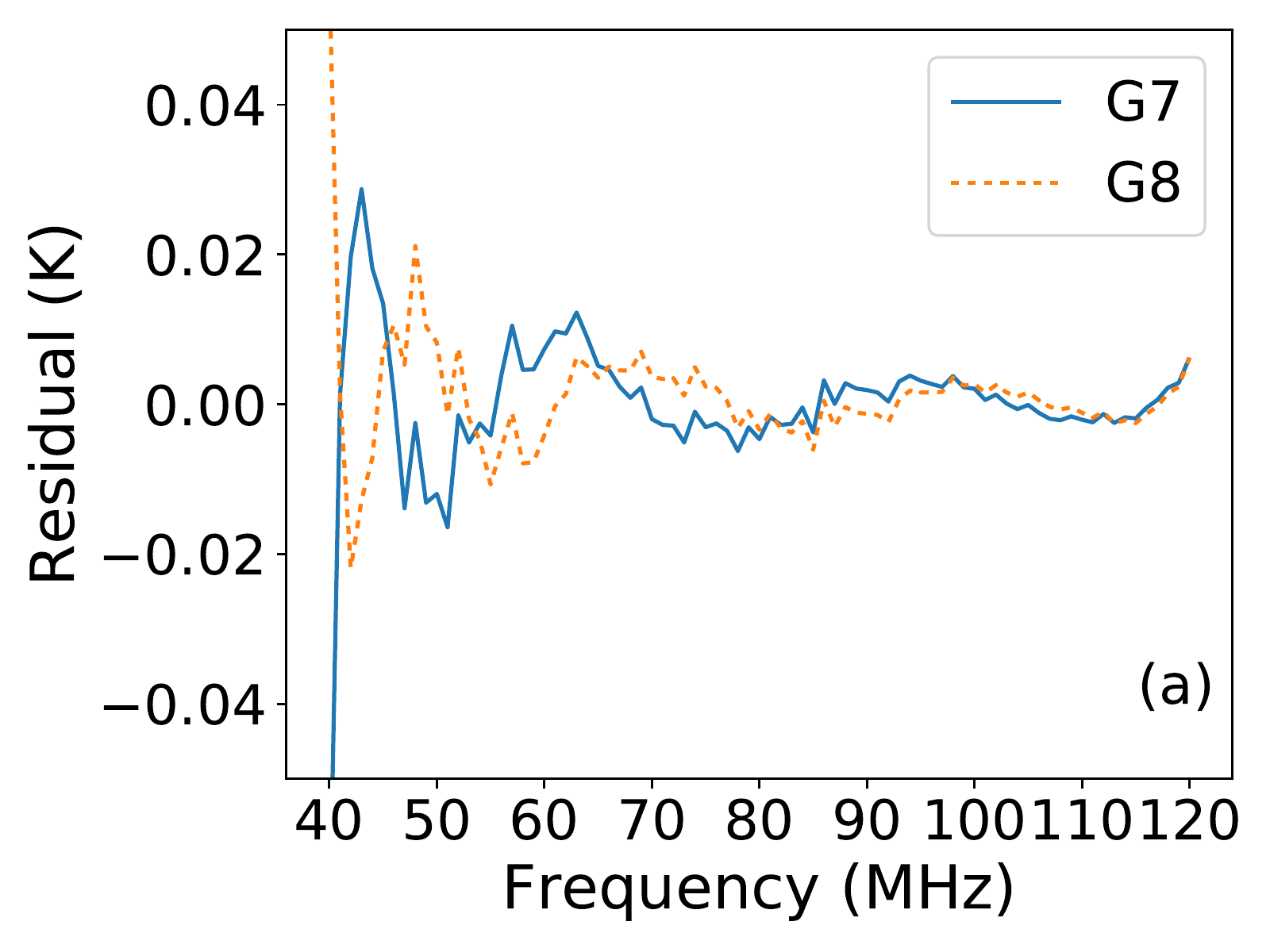}
    \includegraphics[width=.48\columnwidth]{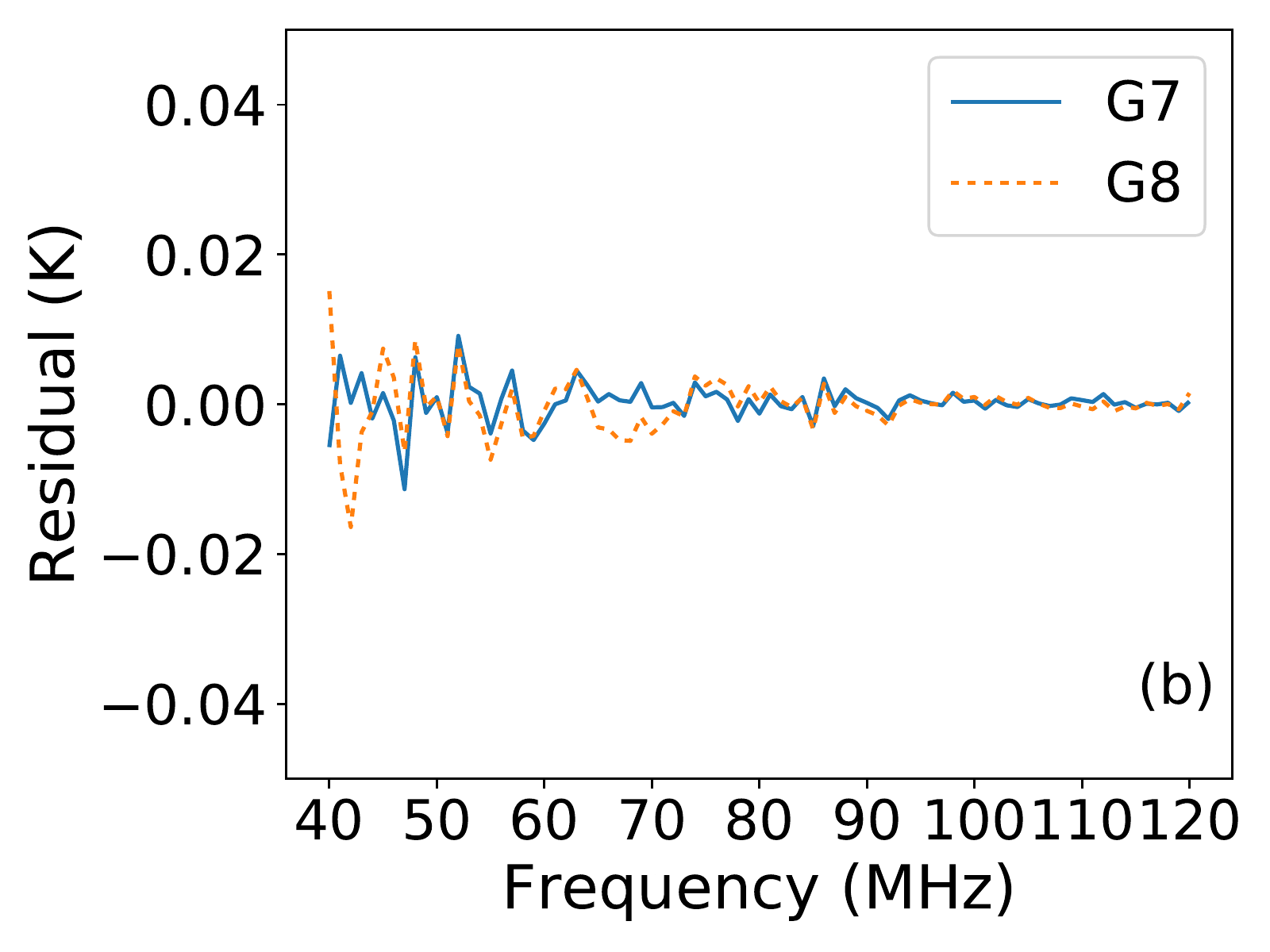}
    \caption{\label{fig:residual_vs_gain}
        Residuals calculated as the total signal and the model predictions for test cases G7 and G8. In Figure (a), parameters for simulating the signals are used, while in Figure (b), optimal parameters (i.e., with the highest posterior probability) are used.
    }
\end{figure}

We are curious about whether the bias is caused by the calibration error, or caused by the malfunction of the sampling method.
In order to discriminate the above two possibilities, we plot the residual calculated with the parameters that are used to simulate the signal and the optimal parameters (i.e., parameters with the maximum posterior probability) in Figure \ref{fig:residual_vs_gain} (a) and (b), respectively. 
Apparently, for both G7 and G8 conditions, the residuals corresponding to optimal parameters are smaller than the residual corresponding to input parameters. 
Meanwhile, for the optimal parameters, the residual of G8 is larger than the G7 condition; this is in agreement with our above results that the G7 condition can lead to an acceptable result, while G8 condition cannot. 
Also, after examining the parameter distribution in parameter space and the plot of parameter values versus sample numbers, we conclude that for both G7 and G8 conditions, the Markov chains have reached convergences. 
So we could safely rule out the possibility of the malfunction of the sampling method. 

Although we use only mathematics-based models to represent calibration problems, instead of actual frontend electronic devices (mainly a series of amplifiers, filters, and in some conditions, frequency mixers), we suggest that there are at least two potential sources could cause frequency-dependent gain calibration error.
One is the broadband amplifier gain flatness calibration error, and the other is the complexity of broadband impedance matching between analog devices.
The latter factor has been addressed in e.g., \cite{2012MNRAS.419.1070H}.
We choose to not calculate the result with the equations in their work because besides the impedance mismatch, and there are other complex factors that will make the result hard to be represented in any analytical form.

\subsubsection{Frequency-Dependent Antenna Beam Pattern}
If in principle, analog frontend gain can always be calibrated with sufficient accuracy, FDAP is inevitable.
According to antenna theory, when the antenna has a size to wavelength ratio $L/\lambda \ll 1 $, it would become the so-called electrically small antenna, the beam pattern of which is nearly omnidirectional.
So it is plausible that in order to suppress effects caused by the FDAP, smaller antennas are preferable.
Besides the antenna size, we also show that the ground effect can cause even more severe FDAP (see Section \ref{ssec:antenna_on_ground}). 
The reason is straightforward---the interval between the antenna and the earth can form a structure with a size comparable to the wavelength. 
This could be a big challenge that prevents ground-based instruments from detecting the global $\HI$ signal. 
In order to make ground-based instruments suitable for global $\HI$ signal detection, a smaller distance between the antenna and the ground plane can be a good choice. 
Actually among several currently running experiments that are devoted to the detection of global 21 cm signal, the antenna design of EDGES \citep[][]{2016MNRAS.455.3890M}, PAPER \citep{2011AAS...21743206P}, and BIGHORNS \citep{2015PASA...32....4S} seems to fulfill the above requirements about the antenna size.
It could be helpful for us to get more realistic results if we can take into account the actual mechanical models of the above antennas in our simulation.

There should be other ways to solve the FDAP problem.
One method is to used digitized beamforming arrays because digitized beamforming arrays can adjust the complex gain (both amplitude and phase) of each antenna element to form the desired antenna beam that is independent of frequency.
However, instrument calibration will become a more significant challenge for such kinds of complex systems.
Another method is to scan as large as possible area of the sky so that the FDAP induced effects could be smeared out, either by putting the instrument in sites with different latitudes or for space-based detectors, pointing the antenna uniformly to different directions.

\subsection{Which Set of Free Parameters to be Used?}
As we have mentioned in section \ref{ssec:ref_model}, the difference between selecting different sets of free parameters can be nontrivial. 
In our above tests, we choose $f^\prime_{\rm esc}$, $f^\prime_{\rm esc, LW}$, and $f^\prime_{\rm X}$ to be free parameters and assume their priori distributions to be a uniform distribution.
On the other hand, there is another possible choice: directly sampling $f_{*}$, $f_{\rm esc}$, and $f_{\rm esc, LW}$ and fix $f_{\rm X}=1$, still assuming their priori distributions to be uniform, then the obtained corresponding distribution of ($f^\prime_{\rm esc}$, $f^\prime_{\rm esc, LW}$, $f^\prime_{\rm X}$) will be definitely not same as the results we have presented above.
The reason is obvious: if $f_{*}$ and $f_{\rm esc}$ (or $f_{\rm esc, LW}$) both independently follow the uniform distribution, their product $f^\prime_{\rm esc}$ (or $f^\prime_{\rm esc, LW}$) will not follow the uniform distribution.

If we do not have enough priori knowledge, assuming the chosen parameters to follow uniform distributions independently seems to be a rather natural decision.
However, this decision can cause a potential problem.
Different choices of free parameters can lead to different estimation results.
In this work, we are not able to answer the question of how to choose the free parameters.

\subsection{Using Priori knowledge from Other Observations}
As as we have pointed out in section \ref{ssec:overfitting}, when $N_{\rm poly}\ge 4$ is used in the  foreground component model, the parameters cannot be tightly constrained. This phenomenon is most apparent for parameter $f_{\rm esc}^\prime$. 
From Figures \ref{fig:bl7}, \ref{fig:bl7nn}, and \ref{fig:bl4nn}, it can be seen that parameter distributions deviate from normal-like distributions.
And by examining the joint distributions of $f_{\rm esc}^\prime$ and $f_{\rm X}^\prime$ in the above figures, one will easily find a strong correlation between these two parameters.
This phenomenon suggests that these two parameters can play some equivalent roles in forming the global $\HI$ spectrum, especially for the absorption trough feature.
It could be possible to use the priori knowledge from other observations as further constraints to decouple these two parameters.
Recalling that $f_{\rm esc}^\prime$ mainly affect the process of ionizing, the CMB optical depth
\begin{gather}
    \tau_{\rm CMB}(z)=n_{\rm H}(0)c\sigma_{\rm T}\int_{z}^{z_{\max}}dz \bar{x}_i(z)\frac{(1+z)^2}{H(z)},
\end{gather}
where $n_{\rm H}(0)$ is the number density of hydrogen atoms at $z=0$, $c$ is the speed of light, $\sigma_{\rm T}$ is the Thomson scattering cross-section, $H(z)$ is the Hubble parameter at redshift $z$, can be used to impose further constraint on this parameter.
This can be verified by examining corresponding $\tau_{\rm CMB}$ values of different $f_{\rm esc}^\prime$.
For example, the reference model value $f_{\rm esc}^{\prime}=0.005$ corresponds to $\tau_{\rm CMB}(10.8)=0.012$, which does not conflict with the Planck result $\tau_{\rm CMB}^{\rm Planck}(0)=0.054\pm0.007$ \citep{2018arXiv180706209P}.
If we increase $f_{\rm esc}^\prime$ by one order of magnitude to $0.05$, then a $\tau_{\rm CMB}(10.8)=0.071>\tau_{\rm CMB}^{\rm Planck}(0)$ is obtained, which can immediately be ruled out.
Increasing $f_{\rm esc}^\prime$ further by another one order of magnitude to $0.5$ will result in a $\tau_{\rm CMB}(10.8)=0.133$.
In this condition, the global $\HI$ signal reaches zero shortly after the absorption peak.
Though we only use the CMB optical depth at $z=0$  from the Planck data as an example here, a refined optical depth at higher redshifts can calculated by properly modeling the electron number density at lower redshifts and subtracting their contribution to the total value, which can be used as a tighter constraint to the global $\HI$ signal model parameters.

\subsection{The Performance of Simplified Gaussian Model}
As was introduced in \citep{2016MNRAS.461.2847B}, a simplified model that is based on Gaussian function can be a possible method to extract information from global 21 cm signal from the cosmic dawn. We test the Gaussian model and compare the results with the previous physical model here.

We use the reference model that we have simulated above, with the parameters listed in Table \ref{tbl:ref_model_param}. Same as \cite{2016MNRAS.461.2847B}, the Gaussian model is expressed as 
\begin{gather}
    \delta T_b(\nu)=A_{\HI} e^{-\frac{(\nu-\nu_{\HI})^2}{2\sigma_{\HI}}},\label{eqn:Gaussian_signal}
\end{gather} 
where the amplitude $A_{\HI}$, peak position $\nu_{\HI}$, and standard deviation $\sigma_{\HI}$ are to be inferred through the MCMC sampling. 
Foreground model is still the 7th-degree log-polynomial same as we used in previous sections. 
For the Gaussian model is only suitable to approximate the 21 cm signal from pre-reionization epoch, it is necessary to set an upper frequency limit $\nu_{\max}$.
We set $\nu_{\max}$ to be $100$, $105$, $110$, and $120$ MHz according to our knowledge about the simulated signal, and perform the test.
    
\begin{figure}
    \centering
    \includegraphics[width=.8\columnwidth]{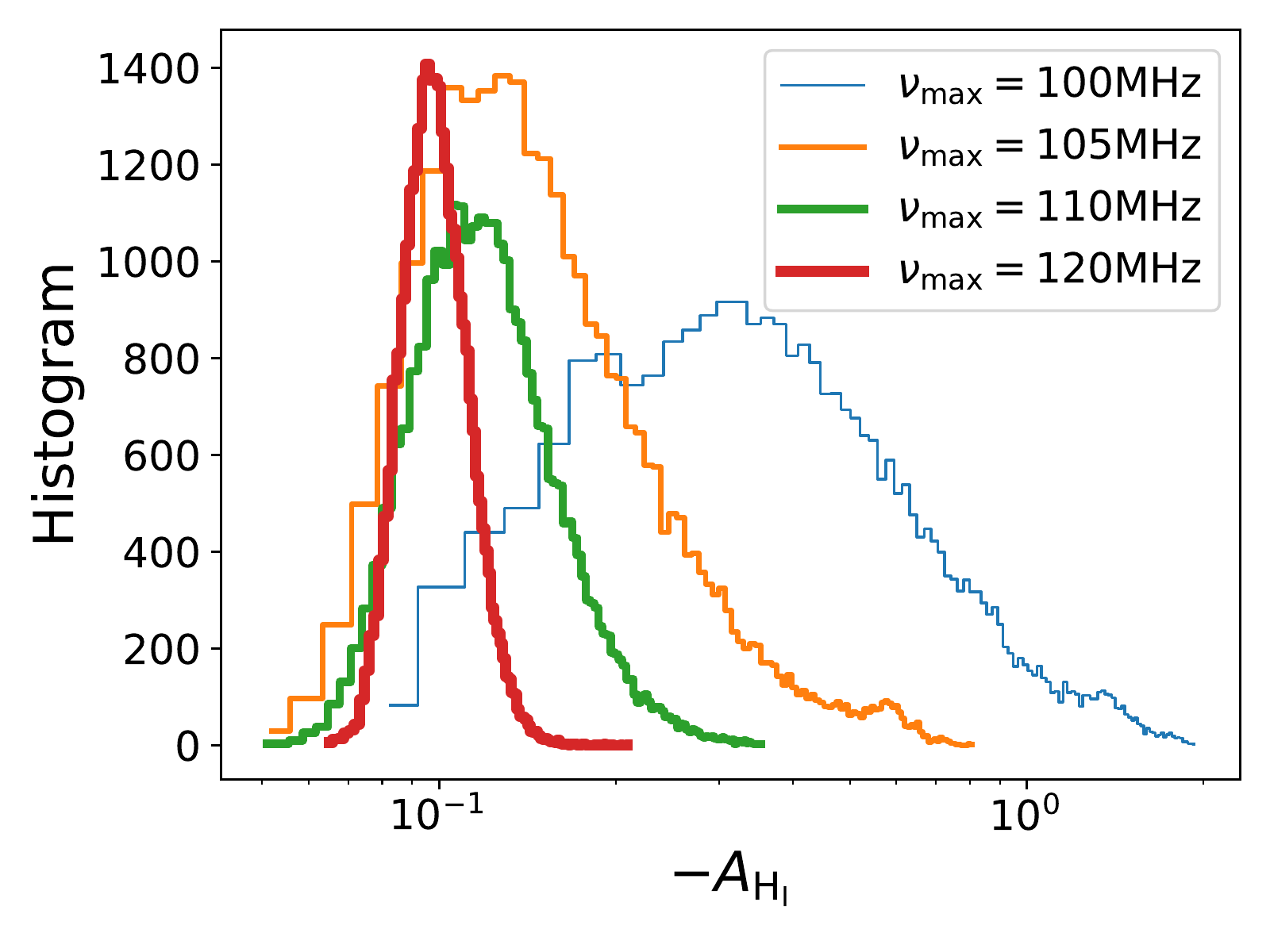}
    \caption{\label{fig:A_hist} The histograms of $A_{\HI}$ in the Gaussian model sampled assuming
    $\nu_{\max}=100, 105, 110$ and $120$ MHz, respectively.}
\end{figure}

\begin{figure}
    \centering
    \includegraphics[width=.48\columnwidth]{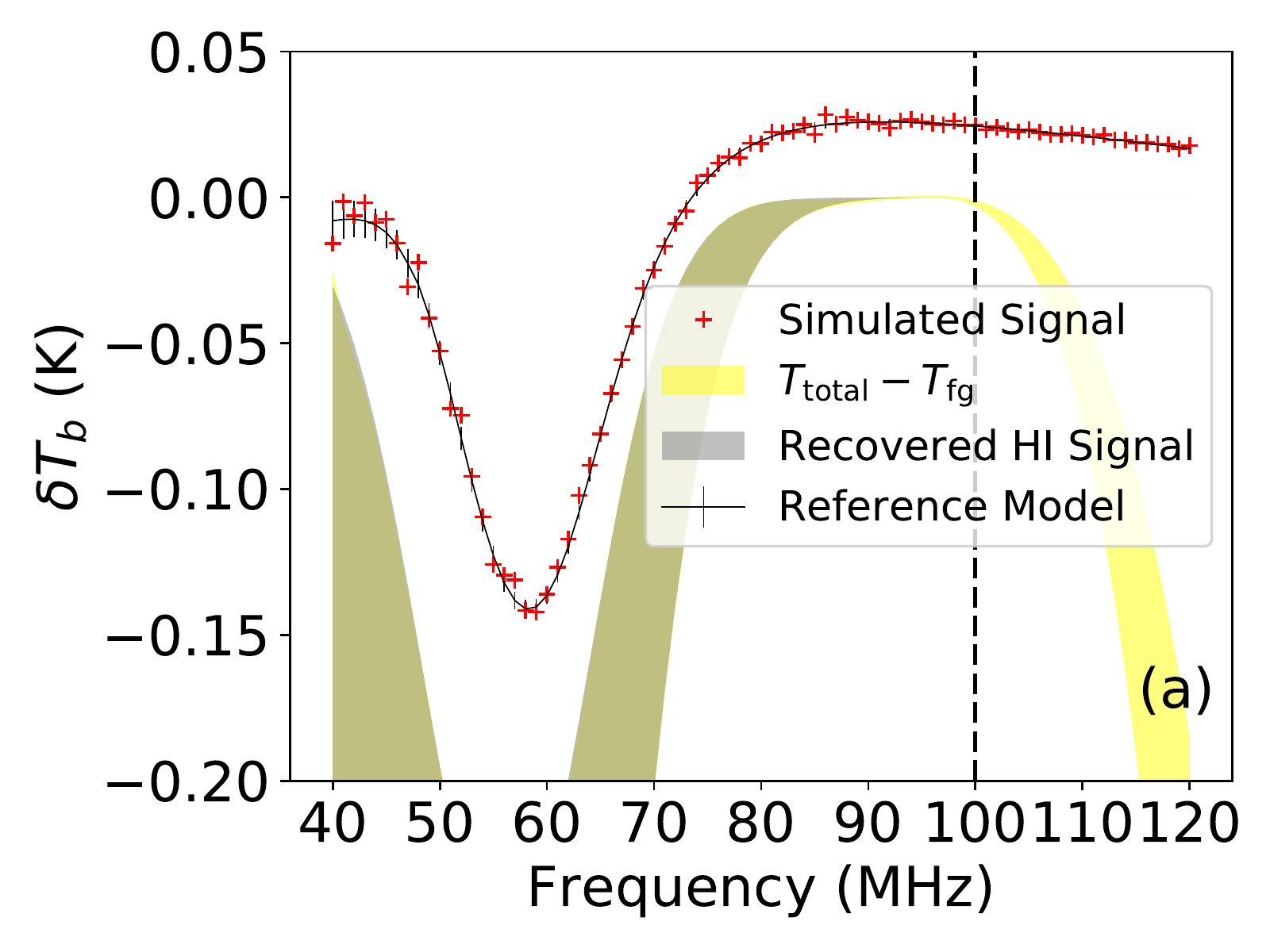}
    \includegraphics[width=.48\columnwidth]{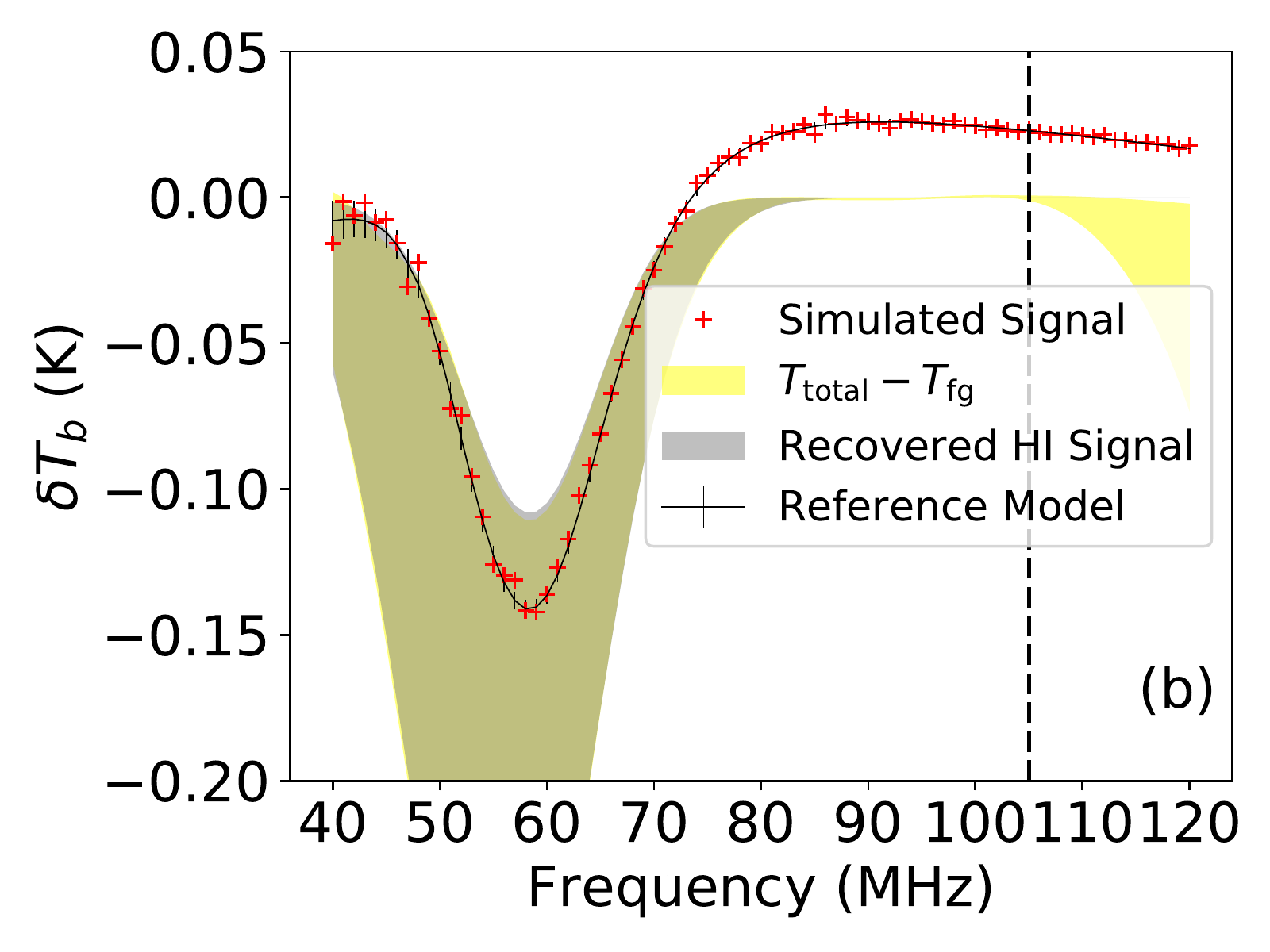}\\
    \includegraphics[width=.48\columnwidth]{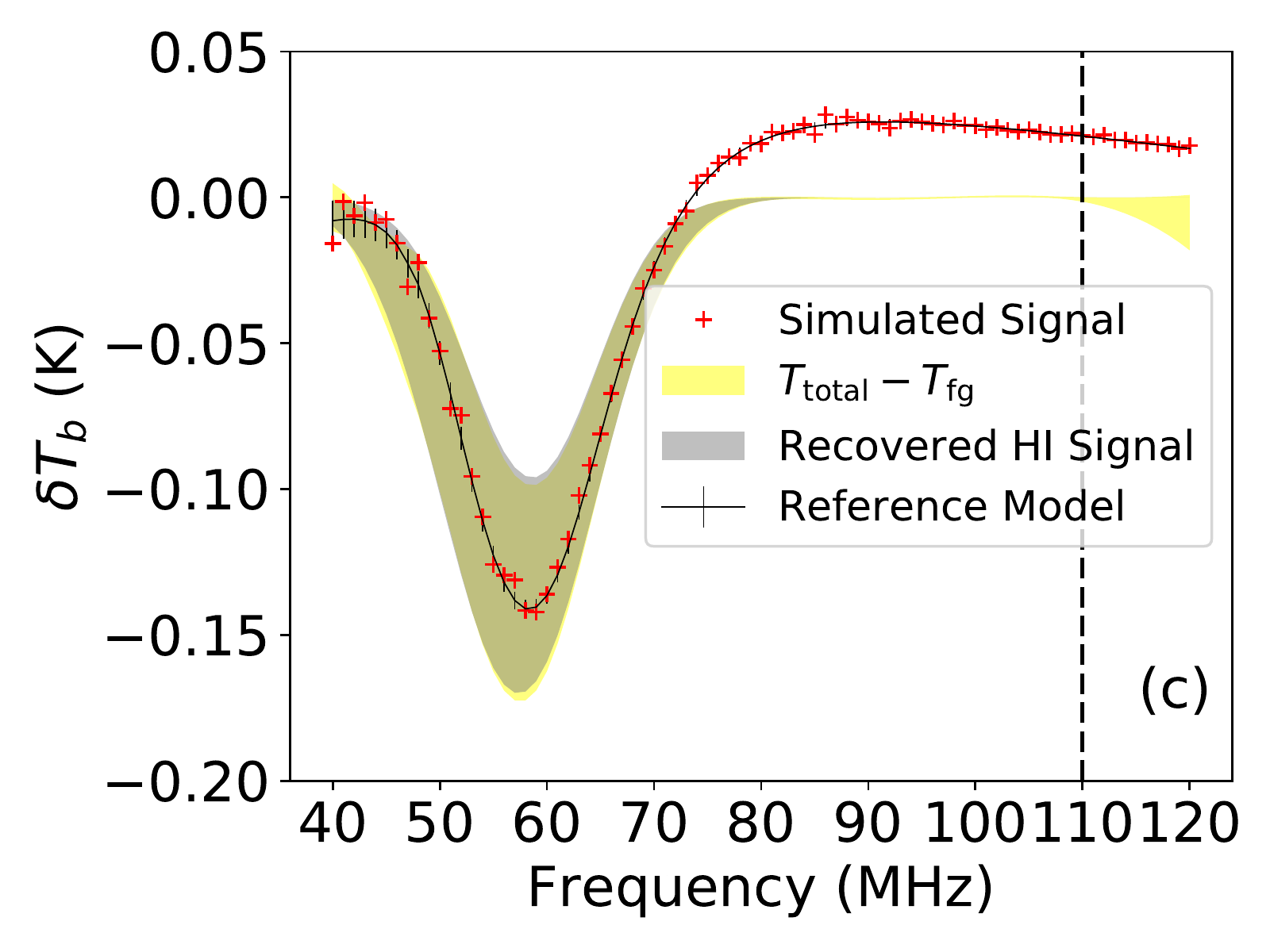}
    \includegraphics[width=.48\columnwidth]{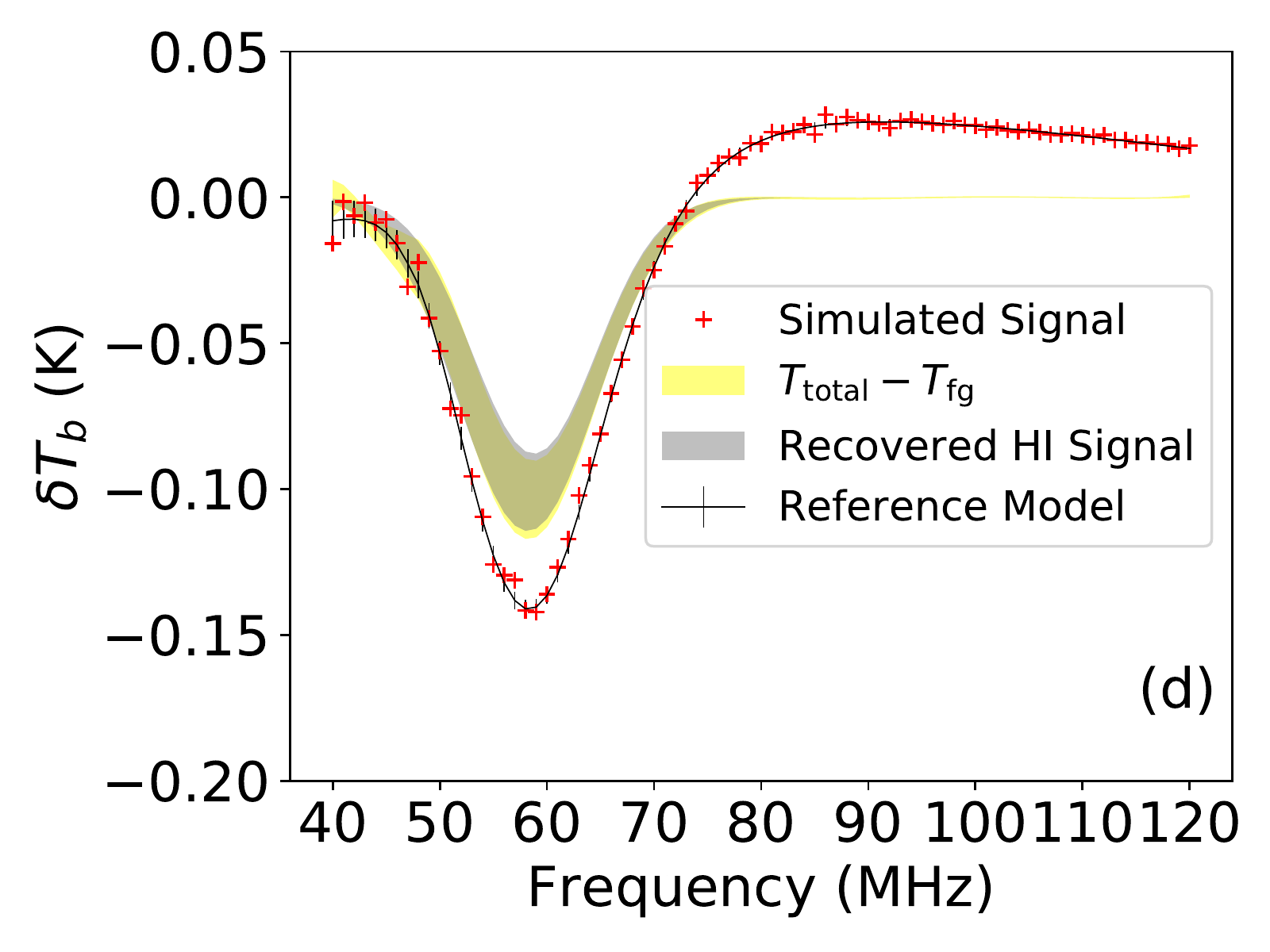}
    \caption{\label{fig:signal_gauss}
        Recovered $\HI$ signals by using Gaussian model assuming $\nu_{\max}=100, 105, 110$ and $120$ MHz, respectively. The vertical dashed lines denote the $\nu_{\max}$. Grey bands are the 68\% confidence ranges of the signals calculated with Equation \ref{eqn:Gaussian_signal} by using the sampled model parameters. Yellow bands are the signals calculated as the differences between total signal and the foreground model values. The black solid lines denote the input global $\HI$ signal model. The red crosses denote the sum of the input signal and the actually simulated noise.
    }
\end{figure}

\begin{figure}
    \centering
    \includegraphics[width=.48\columnwidth]{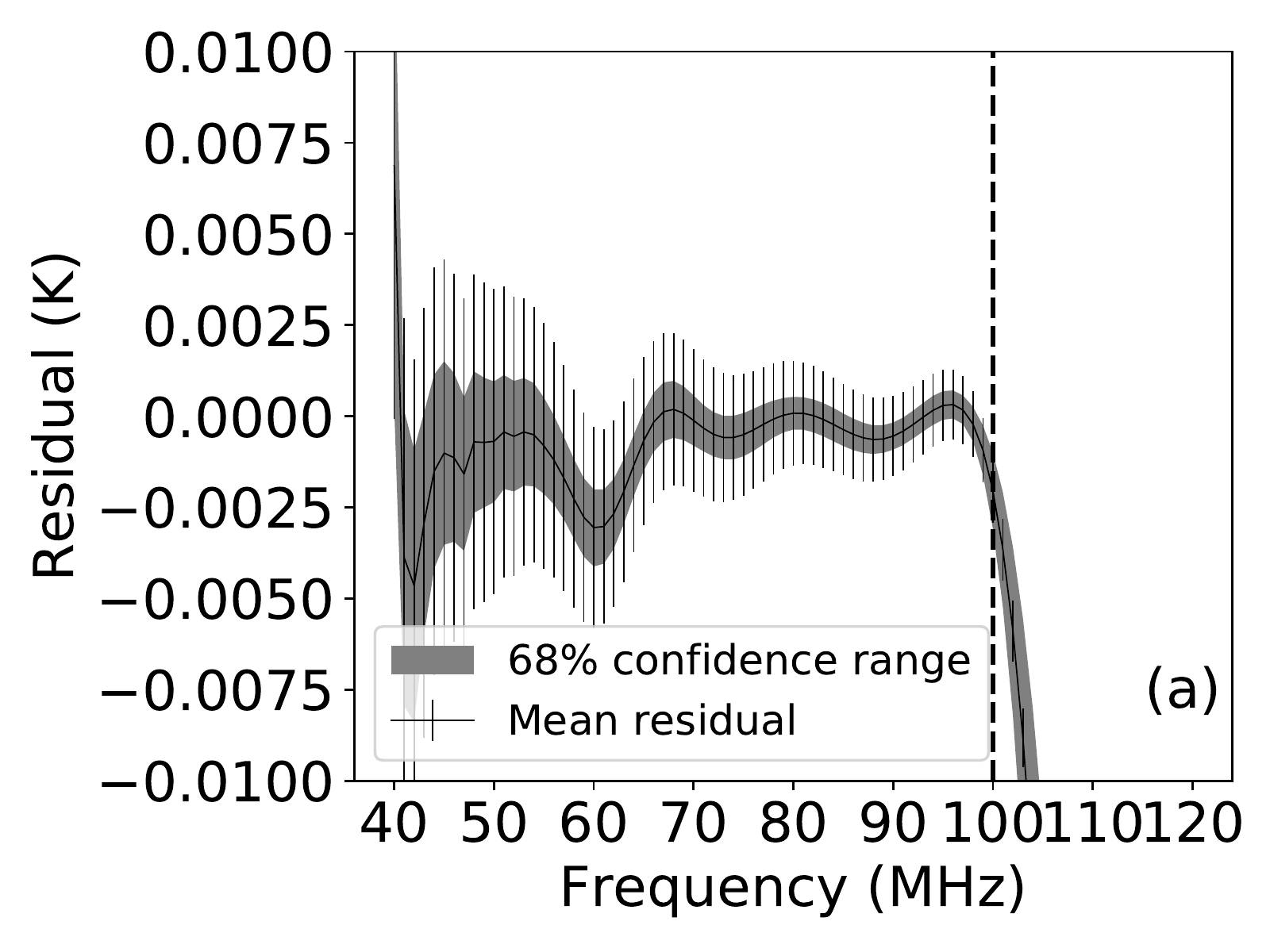}
    \includegraphics[width=.48\columnwidth]{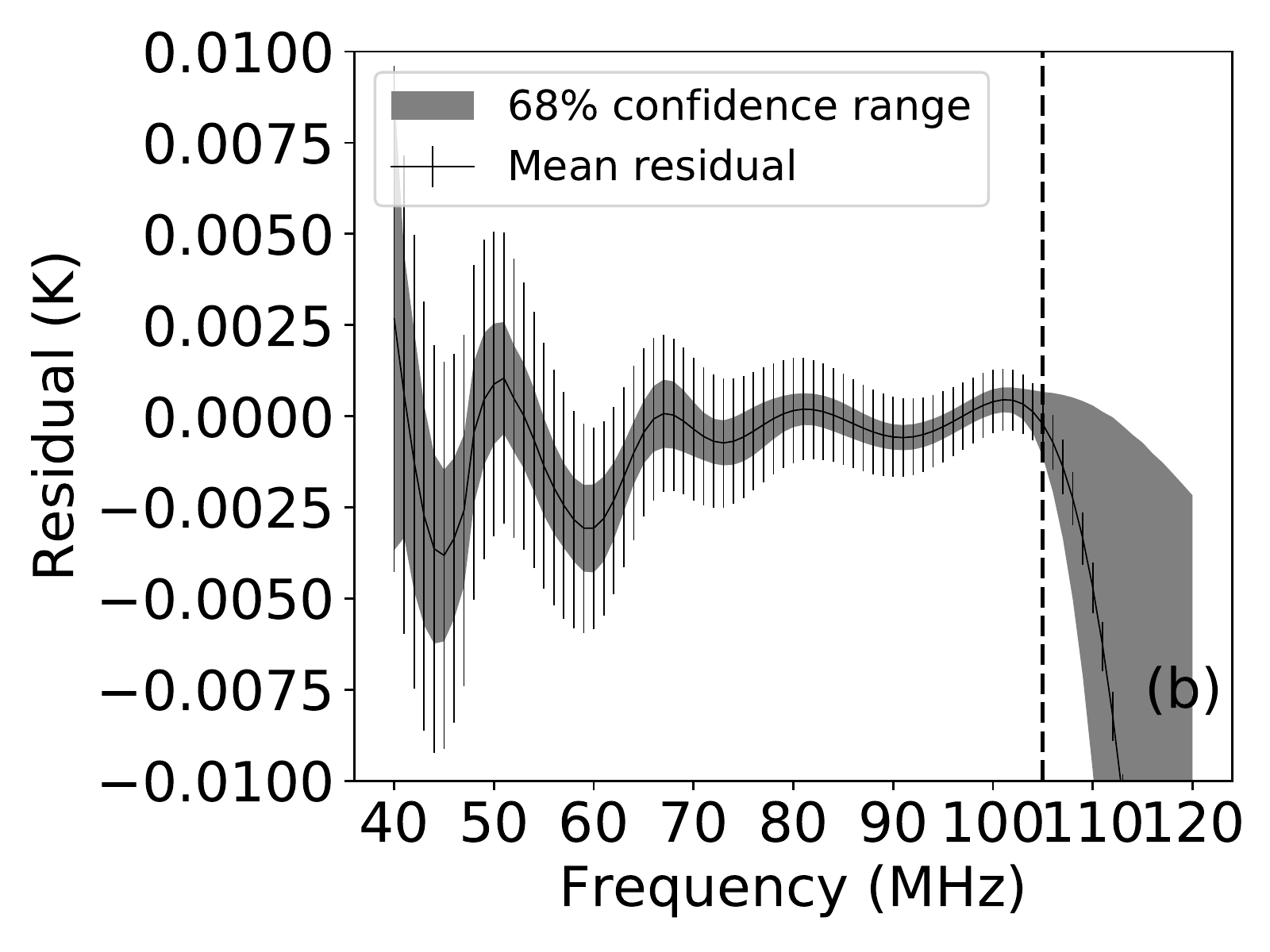}\\
    \includegraphics[width=.48\columnwidth]{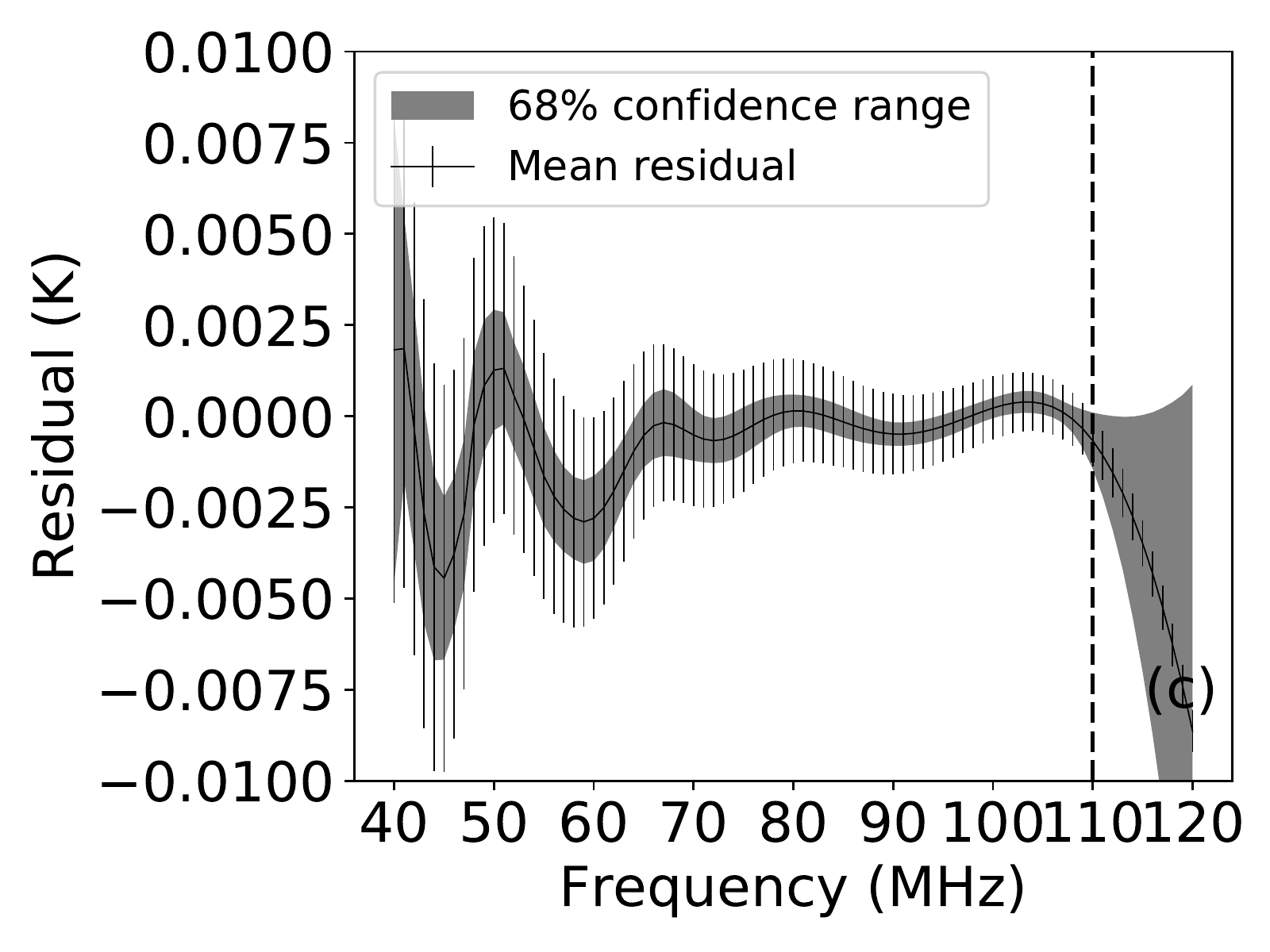}
    \includegraphics[width=.48\columnwidth]{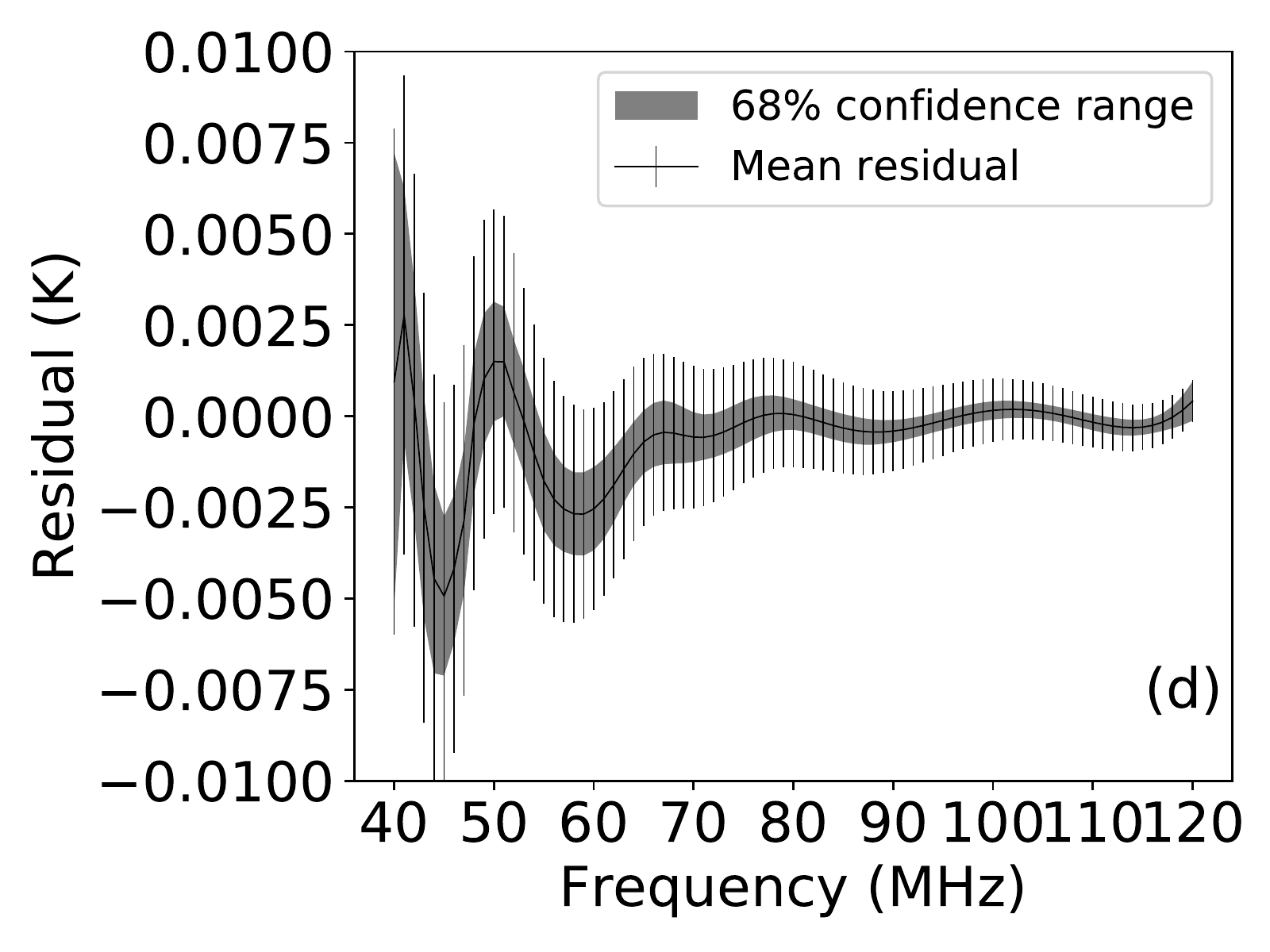}
    \caption{\label{fig:resid_gauss}
    The distributions of the residuals between the total signal and the model predicted values assuming $\nu_{\max}=100, 105, 110$ and $120$ MHz, respectively. The vertical dashed lines denote the $\nu_{\max}$.
    The $\HI$ signal is approximated with a Gaussian model.
    The residuals defined as $T_{\rm total}-T_{\rm fg}-\delta T_{b}$, where $T_{\rm total}$ is the total signal and $\delta T_{b}$ is calculated with Equation \ref{eqn:Gaussian_signal}.
    The solid lines are the mean residual calculated over all the sampled parameters.
    The error bars represent the standard deviation of noise.
    The grey bands denotes the 68\% confidence ranges of the residuals.
    }
\end{figure}

We show the histograms of $A_{\HI}$ in Figure \ref{fig:A_hist}. It can be seen clearly that the distribution of $A_{\HI}$ varies significantly with $\nu_{\max}$.
In other words, the upper frequency limit can heavily affect the level of the absorption feature of $\HI$ signal.
Then we study the recovered $\HI$ signal, which is plotted in Figure \ref{fig:signal_gauss}. 
Obviously, if correctly choosing the upper frequency limit, we can recover the $\HI$ signal with a relatively small error. 
We show the residual between the total signal and the sum of $\HI$ model component and the foreground model component in Figure \ref{fig:resid_gauss}, to check whether we can diagnose the results in actual observations. 
From the figure, we find that within the used frequency ranges, the residuals distribute almost symmetrically, and do not show obvious difference between test cases with different $\nu_{\max}$. 
The residuals of all four $\nu_{\max}$ values also show oscillating features.

\begin{figure}
    \centering
    \includegraphics[width=.48\columnwidth]{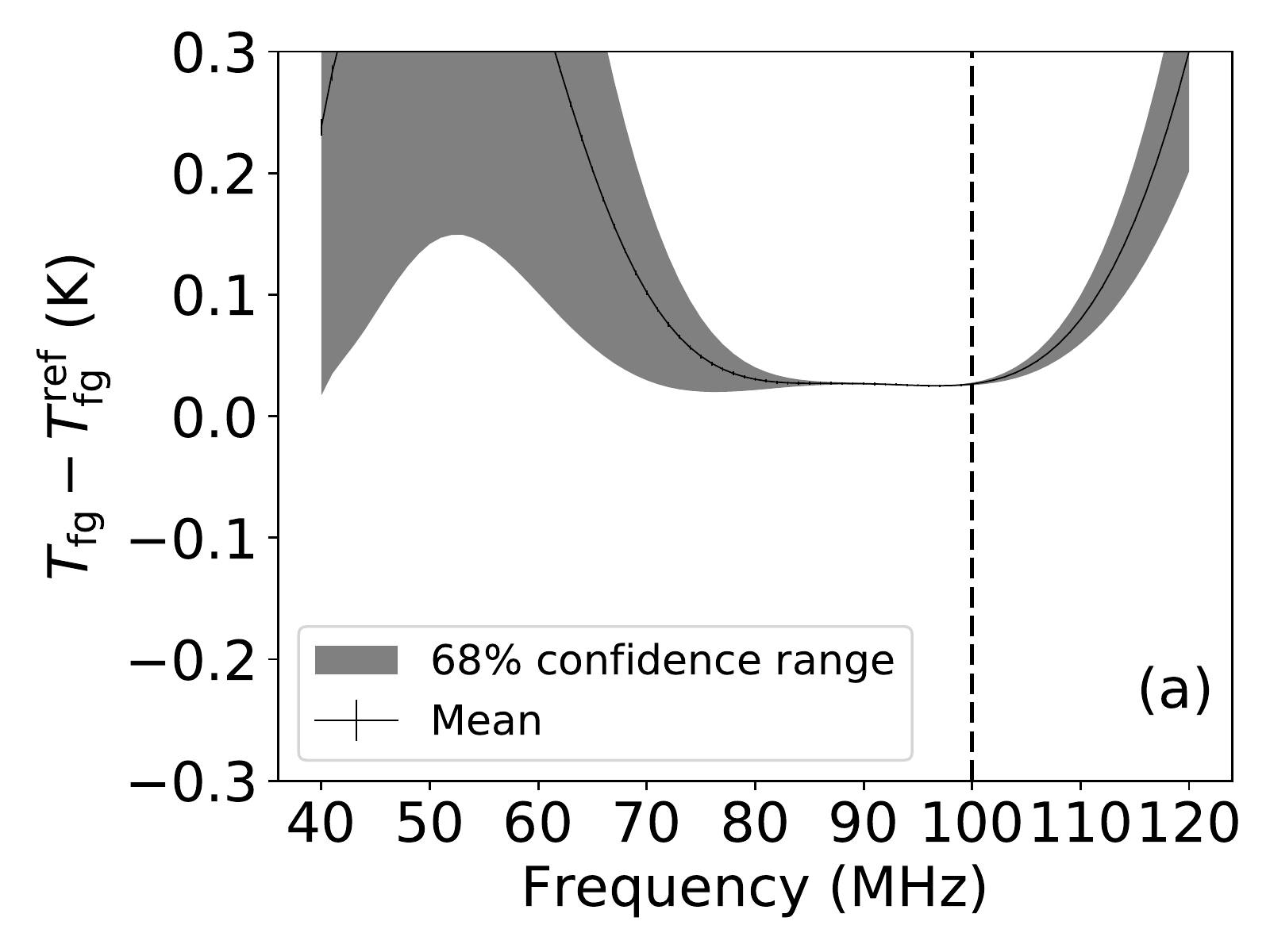}
    \includegraphics[width=.48\columnwidth]{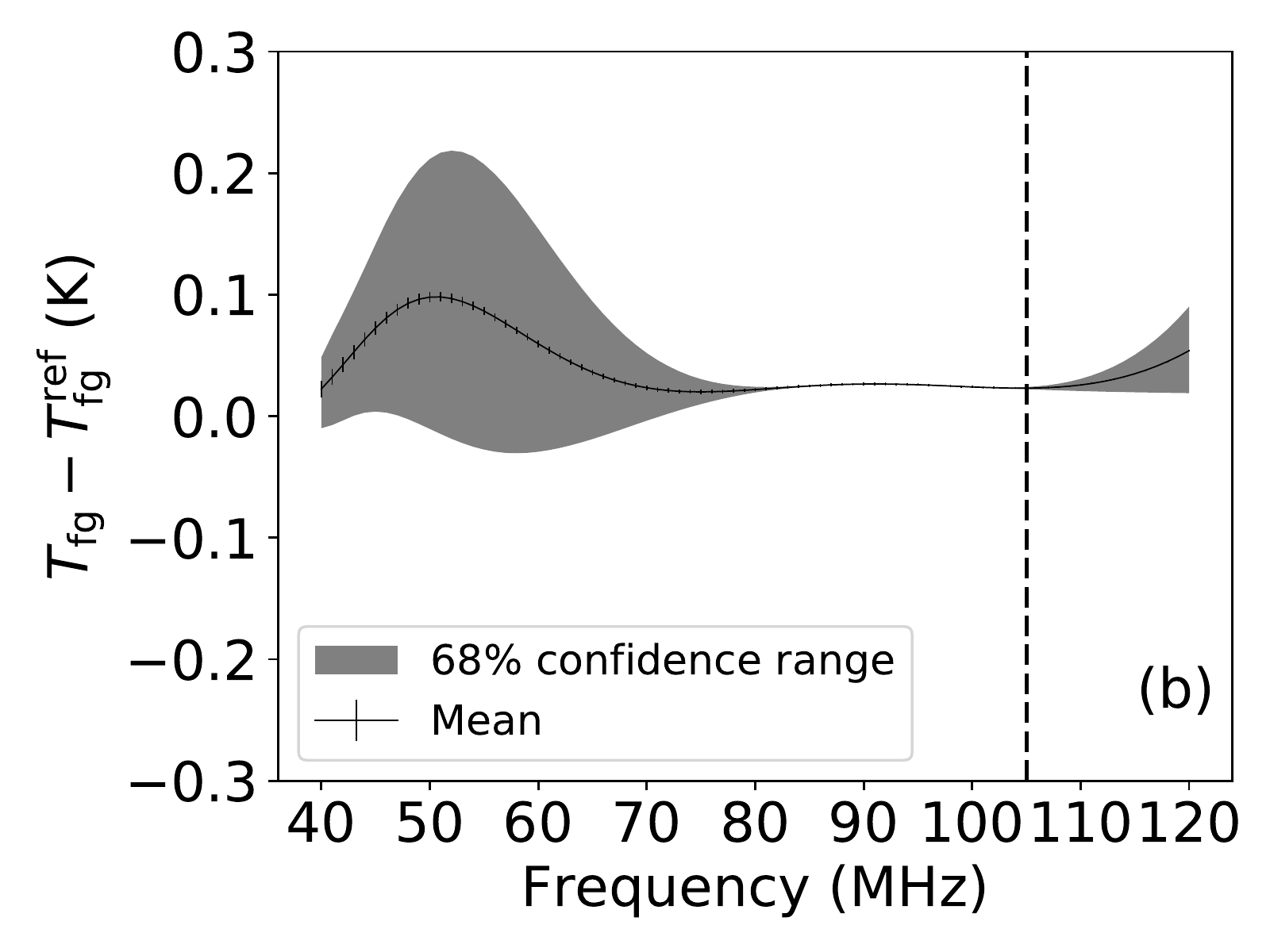}\\
    \includegraphics[width=.48\columnwidth]{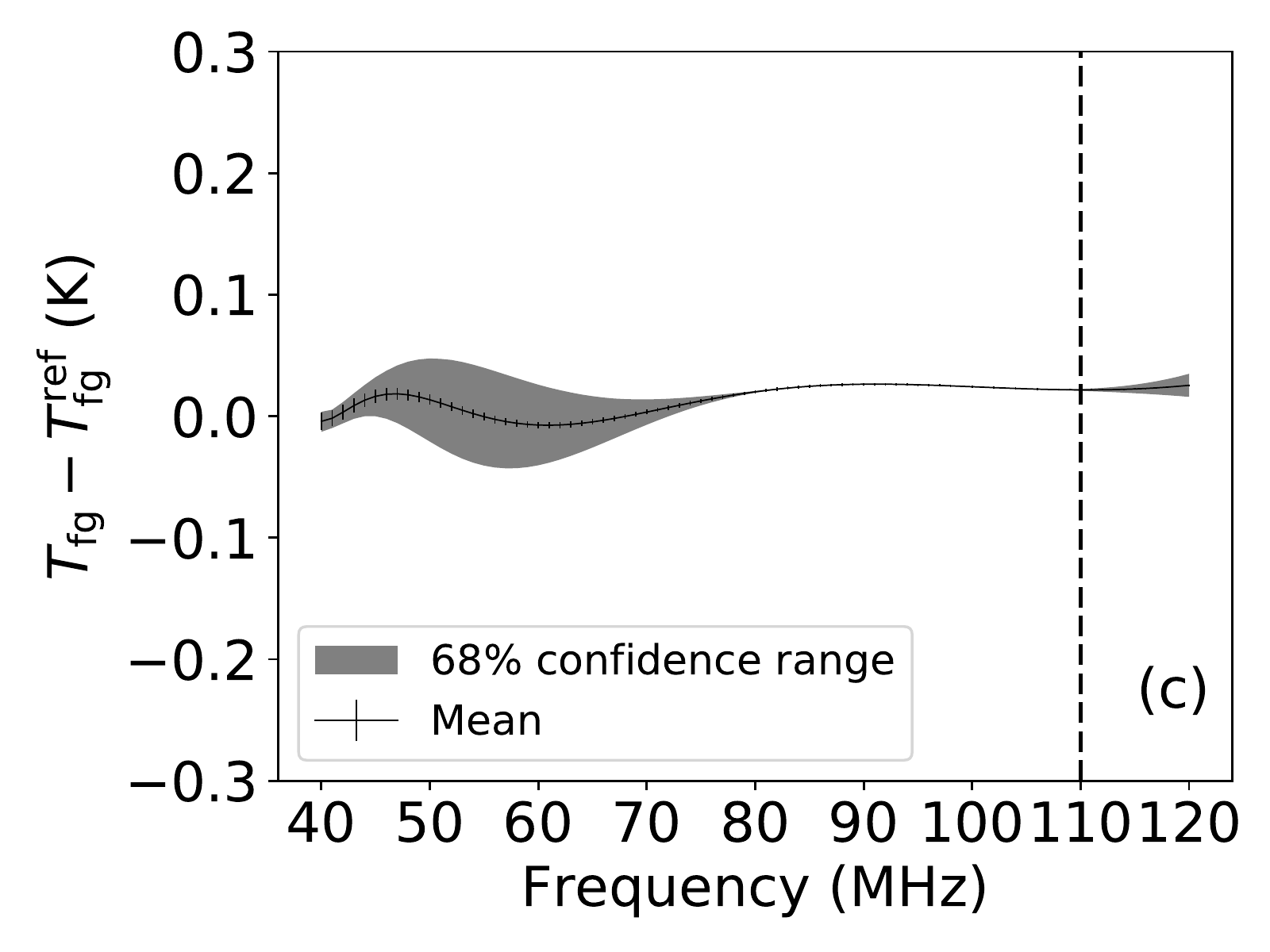}
    \includegraphics[width=.48\columnwidth]{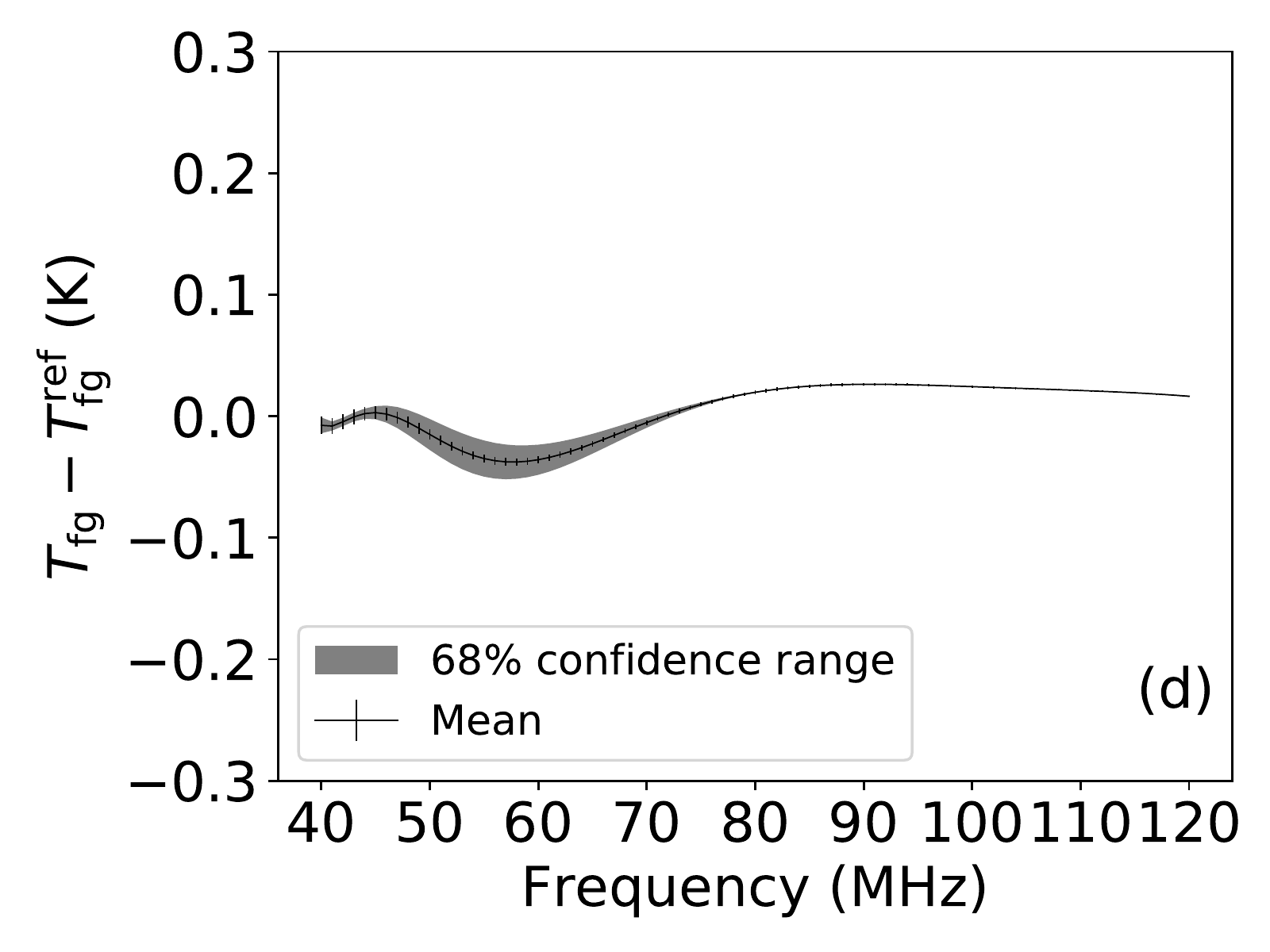}
    \caption{\label{fig:fgdiff_gauss}
        The errors of estimations to the foreground component in the conditions of $\nu_{\max}=100, 105, 110$ and $120$ MHz, respectively. The vertical dashed lines denote the $\nu_{\max}$. The solid lines represent the mean residual and the grey bands denote the 68\% confidence range.
    }
\end{figure}

It would be interesting to check the residual between the recovered foreground component and the input model, which we show in Figure \ref{fig:fgdiff_gauss}. 
Obviously for $\nu_{\max}=100$ and $120$ biased foreground estimations are obtained.
The most significant deviation from the reference value appears right at the frequency of absorption feature of the $\HI$ signal, which happens to compensate for the biased $\HI$ signal estimation and lead to the symmetric residual distributions as are shown in Figure \ref{fig:resid_gauss}. 
Hence, although both the recovered $\HI$ and foreground components are biased, the total recovered model component show no apparent biasing evidence, except for some oscillation features, and the oscillation can also be caused by the instrument calibration inaccuracy. 
So we have to conclude that although modeling the $\HI$ signal in the early stage of EoR can significantly simplify the parameter inference and quickly obtain the results, it is still risky and mostly suitable only for qualitative analyses.

\subsection{A Caveat about this Work}
In the previous sections, we do confirm that with a direct modeling method and proper sampling algorithm, it is possible to obtain unbiased parameter estimation from the global $\HI$ signal.
This also partially answers the question raised by \cite{2016MNRAS.455.3829H} that whether the inference result would be biased if the signal were fitted with the exact the model used to generate it.
However, we shall point out that we cannot test all possible reionization models and there may still exist some reionization models that could not well constrained by the observations to global $\HI$ signals, and imaging detection or power spectra measuring are required to constrain the parameters of those models.
The result of this work should be treated as a test to some specific reionization scenarios.
Nevertheless, the method used in this paper is still useful to test if a direct one-stage MCMC sampling works in the condition of other models that are not covered in this work.

\subsection{Some Practical Suggestions of Performing MCMC Sampling}
Last but not least, we would like to give some practical suggestions for MCMC sampling.
It is true that in an ideal condition when the parameter space is not multimodal, any casually chosen initial parameter can finally lead to convergence. However, our experience shows that in practical conditions, choosing a `good' initial parameter is still essential.
We find that when instrumental gain calibration error or FDAP involves, an arbitrarily chosen initial parameter may require a rather large number of sampling before convergence.
A practical method is to find the foreground polynomial coefficients through an ordinary least square fitting with polynomial fitting to the total signal and use the fitting result as the foreground parameters' initial values.

\section{Conclusions}
We survey the feasibility of directly inferring physical model parameters from global EoR signal data and the influences of some factors that should be considered in actual observations to the model parameter inference.
We find that if only thermal noise is considered, unbiased model parameters can be inferred.
The inaccuracy of analog frontend and frequency-dependent antenna beam can bias the results on different levels.
Even a relative amplifier gain calibration error of $10^{-5}$ can significantly influence the inferred model parameters given that a low degree polynomial cannot well fit the calibration error as a function of frequency.
The FDAP should be carefully considered to prevent it from creating artificial spectral structures.
Geographical position can also cause differences in the difficulties of obtaining correct model parameters. 
The northern hemisphere seems to be a better choice than the southern hemisphere because of the distribution of the brightness of the Milkyway.
Finally, we give some practical suggestions about performing MCMC sampling and parameter inference in global EoR signal detections.

\section*{Acknowledgements}
We are sincerely grateful to the referee for constructive and valuable suggestions and comments, which help us improve the manuscript.

We thank the high-performance computing center of the National Astronomical Observatories, Chinese Academy of Sciences for providing us with computing resources.
We thank the SKA Regional Center prototype developed by Shanghai Astronomical Observatory for providing us with computing resources.
This work was supported by the National Key R\&D Programme of China  (Grant No. 2018YFA0404600) and the Strategic Priority Research Program of Chinese Academy of Sciences (Grant No. XDB23010300).

%%%%%%%%%%%%%%%%%%%%%%%%%%%%%%%%%%%%%%%%%%%%%%%%%%

%%%%%%%%%%%%%%%%%%%% REFERENCES %%%%%%%%%%%%%%%%%%

% The best way to enter references is to use BibTeX:

\bibliographystyle{mnras}
\bibliography{ms}

% Don't change these lines
\bsp	% typesetting comment
\label{lastpage}
\end{document}

%% file: fig_dependences.tex
\scriptsize  
\begin{tikzpicture}[node distance=8mm,  
auto,>=latex',  
thin,  
start chain=going below,  
every join/.style={norm}]
\node [ellipse, draw] (J_nu) {$J_\nu$};
\node [ellipse, draw] (tau_nu) [above=of J_nu] {$\tau_\nu$};
\node [ellipse, draw] (epsilon_X) [below=of J_nu] {$\epsilon_X$};
\node [ellipse, draw] (gammas) [left=of J_nu] {$\Gamma~\&~\gamma$};
\node [ellipse, draw] (xHI) [left=of tau_nu] {$x_{\HI}$};
\node [ellipse, draw] (epsilon_comp) [left=of gammas] {$\epsilon_{\rm comp}$};
\node [ellipse, draw] (AlphaBeta) [left=of xHI] {$\alpha^B~\&~\beta$};
\node [ellipse, draw] (SourceModel) [right=of tau_nu] {source model};
\node [ellipse, draw] (J_alpha) [right=of J_nu] {$J_\alpha$};
\node [ellipse, draw] (x_alpha) [right=of epsilon_X] {$x_\alpha$};
\node [ellipse, draw] (T_K) [below=of epsilon_X] {$T_K$};
\node [ellipse, draw] (T_S) [right=of T_K] {$T_S$};
\node [ellipse, draw] (x_c) [right=of x_alpha] {$x_c$};
\node [ellipse, draw] (dTb) [below=of T_S] {$\delta T_b$};
\draw [->] (J_nu)--(epsilon_X);
\draw [->] (xHI)--(J_nu);
\draw [->] (xHI)--(tau_nu);
\draw [->] (AlphaBeta)--(xHI);
\draw [->] (tau_nu)--(J_nu);
\draw [->] (J_nu)--(gammas);
\draw [->] (gammas)--(xHI);
\draw [->] (xHI)--(epsilon_comp);
\draw [->] (epsilon_comp)--(T_K);
\draw [->] (T_K)--(AlphaBeta);
\draw [->] (epsilon_X)--(T_K);
\draw [->] (SourceModel)--(J_nu);
\draw [->] (SourceModel)--(J_alpha);
\draw [->] (J_alpha)--(x_alpha);
\draw [->] (T_K)--(x_c);
\draw [->] (x_alpha)--(T_S);
\draw [->] (T_K)--(T_S);
\draw [->] (x_c)--(T_S);
\draw [->] (xHI)-- (x_c);
\draw [->] (x_c)--(T_S);
\draw [->] (x_alpha)--(T_S);
\draw [->] (T_S)--(dTb);
\draw [->] (xHI) edge [ bend right=70] (dTb);
\end{tikzpicture}

%% file: fig_flow_chart.tex
\scriptsize  
\begin{tikzpicture}[node distance=8mm,  
auto,>=latex',  
thin,  
start chain=going below,  
every join/.style={norm},]  
\node[format] (n_begin){Begin};
\node[format,below of=n_begin] (n_set_z){set $z=z_{\rm f}$};
\node[test,below of=n_set_z] (n_while){$z>z_{\rm min}$ ?};
\node[format,right of=n_while, xshift=2cm] (n_end){end};
\node[format,below of=n_while, yshift=-0.5cm] (n_Jdz){calculate $\hat{J}_\nu$, $\hat{J}_\alpha$, and $\D z$};

\node[format,below of=n_Jdz] (n_coeff){update $\Gamma$, $\gamma$, etc. }; \node[format,left of=n_coeff,
xshift=-2.5cm] (n_output){output $z,dT_b$};
\node[format,below of=n_coeff]
(n_rate_equation){evolve
$x_{\HI}$, $x_{\HII}$,  $x_i$, and $T_k$
to $z+\D z$}; \node[format,below of=n_rate_equation]
(n_ion_history){update ionization history for RTE};
\node[format,below of=n_ion_history] (n_nextz){$z\leftarrow z+\D z$};

\draw[->] (n_begin.south) -- (n_set_z.north);
\draw[->] (n_set_z) -- (n_while);
\draw[->] (n_while) -- node(Yes){Yes} (n_Jdz);
\draw[->] (n_Jdz) -- (n_coeff);
\draw[->] (n_coeff) -- (n_rate_equation);
\draw[->] (n_rate_equation) -- (n_ion_history);
\draw[->] (n_ion_history) -- (n_nextz);
\draw[->] (n_nextz.west) -| (n_output.south);
\draw[->] (n_output.north) |- (n_while);
\draw[->] (n_while.east) -- node(No){No}(n_end);
\end{tikzpicture}